\documentclass[12pt]{article}
\usepackage{mathrsfs}
\makeatletter \@addtoreset{equation}{section} \makeatother
\renewcommand{\theequation}{\thesection.\arabic{equation}}
\newcommand{\ba}{\begin{array}}
\newcommand{\ea}{\end{array}}
\newcommand{\beq}{\begin{equation}}
\newcommand{\eeq}{\end{equation}}
\newcommand{\bea}{\begin{eqnarray}}
\newcommand{\eea}{\end{eqnarray}}
\def\bce{\begin{center}}
\def\ece{\end{center}}

\def\nonu{\nonumber}

\def\pa{\partial}
\def\al{\alpha}
\def\be{\beta}
\def\ga{\gamma}

\def\de{\delta}

\def\ep{\epsilon}

\def\la{\lambda}

\def\eps6{{\displaystyle \mathop{\epsilon}^{6}}{}}
\def\g6{{\displaystyle \mathop{g}^{6}}{}}

\def\nab6{{\displaystyle \mathop{\nabla}^{6}}{}}
\def\0{{\sst{(0)}}}
\def\1{{\sst{(1)}}}
\def\2{{\sst{(2)}}}
\def\3{{\sst{(3)}}}
\def\4{{\sst{(4)}}}
\def\5{{\sst{(5)}}}
\def\6{{\sst{(6)}}}
\def\7{{\sst{(7)}}}
\def\8{{\sst{(8)}}}

\def\ba{\begin{array}}
\def\ea{\end{array}}
\def\beq{\begin{equation}}
\def\eeq{\end{equation}}
\def\be{\begin{equation}}
\def\ee{\end{equation}}

\def\la{\lambda}
\def\eps{\epsilon}

\def\ba{\begin{array}}
\def\ea{\end{array}}
\def\beq{\begin{equation}}
\def\eeq{\end{equation}}
\def\be{\begin{equation}}
\def\ee{\end{equation}}

\def\la{\lambda}
\def\eps{\epsilon}

\def\eps6{{\displaystyle \mathop{\epsilon}^{6}}{}}

\def\nab6{{\displaystyle \mathop{\nabla}^{6}}{}}

\newcommand{\bean}{\begin{eqnarray*}}
\newcommand{\eean}{\end{eqnarray*}}

\begin{document}
\thispagestyle{empty} \addtocounter{page}{-1}
   \begin{flushright}
%PUPT-2395 \\
%CALT-68-nnnn \\
%{\tt hep-th/yymmnnn}\\
\end{flushright}

\vspace*{1.3cm}
  
\centerline{ \Large \bf
  Worldsheet Free Fields, Higher Spin Symmetry
  and
}
\vspace*{0.3cm}
\centerline{ \Large \bf
 Free ${\cal N}=4$ Super Yang-Mills 
 } 
%and }
\vspace*{1.5cm}
\centerline{ {\bf  Changhyun Ahn}
%\footnote{On leave from the Department of Physics, Kyungpook National University, Taegu
%  702-701, Korea and 
%address until Aug. 31, 2011:
%Department of Physics, Princeton University, Jadwin Hall, 
%Princeton, NJ 08544, USA}
} 
\vspace*{1.0cm} 
\centerline{\it 
 Department of Physics, Kyungpook National University, Taegu
41566, Korea} 
\vspace*{0.5cm}
%^\centerline{\it 
 % $\star$ Institut f$\ddot{u}$r Theoretische Physik,
 % ETH Zurich, 8093 Z$\ddot{u}$rich, Switzerland}
%\vspace*{0.8cm} 
\centerline{\tt ahn@knu.ac.kr
  %\qquad  
} 
\vskip2cm

\centerline{\bf Abstract}
\vspace*{0.5cm}

By using the free field worldsheet realization
described by Gaberdiel and Gopakumar recently, we construct
the nontrivial lowest generators of the higher spin superalgebra
$hs(2,2|4)$. They consist of cubic terms between the bilinears of
ambitwistor-like fields.
We also obtain the worldsheet description for the findings of
Sezgin and Sundell twenty years ago
given by the familiar oscillator construction.
The first order poles of the operator product expansions
(OPEs), between the conformal weight-$1$
generators of Lie superalgebra $PSU(2,2|4)$ and the above
conformal weight-$3$
generators of $hs(2,2|4)$, are determined explicitly
and the additional generators appear in the worldsheet theory.

%\vspace*{4cm}
% \begin{flushright}
%{\it On the occasion of my sixtieth birthday}
%\end{flushright}

\baselineskip=18pt
\newpage
\renewcommand{\theequation}
{\arabic{section}\mbox{.}\arabic{equation}}

\tableofcontents

%%%%%%%%%%%%%%%%%%%%%%%%%%%%%%%%%%%%%%%%%%%%%%%%%%%%%%%%%%%%%%%%%%%%%
%%%%%%%%%%%%%%%%%%%%%%%%%%%%%%%%%%%%%%%%%%%%%%%%%%%%%%%%%%%%%%%%%%%%%%
%%%%%%%%%%%%%%%%%%%%%%%%%%%%%%%%%%%%%%%%%%%%%%%%%%%%%%%%%%%%%%%%%%%%%
%%%%%%%%%%%%%%%%%%%%%%%%%%%%%%%%%%%%%%%%%%%%%%%%%%%%%%%%%%%%%%%%%%%%%%
\section{ Introduction}
%1%%%%%%%%%%%%%%%%%%%%%%%%%%%%%%%%%%%%%%%%%%%%%%%%%%%%%%%%%%%%%%%%%%%%%
%%%%%%%%%%%%%%%%%%%%%%%%%%%%%%%%%%%%%%%%%%%%%%%%%%%%%%%%%%%%%%%%%%%%%

Gaberdiel and Gopakumar have described the worldsheet description for the
$
AdS_5 \times S^5$ string theory
dual to free four dimensional ${\cal N}=4$ super Yang-Mills theory in
\cite{GG2104}. Their free field description is related to the ambitwistor
string theory and the finite set of generalized zero modes
(or wedge modes) in each
spectrally flowed sector are physical. Furthermore, they impose
some residual gauge constraints on the Fock space generated by 
these wedge oscillators, and demonstrate the matching of the
physical spectrum of the string theory with that of free
${\cal N}=4$ super Yang-Mills theory at the planar level
\cite{GG2105}.
See also the relevant works in \cite{EGG1,EGG2,Eberhardt,DGGK} where
the tensionless string theory on $AdS_3 \times S^3$, in the worldsheet
theory
with free fields,
is studied.

At vanishing gauge coupling constant, the Lie superalgebra
$PSU(2,2|4)$ of ${\cal N}=4$ super Yang-Mills theory
gets enhanced to the higher spin superalgebra  $hs(2,2|4)$.
The fundamental unitary irreducible representation of
$hs(2,2|4)$ is the singleton with vanishing central charge
\cite{GM,GM1,GMZ,GMZ1}.
The symmetric tensor product of two singletons yields
the massless $AdS_5$ higher spin gauge fields.
The physical fields after gauging are organized by
the `levels' $l =0, 1, 2, \cdots, \infty$
of $PSU(2,2|4)$ multiplets \cite{SS,SS1}. See also the original paper
\cite{Vasiliev2001} used in \cite{SS}.
In particular, the level $l=0$ multiplet is the
five dimensional  ${\cal N}=8$ gauged supergravity
multiplet \cite{GRW} and the $hs(2,2|4)$ generators
depending on the $U(1)$ charge are classified by the levels
explicitly.
See also some relevant papers on the construction of
the composite operators built out of the singleton \cite{Bdd,HST,AF}.
Moreover,
the spectrum of single trace operators in the free
${\cal N}=4$ super Yang-Mills
theory can be decomposed into the irreducible representations of the
$hs(2,2|4)$ \cite{BBMS}. See also \cite{BMS}.

As pointed out by \cite{GG2104,GG2105},
the worldsheet realization provides the familiar oscillator
construction \cite{GM}
by considering each pair of modes of the free fields.
In this paper, we would like to
determine the worldsheet realization for the
higher spin generators found in \cite{SS}.
The first nontrivial case appears when the level becomes $l=1$
and the higher spin generators
consist of the cubic terms between the bilinears of
ambitwistor-like fields in the worldsheet approach by counting the
number of oscillators \cite{SS,BBMS}. Then the generators of
$PSU(2,2|4)$ have the conformal weight-$1$ while
the higher spin generators of $hs(2,2|4)$
have the conformal weight-$3$.
We will obtain the complete expressions for the
higher spin generators of $hs(2,2|4)$ for the level $l=1$
by using the standard operator product expansions (OPEs)
in two dimensional conformal field theory
\footnote{See Maldacena's comment on Gopakumar's talk in strings 2021.}.

In section $2$,
we review the free field construction of the worldsheet theory
in \cite{GG2104,GG2105}, express the $PSU(2,2|4)$ explicitly and
the stress energy tensor is described.

In section $3$, we obtain the lowest
higher spin generators of $hs(2,2|4)$ by using the free field
construction with the help of two dimensional conformal field theory.

In section $4$, we write down the complete first order poles from the
OPEs between
the generators of $PSU(2,2|4)$ and those of $hs(2,2|4)$.

In section $5$, we summarize the main results of this paper and
the future directions of related works are given. 

In Appendix, some details of the previous sections are presented
explicitly.

%%%%%%%%%%%%%%%%%%%%%%%%%%%%%%%%%%%%%%%%%%%%%%%%%%%%%%%%%%%%%%%%%%%%%%
\section{ Review}
%1%%%%%%%%%%%%%%%%%%%%%%%%%%%%%%%%%%%%%%%%%%%%%%%%%%%%%%%%%%%%%%%%%%%%%
%%%%%%%%%%%%%%%%%%%%%%%%%%%%%%%%%%%%%%%%%%%%%%%%%%%%%%%%%%%%%%%%%%%%%

%%%%%%%%%%%%%%%%%%%%%%%%%%%%
\subsection{Free fields}
%%%%%%%%%%%%%%%%%%%%%%%%%%%%

We consider the weight-$\frac{1}{2}$ conjugate pairs of symplectic
boson \cite{GOW} fields $(\la^{\al}, \mu_{\al}^{\dagger})$
and $(\mu^{\dot{\al}},\la_{\dot{\al}}^{\dagger})$ where
$\al, \dot{\al} =1, 2$
and four weight-$\frac{1}{2}$ complex fermions
$(\psi^a, \psi_a^{\dagger})$ where $a=1,2,3,4$ \cite{GG2104,GG2105}.
The $\al$ and $\dot{\al}$ are spinor indices
with respect to two different $SU(2)$'s and $\psi^a$
transforms in the fundamental representation of $SU(4)$.
Note that the conformal dimension-$\frac{1}{2}$ fields,
$(\la^{\al}, \mu_{\al}^{\dagger})$
and $(\mu^{\dot{\al}},\la_{\dot{\al}}^{\dagger})$, are bosonic
and they satisfy `quasi' statistics. 
We will follow most of the notations presented in \cite{GG2104,GG2105}. 

Their nontrivial operator product expansions (OPEs)
in the left-moving sector of the worldsheet theory
we are describing
are given by
\bea
\la^{\alpha}(z) \, \mu_{\beta}^{\dagger}(w) & = & \frac{1}{(z-w)} \,
\de^{\alpha}_{\beta}+ \cdots,
\nonu \\
\mu^{\dot{\al}}(z) \la_{\dot{\beta}}^{\dagger}(w) & = &
\frac{1}{(z-w)} \,
\de^{\dot{\alpha}}_{\dot{\beta}} + \cdots,
\nonu \\
\psi^{a}(z) \, \psi_b^{\dagger}(w) &=&
\frac{1}{(z-w)} \,
\de^{a}_{b}+ \cdots.
\label{threeOPEs}
\eea
The abbreviated parts in (\ref{threeOPEs})
are the regular terms as usual
in two dimensional conformal field theory.
By introducing the components of ambitwistor fields \cite{Berkovits}
\bea
Z^I \equiv (\la^{\al}, \mu^{\dot{\al}},\psi^a),
\qquad
Y_J \equiv (\mu_{\al}^{\dagger}, \la_{\dot{\al}}^{\dagger},
\psi_a^{\dagger}),
\label{ZY}
\eea
we can  rewrite the above three OPEs (\ref{threeOPEs})
as a single one \cite{Uvarov} alternatively
\bea
Z^I(z) \, Y_J(w) = \frac{1}{(z-w)}\, \de^{I}_J + \cdots.
\label{ZYOPE}
\eea
The upper and lower indices $I, J$
stand for $\al, \dot{\al}$ and $a$. 
For the calculations of any OPEs
containing the multiple of ambitwistor fields (\ref{ZY}),
it is useful to use (\ref{ZYOPE}) rather than (\ref{threeOPEs})
and after that we can specify
the indices $I,J,K \cdots$ of these from (\ref{ZY})
later
\footnote{\label{YZope} If we
interchange the
order of the OPE in (\ref{ZYOPE}), then we have
$Y_J(z) \, Z^I(w) = \frac{1}{(z-w)} \, (-1)^{d_I \, d_J +1}\, \de^I_J +
\cdots $ where the grading $d_I=2$ for the bosonic fields and
$d_I=1$ for the fermionic fields \cite{DTH,FL,Bowcock,AIS}.
In other words, the additional factor $(-1)^{d_I \, d_J }$
arises. Note that the components $Z^a=\psi^a$
and $Y_a=\psi_a^{\dagger}$ are fermionic.}.

By constructing the quadratic terms \cite{Berkovits,GG2104,GG2105}
\bea
J^I_{\,\,\, J} \equiv Y_J \, Z^I,
\label{J}
\eea
the current algebra version of the oscillator construction \cite{GM} of
Lie superalgebra $U(2,2|4)$ can be described by
i) the generators of Lorentz symmetry, ${\cal L}^{\al}_{\,\,\,\beta}$
and $\dot{{\cal L}}^{\dot{\al}}_{\,\,\,\dot{\beta}}$,
ii) the generator of $R$ symmetry, ${\cal R}^{a}_{\,\,\, b}$,
iii) the generators of super translations, ${\cal Q}^{a}_{\,\,\, \al},
\dot{\cal Q}^{\dot{\al}}_{\,\,\,a}$ and ${\cal P}^{\dot{\al}}_{\,\,\, \beta}$.
Moreover, the ${\cal N}=4$ super Poincare algebra obtained by these
generators can be enlarged by the generators of super conformal
boosts, ${\cal S}^{\al}_{\,\,\,a}, \dot{{\cal S}}^{a}_{\,\,\,\dot{\al}}$
and ${\cal K}^{\al}_{\,\,\,\dot{\beta}}$.
There exist also
the $U(1)$ hyper charge ${\cal B}$, the central charge
${\cal C}$ and the dilatation generator ${\cal D}$.
Then the generators \cite{Beisert} of Lie superalgebra $U(2,2|4)$
can be extended by the
following generators in terms of ambitwistor fields \cite{GG2104,GG2105}
\bea
{\cal L}^{\al}_{\,\,\,\beta} &= & Y_{\beta}\, Z^{\al}-\frac{1}{2}\,
\de^{\al}_{\beta} \, Y_{\ga} \, Z^{\ga},
\qquad
\dot{{\cal L}}^{\dot{\al}}_{\,\,\,\dot{\beta}} = 
Y_{\dot{\beta}}\, Z^{\dot{\al}}-\frac{1}{2}\,
\de^{\dot{\al}}_{\dot{\beta}} \, Y_{\dot{\ga}} \, Z^{\dot{\ga}},
\qquad
{\cal R}^a_{\,\,\, b}  =  Y_{b}\, Z^{a}-\frac{1}{4}\,
\de^{a}_{b} \, Y_{c} \, Z^{c},
\nonu \\
{\cal Q}^a_{\,\,\, \al}  & = &  Y_{\al}\, Z^a,
\qquad
\dot{{\cal Q}}^{\dot{\al}}_{\,\,\, a}  =  Y_a \, Z^{\dot{\al}},
\qquad
{\cal P}^{\dot{\al}}_{\,\,\, \beta}  =  Y_{\beta} \, Z^{\dot{\al}},
\nonu \\
{\cal S}^{\al}_{\,\,\, a}  & = &  Y_a \, Z^{\al},
\qquad
\dot{{\cal S}}^{a}_{\,\,\, \dot{\al}}  =  Y_{\dot{\al}}\, Z^a,
\qquad
{\cal K}^{\al}_{\,\,\, \dot{\beta}}  =  Y_{\dot{\beta}} \, Z^{\al},
\nonu \\
{\cal B} &=& \frac{1}{2} \, (Y_{\al}\, Z^{\al} + Y_{\dot{\al}} \,
Z^{\dot{\al}}),
\qquad {\cal C} =
\frac{1}{2} \, (Y_{\al}\, Z^{\al} + Y_{\dot{\al}} \,
Z^{\dot{\al}} + Y_a\, Z^a),
\nonu \\
{\cal D} & = & \frac{1}{2} \, (Y_{\al}\, Z^{\al}
- Y_{\dot{\al}} \,
Z^{\dot{\al}}).
\label{12generators}
\eea
As usual, the repeated indices are summed over the corresponding
indices.
As noted in \cite{GG2104,GG2105},
each pair of modes of the free fields provides
two copies of the usual oscillator construction.
Therefore, once we restrict to the zero modes
of (\ref{12generators}) in their (anti)commutator relations,
the known Lie superalgebra
$U(2,2|4)$ \cite{Beisert} can be obtained.
We present their complete OPEs in Appendix $A$
in the worldsheet theory \footnote{We use
the Thielemans package \cite{Thielemans} with a mathematica
\cite{mathematica}. Note that the group indices $\al, \dot{\al}$ and
$a$ are fixed. All the coefficients appearing in the right hand sides
of the OPEs are numerical values.
Once we identify the group index structures both sides of the OPEs, then
it is straightforward to calculate all these coefficients inside a
Package explicitly due to the free fields.}. 

It is useful to introduce the following $U(1)$ generators
which appear in the above ${\cal B}, {\cal C}$ and ${\cal D}$
generators
\bea
{\cal U} \equiv  Y_{\ga} \, Z^{\ga}, \qquad
\dot{{\cal U}} \equiv      Y_{\dot{\ga}} \, Z^{\dot{\ga}},
\qquad
{\cal V} \equiv  Y_{c} \, Z^{c}.
\label{uudotv}
\eea
Note that the ${\cal V}$ appears in the second term of
${\cal R}^a_{\,\,\,b}$ in (\ref{12generators}) which is
traceless: ${\cal R}^a_{\,\,\,a}=0$.

\begin{itemize}
  \item[]
In particular, the nonzero
${\cal V}$-charge for ${\cal Q}^a_{\,\,\,\al}$
is equal to $-1$ and
the nonzero ${\cal V}$-charge for $\dot{{\cal Q}}^{\al}_{\,\,\,a}$
is equal to $1$ from the observation of Appendix (\ref{opewithuudotv}).
This corresponds to $Y$-charge in \cite{SS} up to sign.
By simply counting the number of
supersymmetry generators in the multiple product of the generators
of (\ref{12generators}), we can determine the ${\cal V}$-charge.
The remaining ten generators have vanishing ${\cal V}$-charges.
%\label{vcharge}
\end{itemize}

Note that the ordering of two operators
in (\ref{J}) or (\ref{12generators}) is important
because sometimes we will have additional minus sign
when we interchange the ambitwistor fields each other.

%%%%%%%%%%%%%%%%%%%%%%%%%%%%%%%%%%%%%%%%%%%%%%%%%%
\subsection{The Lie superalgebra $PSU(2,2|4)$}
%%%%%%%%%%%%%%%%%%%%%%%%%%%%%%%%%%%%%%%%%%%%%%%%%%

We can calculate the OPEs between the conformal weight-$1$ currents
in (\ref{J}) by using the defining relation in (\ref{ZYOPE})
with the help of the footnote \ref{YZope}
and it turns out that 
\bea
J^{I}_{\,\,\,J}(z) \, J^{K}_{\,\,\,L}(w) & = &
-\frac{1}{(z-w)^2}\, (-1)^{d_J \, d_K} \, \de^{I}_{L}\, \de^{K}_{J}
+ \frac{1}{(z-w)}\, \Bigg[ \de^{I}_{L} \, J^{K}_{\,\,\,J}
\nonu \\
& + &
(-1)^{(d_L+d_K)(d_I+d_J)+1}\, \de^{K}_{\,\,\,J} \, J^{I}_{\,\,\,L}
\Bigg](w) +\cdots. 
\label{jjope}
\eea
The grading $d_I$ is defined in the footnote \ref{YZope}.
We can also check, from (\ref{jjope}), that the second order pole of the
OPE between $J^+ \equiv {\cal L}^1_{\,\,\,2}$,
$J^- \equiv {\cal L}^2_{\,\,\,1}$ and $J^3 \equiv \frac{1}{2}
({\cal L}^2_{\,\,\,2}-{\cal L}^1_{\,\,\,1})$
implies that the level is equal to $-1$.
Similarly,
the OPE between $\dot{J}^+ \equiv \dot{{\cal L}}^{\dot{1}}_{\,\,\,\dot{2}}$,
$\dot{J}^- \equiv \dot{{\cal L}}^{\dot{2}}_{\,\,\,\dot{1}}$
and $\dot{J}^3 \equiv \frac{1}{2}
(\dot{{\cal L}}^{\dot{2}}_{\,\,\,\dot{2}}-
\dot{{\cal L}}^{\dot{1}}_{\,\,\,\dot{1}})$
leads to the fact  that the level is also equal to $-1$.
We obtain Appendix $A$ from this defining relation (\ref{jjope})
by specifying the indices explicitly. The OPEs between the $U(1)$
generator ${\cal C}$ appearing in (\ref{12generators})
and other generators of $U(2,2|4)$
do not have any singular terms 
in Appendix (\ref{psualgebra}) except the OPE ${\cal B}(z) \,
{\cal C}(w)$. We are left with $PSU(2,2|4)$ after
the $U(1)$ generator ${\cal C}$ is `quotiented' \cite{GG2104,GG2105}.

We can calculate the OPEs between
the single $J^{I}_{\,\,\,J}(z)$ and
the quadratic term
$ J^{K}_{\,\,\,L}\, J^{M}_{\,\,\,N}(w)$
and the OPEs
between
the single $J^{I}_{\,\,\,J}(z)$ and
the cubic term $ J^{K}_{\,\,\,L}\, J^{M}_{\,\,\,N} \,
J^{P}_{\,\,\,Q}(w)$ but we do not present them in this paper
because they have long expressions due to the presence of
various gradings. Later we will present the first order pole
of the latter explicitly in next section.

%%%%%%%%%%%%%%%%%%%%%%%%%%%%%%%%%%%%%%%%%%%%%%%%%
\subsection{The stress energy tensor}
%%%%%%%%%%%%%%%%%%%%%%%%%%%%%%%%%%%%%%%%%%%%%%%%%%

By requiring that the ambitwistor fields (\ref{ZY})
are weight-$\frac{1}{2}$ primary
and the generators (\ref{12generators})
are weight-$1$ primary (See also the footnote \ref{opewithT}),
we can determine
the stress energy tensor from the possible quadratic terms
from (\ref{12generators}) completely
and it is given by
\bea
T & = & \frac{1}{2} (\la^{\al} \, \pa \mu_{\al}^{\dagger} +
\mu^{\dot{\al}} \, \pa \la^{\dagger}_{\dot{\al}}- \psi^a \, \pa \,
\psi_a^{\dagger}
-\pa \, \la^{\al} \,  \mu_{\al}^{\dagger} -
\pa \, \mu^{\dot{\al}} \,  \la^{\dagger}_{\dot{\al}}+ \pa \,
\psi^a \,
\psi_a^{\dagger}
)
\nonu \\
& = &  \frac{1}{2}\, (-1)^{d_I} \, ( Z^I \, \pa \, Y_I -\pa \,
Z^I \, Y_I).
\label{stressenergy}
\eea
As before, the repeated indices are summed.
Note that there is an additional factor for the grading
when we change the
order between the ambitwistor fields in
the second expression of (\ref{stressenergy}).
This stress energy tensor satisfies the usual standard OPE
$T(z) \, T(w)$ and the central charge is equal to zero.
We will use the explicit expression (\ref{stressenergy})
in order to calculate the possible (quasi)primary operators in next
section \footnote{\label{opewithT} Therefore, we have
$T(z) \, Z^I(w) = \frac{1}{(z-w)^2}\, \frac{1}{2} \,
  Z^I(w) + \frac{1}{(z-w)}\, \pa \, Z^I(w) + \cdots $,
$T(z) \, Y_I(w) = \frac{1}{(z-w)^2}\, \frac{1}{2} \,
  Y_I(w) + \frac{1}{(z-w)}\, \pa \, Y_I(w) + \cdots $,
  $T(z) \, J^{I}_{\,\,\,J}(w) =\frac{1}{(z-w)^2}\, J^I_{\,\,\,J}(w) +
  \frac{1}{(z-w)} \, \pa \, J^I_{\,\,\,J}(w) +\cdots$
  and  from these we can calculate the following OPE
  $T(z) \, J^{I}_{\,\,\,J}\, J^{K}_{\,\,\,L}(w)=\frac{1}{(z-w)^4}\, (-1)^{
d_J \, d_K +1}\, \de^I_L\, \de^K_J  +\frac{1}{(z-w)^3}\, \Bigg[
\de^I_L \, J^{K}_{\,\,\,J} +(-1)^{(d_L+d_K)(d_I+d_J)+1}\, \de^{K}_J \,
J^{I}_{\,\,\,L} \Bigg](w)+\frac{1}{(z-w)^2} \, 2 \,
J^{I}_{\,\,\,J}\, J^{K}_{\,\,\,L}(w)+
\frac{1}{(z-w)}\, \pa \,  (J^{I}_{\,\,\,J}\, J^{K}_{\,\,\,L})(w)+
\cdots$ which implies that this does not produce the (quasi)primary
operator in general. We can check whether this
is really (quasi)primary or not after specifying the indices
explicitly.}.

In this section,
we summarize the `extension' of the Lie superalgebra $PSU(2,2|4)$
generated by (\ref{12generators}) in the worldsheet theory.
Implicitly it is given by (\ref{jjope}) or explicitly
it is also given by Appendix (\ref{psualgebra}).
If we focus on the zero modes for these generators,
then this will lead to the standard
(anti)commutator relations \cite{Beisert}.

%%%%%%%%%%%%%%%%%%%%%%%%%%%%%%%%%%%%%%%%%%%%%%%%%%%%%%%%%%%%%%%%%%%%%%
\section{ Construction of the lowest generators of
the higher spin superalgebra $hs(2,2|4)$}
%1%%%%%%%%%%%%%%%%%%%%%%%%%%%%%%%%%%%%%%%%%%%%%%%%%%%%%%%%%%%%%%%%%%%%%
%%%%%%%%%%%%%%%%%%%%%%%%%%%%%%%%%%%%%%%%%%%%%%%%%%%%%%%%%%%%%%%%%%%%%

We would like to construct the worldsheet description for the
higher spin generators of $hs(2,2|4)$ found in \cite{SS,SS1}.
We have seen the conformal weight-$1$ generators which are primary
under the stress energy tensor (\ref{stressenergy}).
According to the results of \cite{SS}, the nontrivial
lowest generators consist of cubic terms in the above
weight-$1$ generators corresponding to the level $l=1$ case
(For $l=0$ case, they are linear in the weight-$1$ generators
while for $l=2$ case they are quintic in the weight-$1$ generators).

\begin{itemize}
\item[]
We observe that
the $(2l +1)$ can be identified with  the conformal dimension
(or weight or spin) under
(\ref{stressenergy}) in the worldsheet theory.
\end{itemize}

From the conformal field theory analysis
\cite{Blumenhagenetal,CFT,Ahn1211},
it is known that in the OPE between the weight-$1$ operator
(which is a primary) and
the weight-$3$ (quasi)primary operator,
in principle, there appear a (new)
weight-$1$ operator
in the third order pole and a (new) weight-$2$ operator
in the second order
pole.
By simple counting the relative coefficients for the descendant
operators of these operators
which will appear in the second and first order poles, 
they do not appear in the first order pole.

\begin{itemize}
  \item[]
Therefore, we will focus on the first order pole in 
the OPE between the weight-$1$ operator and
the weight-$3$ operator. This first order pole provides
a new (quasi)primary operators.
In doing this, we should check that
the weight-$3$ operator should be (quasi) primary.
That is, at least the third order pole of the
OPE between the stress energy tensor
and this weight-$3$ operator should vanish.
\end{itemize}

We have the following first order pole
in the OPE between  $J^I_{\,\,\, J}(z)$ and
$J^{K}_{\,\,\,L}\, J^{M}_{\,\,\,N}\,
J^{P}_{\,\,\,Q}(w)$ by using (\ref{jjope}) successively as follows:
\bea
&& J^I_{\,\,\, J}(z) \, J^{K}_{\,\,\,L}\, J^{M}_{\,\,\,N}\,
J^{P}_{\,\,\,Q}(w)\Bigg|_{\frac{1}{(z-w)}} =
\de^{I}_{L}\, J^{K}_{\,\,\,J}\, J^M_{\,\,\, N}\, J^{P}_{\,\,\,Q}(w)
+ (-1)^{(d_I+d_K)(d_I+d_J)+1}\, \de^{K}_J \,  J^{I}_{\,\,\,L}\, J^M_{\,\,\, N}\,
J^{P}_{\,\,\,Q}(w) \nonu \\
&& + (-1)^{(d_I+d_J)(d_K+d_L)}\, J^{K}_{\,\,\,L}\, \Bigg[
\de^{I}_{N}\,  J^M_{\,\,\, J}\, J^{P}_{\,\,\,Q}
+ (-1)^{(d_N+d_M)(d_I+d_J)+1}\, \de^{M}_J \, J^{I}_{\,\,\, N}
\, J^{P}_{\,\,\,Q}\,
\nonu \\
&& + (-1)^{(d_I+d_J)(d_N+d_M)}\, \de^{I}_Q \, J^{M}_{\,\,\, N}
\, J^{P}_{\,\,\,J}+
(-1)^{(d_I+d_J)(d_N+d_M+d_P+d_Q)+1}\, \de^{P}_J \, J^{M}_{\,\,\, N}
\, J^{I}_{\,\,\,Q}  
\Bigg](w).
\label{pole1}
\eea
Let us emphasize that the right hand side of (\ref{pole1})
is a (quasi)primary operator as before as long as
the third order pole of Appendix (\ref{tjjj}) vanishes.
We obtain all the information on the higher spin
generators in this section from this (implicit) OPE (\ref{pole1})
by imposing the explicit indices on (\ref{pole1}).
In other words, the first order pole can be written in terms of the
known operators by collecting them
appropriately or if not, then there appears in the new
(quasi)primary operator. We do not have to subtract the contributions
from the descendant operators as we mentioned before. 
Of course,
there are also fourth, third and second order poles in the above OPE.

We focus on the tables $4$ and $5$ of \cite{SS} with $l=1$ case
and $s=1, \frac{3}{2}, 2, \frac{5}{2}, 3, \frac{7}{2}$ and $4$.
Their $l$ is related to the numbers of bosonic and fermionic
oscillators and is given by
the equation $(3.6)$ in \cite{SS}
and their $s$ is related to the numbers of
bosonic oscillators and is given around equation $(3.18)$ in \cite{SS}.
Furthermore, their equation $(3.19)$ contains all the information on the
above two tables although it is not easy to
read off the relevant quantities properly
\footnote{
\label{fiveweightone}
For $l=0$ in table $4$ of \cite{SS},
there are generators ${\cal R}^a_{\,\,\, b}, {\cal Q}^a_{\,\,\,\al},
\dot{{\cal Q}}^{\dot{\al}}_{\,\,\,a}$ and ${\cal P}^{\dot{\al}}_{\,\,\,\beta}$
corresponding to ${\bf 15}_0, {\bf 4}_{-1}, \overline{\bf 4}_{1}$
and ${\bf 1}_0$ respectively. It is easy to see that
they are closed by themselves in Appendix (\ref{psualgebra}).
In the oscillator construction,
the remaining generators of
$PSU(2,2|4)$ acting on the physical vacuum state
vanish \cite{Beisert,Aoyama}.
We will calculate the OPEs between these weight-$1$ operators including
the $U(1)$ operator ${\cal V}$ relevant to
${\cal R}^a_{\,\,\,b}$ and the weight-$3$ operators
in next section. The algebra from these
five weight-$1$ operators is closed.}.

%\begin{itemize}
%\item[]
%Note that
%the spin $s=1$ for
%${\cal V}$ and ${\cal R}^a_{\,\,\,b}$,
%the spin $s=\frac{3}{2}$ for ${\cal Q}^a_{\,\,\,\al}$ and $
%\dot{\cal Q}^{\dot{\al}}_{\,\,\,a}$(${\cal S}^{\al}_{\,\,\,a}$ and
%$\dot{\cal S}^{a}_{\,\,\,\dot{\al}}$) and the spin $s=2$ for
%${\cal P}^{\dot{\al}}_{\,\,\,\beta}$ (${\cal L}^{\al}_{\,\,\,\beta}$,
%$\dot{\cal L}^{y\dot{\al}}_{\,\,\,\dot{\beta}}$, ${\cal B}$, ${\cal D}$,
%${\cal U}$, $\dot{\cal U}$ and
%${\cal K}^{\al}_{\,\,\,\dot{\beta}}$).
%\end{itemize}

%%%%%%%%%%%%%%%%%%%%%%%%%%%%%%%%%
\subsection{ The $s=1$ case:  ${\bf 1}_0$ and ${\bf 15}_0$}
%%%%%%%%%%%%%%%%%%%%%%%%%%%%%%%%%

Because their $X$ appearing in equation $(2.4)$
in \cite{SS} corresponds to
our ${\cal V}$ up to sign and normalization,
we can observe that the $SU(4)$ singlet is a cubic
in ${\cal V}$ which has vanishing ${\cal V}$-charge from
Appendix (\ref{opewithuudotv}).
Moreover, the $SU(4)$ nonsinglet contains
the quadratic in $X$ and we can identify this
as a quadratic in ${\cal V}$ together with
${\cal R}^a_{\,\,\, b}$ which is a ${\bf 15}$ representation
of $SU(4)$. Note that by construction of (\ref{12generators}),
we observe the fact that ${\cal R}^a_{\,\,\, a}$ vanishes.
In the tensor product of ${\bf 4} \otimes \overline{{\bf 4}}=
{\bf 1} \oplus {\bf 15}$ \cite{Slansky,FKS},
after subtracting the ${\cal V}$ part,
we are left with the representation ${\bf 15}$.
Once again, the ${\cal V}$-charge in the cubic of
${\cal V}\, {\cal V} \, {\cal R}^a_{\,\,\,b}$ vanishes.

Therefore we identify the following higher spin generators
corresponding to the representations ${\bf 1}_0$ and ${\bf 15}_0$
respectively as follows \footnote{We denote the higher spin
generators as the letter ${\cal W}$ with appropriate group indices.
For the additional `new' higher spin generators we put a hat 
on ${\cal W}$
with some indices.}:
\bea
{\cal W} & \equiv & {\cal V}\, {\cal V}\, {\cal V},
\nonu \\
{\cal{W}}^{a}_{\,\,\, b} & \equiv &  {\cal V}\, {\cal V}\,
{\cal R}^a_{\,\,\,b} +  {\cal V}\,
{\cal R}^a_{\,\,\,b}\,  {\cal V}\, +
{\cal R}^a_{\,\,\,b}\, {\cal V}\,
{\cal V}.
\label{s-one}
\eea
We can check these higher spin generators
in (\ref{s-one}) are quasiprimary operators
under the stress energy tensor (\ref{stressenergy}).
In other words, the OPEs between the stress energy tensor
and these generators contain nonzero fourth order poles
although the third order poles become zero according to Appendix
(\ref{tjjj}) by specifying the indices correctly. 

Although the OPE between
${\cal V}$ and ${\cal R}^a_{\,\,\, b}$
is regular and they are commuting operators (the second and the third
terms in the right hand side of
${\cal W}^a_{\,\,\,b}$ are the same as the first one),
we will keep its form in symmetrical way as in (\ref{s-one}). 
When we act the supersymmetry generators on
the ${\cal W}^a_{\,\,\,b}$, then we will observe that each three terms
contributes differently due to the normal ordering.

In next subsections, we will determine the remaining higher spin
generators by acting the supersymmetry generators
${\cal Q}^a_{\,\,\, \al}$ and $\dot{{\cal Q}}^{\dot{\al}}_{\,\,\,a}$ on 
(\ref{s-one}) successively.

%%%%%%%%%%%%%%%%%%%%%%%%%%%%%%%%%%%%%%%
\subsection{ The $s=\frac{3}{2}$ case: ${\bf 4}_{-1}, \overline{{\bf 4}}_1,
{\bf 20}_{-1}$ and $\overline{{\bf 20}}_{1}$}
%%%%%%%%%%%%%%%%%%%%%%%%%%%%%%%%%%%%%%%

Now we move on the next column of the table $4$ with $l=1$ of \cite{SS}.
Eventually we will present all the first order poles in the OPEs between
some weight-$1$ operators and the weight-$3$ operators in next section
with Appendix $C$.
However, in this section, we will focus on some of them which determine
the higher spin generators completely. One way to determine these
particular higher spin generators is to consider that
we can calculate the first order pole
in the OPE between
the supersymmetry generator ${\cal Q}^{a}_{\,\,\,\al}$
which is fermionic and ${\cal{W}}^{b}_{\,\,\, c}$
which is introduced in previous subsection (\ref{s-one}).
Either we can use Appendix (\ref{psualgebra}) or the previous OPE
(\ref{pole1}) can be used by selecting the corresponding indices
for this particular OPE.

It turns out that by antisymmetrizing the upper indices
\footnote{In this paper, the
(anti)symmetric notations are for $SU(4)$ indices. The bracket
$\left[ \right]$ stands for antisymmetric one and
the bracket
$\left( \right)$ stands for symmetric one without any overall numerical
factors.}
\bea
{\cal Q}^{\left[a \right.}_{\,\,\,\al}(z) \, {\cal{W}}^{
\left. b \right]}_{\,\,\, c}(w)
\Bigg|_{\frac{1}{(z-w)}} =
{\cal W}^{\left[a b\right]}_{\,\,\, c\, \al}(w) + \de^{\left[ a \right.}_{c} \,
{\cal W}^{ \left. b \right]}_{\,\,\, \al}(w) -\frac{1}{4} \,
\de^{\left[ b \right.}_{c} \, {\cal W}^{\left. a \right]}_{\,\,\, \al}(w),        
\label{s-threehalfope}
\eea
where the right hand side of (\ref{s-threehalfope}) consists of two kinds
of higher spin generators as follows:
\bea
{\cal W}^{a}_{\,\,\, \al} & \equiv &
{\cal V}\, {\cal V}\,
{\cal Q}^a_{\,\,\,\al} +  {\cal V}\,
{\cal Q}^a_{\,\,\,\al}\,  {\cal V}\, +
{\cal Q}^a_{\,\,\,\al}\, {\cal V}\,
{\cal V},      
\nonu \\
{\cal W}^{\left[a b\right]}_{\,\,\, c\, \al} & \equiv &
{\cal V} \, {\cal Q}^{\left[a \right.}_{\,\,\,\al} \, {\cal R}^{\left. b \right]}_{\,\,\,c}
+ {\cal Q}^{\left[a\right.}_{\,\,\,\al} \, {\cal V}\, {\cal R}^{\left. b\right]}_{\,\,\,c}
+ {\cal Q}^{\left[a \right.}_{\,\,\,\al} \, {\cal R}^{\left. b \right]}_{\,\,\,c}\,
{\cal V} 
\nonu \\
&+&
{\cal V} \, {\cal R}^{\left[b \right.}_{\,\,\,c} \, {\cal Q}^{\left. a \right]}_{\,\,\,\al}
+ {\cal R}^{\left[b\right.}_{\,\,\,c} \, {\cal V}\, {\cal Q}^{\left. a\right]}_{\,\,\,\al}
+ {\cal R}^{\left[b \right.}_{\,\,\,c} \, {\cal Q}^{\left. a \right]}_{\,\,\,\al}\,
{\cal V}.
\label{s-threehalf}
\eea
Note that the first one in (\ref{s-threehalf}) is a quasiprimary operator
while the second one in (\ref{s-threehalf}) is a primary operator
according to Appendix (\ref{tjjj}).
Note that the second one is antisymmetric in the upper indices.
As mentioned before, the weight-$1$ operator
${\cal Q}^a_{\,\,\,\al}$ has nontrivial OPE with ${\cal V}$ (See also
Appendix (\ref{opewithuudotv})) and 
the ordering between them is not trivial and if we interchange them,
there appears a derivative term of weight-$1$ operator.
The quasiprimary condition of the first operator requires all of three
terms (this is the reason why
we have three terms in (\ref{s-one}))
and we can easily observe that the first operator corresponds to
the representation ${\bf 4}_{-1}$ because it contains a single
weight-$1$ operator which has ${\cal V}$-charge $-1$ (Of course,
the ${\cal V}$-charge of ${\cal V}$ is equal to zero) and it has
upper index $a$ which transforms as a fundamental representation
of $SU(4)$.

In the tensor product of $\overline{\bf 6} \otimes \overline{\bf 4}=
{\bf 20} \oplus {\bf 4}$ \cite{Slansky,FKS},
we obtain the representation ${\bf 20}$ by subtracting the fundamental
representation ${\bf 4}$. The second higher spin generator in
(\ref{s-threehalf})
consists of the upper antisymmetric combination and the lower
antifundamental one. Therefore, in total, it provides
the tensor product
$\overline{\bf 6} \otimes \overline{\bf 4}$. Now we consider
the contracted one which is given by $
{\cal W}^{a b}_{\,\,\, a\, \al}$
which transforms as a fundamental representation ${\bf 4}$ of $SU(4)$.
Then after subtracting this representation from
$\overline{\bf 6} \otimes \overline{\bf 4}$, we will eventually
obtain the representation ${\bf 20}_{-1}$. Furthermore, it has
${\cal V}$-charge $-1$ also because there exists a single
${\cal Q}^a_{\,\,\,\al}$ and the operator ${\cal R}^a_{\,\,\, b}$
has a vanishing ${\cal V}$-charge.
Note that the expression without the antisymmetric bracket in the
second higher spin generator in (\ref{s-threehalf}) is itself a
primary operator and it is obvious to see that
the higher spin generator
${\cal W}^{a b}_{\,\,\, a\, \al}$ also transforms as a
primary operator after taking antisymmetric combination.

Therefore, we should consider the particular antisymmetric
combination in the OPE of (\ref{s-threehalfope}).
Without it, we would not obtain the corresponding right higher spin
generator which transforms properly. In other words, the antisymmetric
combination in the indices $a$ and $b$ is crucial
for the presence of  the representation ${\bf 20}_{-1}$ in the oscillator
construction in \cite{SS} \footnote{\label{conjugat-1}
Similarly, we obtain
$
\dot{{\cal Q}}_{\,\,\,\left[a \right.}^{\dot{\al}}(z) \, {\cal{W}}^{
b }_{\,\,\, \left. c \right]}(w)
\Bigg|_{\frac{1}{(z-w)}} =
-\dot{{\cal W}}^{b\, \dot{\al}}_
{\,\,\,  \left[ a c \right]}(w) - \de^{ b }_{\left[ a \right.} \,
\dot{{\cal W}}^{ \dot{\al}}_{\,\,\, \left. c \right]}(w) +\frac{1}{4} \,
\de^{b }_{\left[ c \right.} \, \dot{{\cal W}}^{\dot{\al}}_{\,\,\, \left. a \right]}(w),        
%\label{s-threehalfope1}
$
where the right hand side  has the following
higher spin generators
$
\dot{{\cal W}}^{\dot{\al}}_{\,\,\, a}  \equiv 
{\cal V}\, {\cal V}\,
\dot{{\cal Q}}^{\dot{\al}}_{\,\,\,a} +  {\cal V}\,
\dot{{\cal Q}}^{\dot{\al}}_{\,\,\,a}\,  {\cal V}\, +
\dot{{\cal Q}}^{\dot{\al}}_{\,\,\,a}\, {\cal V}\,
{\cal V}$ corresponding to the representation $\overline{\bf 4}_1$,
and $
\dot{{\cal W}}^{ b \, \dot{\al}}_{\,\,\, \left[ a c \right]}   \equiv 
{\cal V} \, \dot{{\cal Q}}^{\dot{\al}}_{\,\,\,\left[ a \right.} \,
{\cal R}^{ b }_{\,\,\, \left. c \right]}
+ \dot{{\cal Q}}^{\dot{\al}}_{\,\,\,\left[ a \right.} \,
{\cal V}\, {\cal R}^{b}_{\,\,\, \left. c \right]}
+ \dot{{\cal Q}}^{\dot{\al}}_{\,\,\,\left[ a \right.} \,
{\cal R}^{ b }_{\,\,\, \left. c \right]}\,
{\cal V} 
+
{\cal V} \, {\cal R}^{b }_{\,\,\, \left[ c \right.} \,
\dot{{\cal Q}}^{\dot{\al}}_{\,\,\,\left. a \right]}
+ {\cal R}^{b}_{\,\,\, \left[ c \right.} \,
{\cal V}\, \dot{{\cal Q}}^{\dot{\al}}_{\,\,\,\left. a \right]}
+ {\cal R}^{b }_{\,\,\, \left[ c \right.} \,
\dot{{\cal Q}}^{\dot{\al}}_{\,\,\,\left. a \right]}\,
{\cal V} $ corresponding to the representation $\overline{\bf 20}_1$
from the analysis of the tensor product
${\bf 6} \otimes {\bf 4}=
\overline{\bf 20} \oplus \overline{\bf 4}$.}. 

%%%%%%%%%%%%%%%%%%%%%%%%%%%%%%%%%%%
\subsection{ The $s=2$ case: ${\bf 1}_0, {\bf 15}_0,
  {\bf 20'_{\rm{0}}}, {\bf 6}_{-2}, {\bf 6}_2,  {\bf 10}_{-2}$ and
  $\overline{{\bf 10}}_2$}
%%%%%%%%%%%%%%%%%%%%%%%%%%%%%%%%%%%

Let us consider the next column of
the tables $4$ and $5$ with $l=1$ of \cite{SS}.
Again, we can use
either (\ref{pole1}) or Appendix (\ref{psualgebra}).
We can calculate the OPEs between the supersymmetry generators and
the higher spin generators found in previous subsection.

It turns out, from (\ref{s-threehalf}),  that we have 
\bea
\dot{{\cal Q}}^{\dot{\al}}_{\,\,\,a}(z) \, {\cal{W}}^{
b }_{\,\,\, \beta}(w)
\Bigg|_{\frac{1}{(z-w)}} =
\de^{b}_{a} \,{\cal W}^{\dot{\al}}_{\,\,\,   \beta}(w)+
\widehat{{\cal W}}^{b \, \dot{\al}}_{\,\,\, a \, \beta},        
\label{s-2ope}
\eea
where the right hand side of (\ref{s-2ope}) contains the following
higher spin generators 
\bea
{\cal W}^{\dot{\al} }_{\,\,\,   \beta} & \equiv &
{\cal V} \, {\cal V} \, {\cal P}^{\dot{\al}}_{\,\,\,\beta}      
+{\cal V} \,  {\cal P}^{\dot{\al}}_{\,\,\,\beta}\, {\cal V} 
+ {\cal P}^{\dot{\al}}_{\,\,\,\beta}  \, {\cal V} \, {\cal V},
\nonu \\
\widehat{{\cal W}}^{b \, \dot{\al}}_{\,\,\, a \, \al} & \equiv &
{\cal Q}^{ b }_{\,\,\, \al} \, \dot{{\cal Q}}^{\dot{\al}}_
{\,\,\,a } \, {\cal V}
+{\cal Q}^{ b }_{\,\,\, \al} \,  {\cal V}
\, \dot{{\cal Q}}^{\dot{\al}}_
{\,\,\, a}+
{\cal V} \,
{\cal Q}^{ b }_{\,\,\, \al} \, \dot{{\cal Q}}^{\dot{\al}}_
{\,\,\, a}
\nonu \\
&-&  \dot{{\cal Q}}^{\dot{\al}}_
{\,\,\,a }\,
{\cal Q}^{ b }_{\,\,\, \al} \, {\cal V}-
\dot{{\cal Q}}^{\dot{\al}}_
{\,\,\, a} \,  {\cal V}
\,{\cal Q}^{ b }_{\,\,\, \al} -
{\cal V} \,
\dot{{\cal Q}}^{\dot{\al}}_
{\,\,\, a }\, {\cal Q}^{ b }_{\,\,\, \al}. 
\label{s-2}
\eea
Compared with the previous OPE, there is no (anti)symmetric
combination in the $SU(4)$ indices.
The first higher spin generator of (\ref{s-2})
is a quasiprimary operator
by using Appendix (\ref{tjjj}).
Because there is no $SU(4)$ index, 
the ${\cal V}$-charge vanishes and moreover the quadratic
expression in ${\cal V}$ arises from the oscillator
construction, we can identify this
as ${\bf 1}_0$ in \cite{SS} \footnote{Note that
the OPEs between ${\cal P}^{\dot{\al}}_{\,\,\,\beta}$
and the weight-$1$ operators are regular except ${\cal L}^{\al}_{\,\,\,
\beta}$, $\dot{\cal L}^{\dot{\al}}_{\,\,\,\dot{\beta}}$, ${\cal D}$,
${\cal S}^{\al}_{a}$, $\dot{{\cal S}}^{a}_{\,\,\,\dot{\al}}$, ${\cal K}^
{\al}_{\dot{\beta}}$, ${\cal U}$ and ${\dot{\cal U}}$ from
Appendix (\ref{psualgebra}). In other words, the OPEs between
${\cal P}^{\dot{\al}}_{\,\,\,\beta}$ and 
the five weight-$1$ operators appearing in the footnote
\ref{fiveweightone} do not have the singular terms.
\label{opeproperty}}.

Let us look at the second higher spin generator in
(\ref{s-2}) which is a primary operator under the stress
energy tensor (\ref{stressenergy}).
We can view this as the tensor product of
the representation ${\bf 4}$ corresponding to the upper index
and the representation
$\overline{{\bf 4}}$ corresponding to the lower index
and moreover
its ${\cal V}$-charge vanishes because
there appear two kinds of supersymmetry generators.
We do not find this higher spin generator from the tables $4$ and $5$
of \cite{SS}. As mentioned before, we put a hat on this generator because
this is a `new' primary operator \footnote{We have similar relation
${\cal Q}^{a}_{\,\,\,\al}(z) \, \dot{{\cal W}}^{
\dot{\beta} }_{\,\,\, b}(w)
\Bigg|_{\frac{1}{(z-w)}} =
\de^{a}_{b} \, {\cal W}^{\dot{\beta}}_{\,\,\,   \al}(w)+
\widehat{{\cal W}}^{a \, \dot{\beta}}_{\,\,\, b \, \al}(w)$
where the generators of right hand side are given by (\ref{s-2}).
The $SU(4)$ indices appear separately.}.

Let us move on the following first order pole
in the OPE between the supersymmetry generator and the second higher
spin generator in (\ref{s-threehalf})
after antisymmetrizing for the lower two indices
\bea
\dot{{\cal Q}}^{\dot{\al}}_{\,\,\, \left[ a \right.}(z) \, {\cal{W}}^{
\left[ b c \right]}_{\,\,\, \left. d \right] \, \beta}(w)
\Bigg|_{\frac{1}{(z-w)}} =
\de^{\left[ b\right.}_{\left[ a \right.}\, {\cal W}^{\left. c\right]
\, \dot{\al} }_{\,\,\, \left. d
\right] \,   \beta}(w)
+ {\cal W}^{\left[b c \right]\, \dot{\al}  }_{\,\,\, \left[ a d \right] \,
  \beta }(w)+
\de^{\left[c \right.}_{\left[a \right.} \,
\widehat{{\cal W}}^{ \left. b \right] \, \dot{\al}}_{\,\,\, \left.
d\right] \, \beta}(w)     
-\frac{1}{4} \, \de^{\left[ c \right.}_{\left[ d \right.}\,       
\widehat{{\cal W}}^{ \left. b \right] \, \dot{\al}}_{\,\,\, \left.
a \right] \, \beta}(w),
\label{s-2opeother}
\eea
where the right hand side of (\ref{s-2opeother}) contains
the following higher spin generators together with the previous
operator in (\ref{s-2})
\bea
{\cal W}^{a \, \dot{\al} }_{\,\,\, b \, \beta} & \equiv &
{\cal V} \, {\cal P}^{\dot{\al}}_{\,\,\,\beta} \, {\cal R}^a_{\,\,\, b}+
{\cal P}^{\dot{\al}}_{\,\,\,\beta} \, {\cal V} \,{\cal R}^a_{\,\,\, b}+
{\cal P}^{\dot{\al}}_{\,\,\,\beta} \, {\cal R}^a_{\,\,\, b}\,
{\cal V}
\nonu \\
&+& {\cal V} \, {\cal R}^a_{\,\,\, b}\,  {\cal P}^{\dot{\al}}_{\,\,\,\beta} \,+
{\cal R}^a_{\,\,\, b}\, {\cal V} \, {\cal P}^{\dot{\al}}_{\,\,\,\beta}+
{\cal R}^a_{\,\,\, b}\,{\cal P}^{\dot{\al}}_{\,\,\,\beta} \,
{\cal V}, 
\nonu \\
{\cal W}^{ \left[ a b \right] \, \dot{\al}}_{\,\,\, \left[ c d \right] \, \al}
& \equiv &
{\cal Q}^{\left[ a \right.}_{\,\,\, \al} \, \dot{{\cal Q}}^{\dot{\al}}_
{\,\,\,\left[c \right.} \, {\cal R}^{\left. b \right]}_{\,\,\, \left. d\right]}
+{\cal Q}^{\left[ a \right.}_{\,\,\, \al} \,  {\cal R}^{\left. b \right]}_{\,\,\,
\left[ d\right.} \, \dot{{\cal Q}}^{\dot{\al}}_
{\,\,\,\left. c \right]}+
{\cal R}^{\left[ b \right.}_{\,\,\, \left[ d\right.}\,
{\cal Q}^{\left. a \right]}_{\,\,\, \al} \, \dot{{\cal Q}}^{\dot{\al}}_
{\,\,\,\left. c \right]}
\nonu \\
&-&  \dot{{\cal Q}}^{\dot{\al}}_
{\,\,\,\left[c \right.}\,
{\cal Q}^{\left[ a \right.}_{\,\,\, \al} \, \, {\cal R}^{\left. b \right]}_{\,\,\, \left. d\right]}
- \dot{{\cal Q}}^{\dot{\al}}_
{\,\,\,\left[ c \right.} \,  {\cal R}^{\left[ b \right.}_{\,\,\,
\left. d\right]} \,{\cal Q}^{\left. a \right]}_{\,\,\, \al} -
{\cal R}^{\left[ b \right.}_{\,\,\, \left[ d\right.}\,
 \dot{{\cal Q}}^{\dot{\al}}_
{\,\,\,\left. c \right]}\, {\cal Q}^{\left. a \right]}_{\,\,\, \al}. 
\label{s-2other}
\eea
We can easily identify the first operator of (\ref{s-2other})
which is a primary 
as the representation ${\bf 15}_0$. We have already observed that
the weight-$1$ operator ${\cal R}^a_{\,\,\,b}$ transforms as this
representation under the $SU(4)$. Moreover, there is a
single ${\cal V}$ in this expression (again from the result of
\cite{SS}) and it is obvious that
the ${\cal V}$-charge is equal to zero.

There are two antisymmetric combinations
between the upper indices and lower indices from the second operator of
(\ref{s-2other}).
It is known that in $SU(4)$, we have ${\bf 6}=\overline{\bf 6}$.
In the tensor product of
${\bf 6} \otimes {\bf 6} = {\bf 1} \oplus {\bf 15} \oplus {\bf 20'}$
\cite{Slansky,FKS},
after subtracting the first two representations, we obtain 
the representation ${\bf 20'}$.
That is, we observe that when we contract
one index from ${\cal W}^{\left[ a b \right] \, \dot{\al}}_{\,\,\,
  \left[ c d \right]\, \al}$, then
the representation ${\bf 15}$ corresponds to
${\cal W}^{\left[a b\right] \, \dot{\al}}_{\,\,\, \left[a d\right]\, \al}$.
Further contraction will give us
${\cal W}^{\left[a b\right] \, \dot{\al}}_{\,\,\, \left[a b\right] \, \al}$
which has a representation
${\bf 1}$. Therefore, we obtain the representation
${\bf 20'}$ by restricting to these two
conditions. It is easy to see that the ${\cal V}$-charge vanishes.
We can check this operator is a primary under the stress energy tensor
(\ref{stressenergy})
\footnote{The conjugated version of (\ref{s-2opeother}) appears
as follows:
${\cal Q}^{\left[ a \right.}_{\,\,\, \al}(z) \, \dot{{\cal{W}}}^{
\left. c\right] \, \dot{\beta} }_{\,\,\, \left[ b d \right] }(w)
\Bigg|_{\frac{1}{(z-w)}} =
\de^{\left[ a \right.}_{\left[ b \right.
}\, {\cal W}^{\left. c\right] \, \dot{\beta} }_{\,\,\, \left. d \right] \,
\al}(w)
+ {\cal W}^{ \left[ a c \right]\, \dot{\beta}  }_{\,\,\, \left[ b d\right] \, \al  }(w)-
\de^{\left[a \right.}_{\left[d \right.} \,
\widehat{{\cal W}}^{ \left. c \right] \, \dot{\beta}}_{\,\,\, \left.
b \right] \, \al}(w)     
-\frac{1}{4} \, \de^{\left[ c \right.}_{\left[ d \right.}\,       
\widehat{{\cal W}}^{ \left. a \right] \, \dot{\beta}}_{\,\,\, \left.
    b \right] \, \al}(w)$ together with 
the footnote \ref{conjugat-1},  and the
relations (\ref{s-2}) and (\ref{s-2other}).}.

We continue to analyze the next higher spin generators
which have nonzero ${\cal V}$-charges.
We can calculate the following OPE and obtain
the first order pole, from (\ref{s-threehalf}),  as
follows:
\bea
{\cal Q}^{a}_{\,\,\, \al}(z) \, {\cal{W}}^{
 \left[ b c \right]}_{\,\,\, a \, \beta}(w)
\Bigg|_{\frac{1}{(z-w)}} =
{\cal W}^{\left[b c\right] a }_{\,\,\, a \,   \beta \al}(w) -\frac{15}{4}\,
\widehat{\cal W}^{[b c]  }_{\,\,\,   \beta \al}(w),        
\label{s-2otherope}
\eea
where the right hand side of (\ref{s-2otherope})
consists of the following higher spin generators
\bea
{\cal W}^{\left[ a b \right] c }_{\,\,\, c \, \al  \beta} & \equiv &
-{\cal Q}^{\left[ a \right.}_{\,\,\, \al}\, {\cal R}^{\left. b \right]}_{\,\,\, c} \,
{\cal Q}^c_{\,\,\,\beta}-
{\cal R}^{\left[ b \right.}_{\,\,\, c} \,
{\cal Q}^{\left. a \right]}_{\,\,\, \al}\,
{\cal Q}^c_{\,\,\,\beta}-
{\cal Q}^{\left[ a \right.}_{\,\,\, \al}\,
{\cal Q}^c_{\,\,\,\beta} \, {\cal R}^{\left. b \right]}_{\,\,\, c} 
\nonu \\
&+&{\cal Q}^c_{\,\,\,\beta}\, {\cal R}^{\left[ b \right.}_{\,\,\, c} \,
{\cal Q}^{\left. a \right]}_{\,\,\, \al}+
{\cal R}^{\left[ b \right.}_{\,\,\, c} \,
\,
{\cal Q}^c_{\,\,\,\beta} \, {\cal Q}^{\left. a \right]}_{\,\,\, \al}+
{\cal Q}^c_{\,\,\,\beta} \, {\cal Q}^{\left[ a \right.}_{\,\,\, \al}\,
{\cal R}^{\left. b \right]}_{\,\,\, c}, 
\nonu \\
\widehat{\cal W}^{\left[a b\right] }_{\,\,\,  \al  \beta} & \equiv &
{\cal V} \, {\cal Q}^{\left[a \right. }_{\,\,\, \al} \,
{\cal Q}^{ \left. b \right] }_{\,\,\, \beta}+
{\cal Q}^{ \left[ a \right. }_{\,\,\, \al} \,{\cal V} \,
{\cal Q}^{ \left. b \right] }_{\,\,\, \beta}
+ {\cal Q}^{ \left[ a \right. }_{\,\,\, \al} \,
{\cal Q}^{ \left. b \right] }_{\,\,\, \beta} \, {\cal V}
\nonu \\     
&-& {\cal V} \, \,
{\cal Q}^{ \left[ b \right. }_{\,\,\, \beta} \,
{\cal Q}^{ \left. a \right] }_{\,\,\, \al}-
{\cal Q}^{ \left[ b \right. }_{\,\,\, \beta} \,{\cal V} \,
{\cal Q}^{ \left. a \right] }_{\,\,\, \al} - 
{\cal Q}^{ \left[ b \right. }_{\,\,\, \beta} \,
{\cal Q}^{ \left. a \right] }_{\,\,\, \al} \,{\cal V}.
\label{s-2other1}
\eea
Note that the upper and lower index $a$ is summed
in the left hand side of the OPE of (\ref{s-2otherope}).
We can identify the first operator of (\ref{s-2other1}) as ${\bf 6}_{-2}$
because the two upper $SU(4)$ indices are antisymmetric  
together with the contraction for other two and
due to the two supersymmetric generators, the ${\cal V}$-charge
becomes $-2$ as before. On the other hands,
the second operator of (\ref{s-2other1}), which has
also ${\cal V}$-charge $-2$ and
consists of the tensor product of ${\bf 4}$ and ${\bf 4}$
(again ${\bf 6}_{-2}$)
of $SU(4)$,
can be regarded as a `new' primary operator
which is not present in \cite{SS}.
We can check this is a primary operator from Appendix
(\ref{tjjj}). As done before, we can
obtain the conjugated version of (\ref{s-2otherope}) with the footnote
\ref{conjugat-1} and there exists a relevant generator \footnote{
\label{compother2}
  That is, we have the following
first order pole, from the footnote \ref{conjugat-1}, $\dot{{\cal Q}}^{\dot{\al}}_{\,\,\, a}(z) \,
\dot{{\cal{W}}}^{
a \, \dot{\beta}}_{\,\,\,  \left[ b d \right] }(w)
\Bigg|_{\frac{1}{(z-w)}} =
-\dot{{\cal W}}^{ a \, \dot{\beta} \dot{\al}}_{\,\,\, \left[b  d\right] a}(w) +\frac{15}{4}\,
\dot{\widehat{\cal W}}^{\dot{\beta}\dot{\al}  }_{\,\,\,   \left[ b d\right]}(w)$ where
$\dot{{\cal W}}^{c \, \dot{\al} \dot{\beta} }_{\,\,\, \left[a b\right] c \, }  \equiv 
-\dot{{\cal Q}}^{\dot{\al}}_{\,\,\, \left[ a \right.}\,
{\cal R}^{ c }_{\,\,\,  \left. b \right]} \,
\dot{{\cal Q}}^{\dot{\beta}}_{\,\,\,c     }-
{\cal R}^{c }_{\,\,\, \left[ b \right.} \,
\dot{{\cal Q}}^{\dot{\al}}_{\,\,\, \left. a \right]}\,
\dot{{\cal Q}}^{\dot{\beta}}_{\,\,\,c}-
\dot{{\cal Q}}^{\dot{\al}}_{\,\,\, \left[ a \right.}\,
\dot{{\cal Q}}^{\dot{\beta}}_{\,\,\,c} \,
{\cal R}^{ c }_{\,\,\, \left. b \right]} 
+\dot{{\cal Q}}^{\dot{\beta}}_{\,\,\,c}\, {\cal R}^{c }_{\,\,\,
\left[ b \right.} \,\dot{{\cal Q}}^{\dot{\al}}_{\,\,\, \left. a \right]}+
{\cal R}^{c }_{\,\,\, \left[ b \right.} \,
\,
\dot{{\cal Q}}^{\dot{\beta}}_{\,\,\,c} \, \dot{{\cal Q}}^{\dot{\al}}_
{\,\,\, \left. a \right]}+
\dot{{\cal Q}}^{\dot{\beta}}_{\,\,\,c} \,
\dot{{\cal Q}}^{\dot{\al}}_{\,\,\, \left[ a \right.}\,
  {\cal R}^{c}_{\,\,\, \left. b\right]}$ corresponding to the
representation $\overline{\bf 6}_2$ and
the new higher spin generator 
$\dot{\widehat{\cal W}}^{\dot{\al}\dot{\beta} }_{\,\,\,  \left[a  b\right]}  \equiv 
{\cal V} \, \dot{{\cal Q}}^{\dot{\al} }_{\,\,\, \left[ a \right.} \,
\dot{{\cal Q}}^{ \dot{\beta} }_{\,\,\, \left. b \right]}+
\dot{{\cal Q}}^{ \dot{\al} }_{\,\,\, \left[ a \right.} \,{\cal V} \,
\dot{{\cal Q}}^{ \dot{\beta} }_{\,\,\, \left. b \right]}
+ \dot{{\cal Q}}^{ \dot{\al} }_{\,\,\, \left[ a \right.} \,
\dot{{\cal Q}}^{ \dot{\beta} }_{\,\,\, \left. b \right]} \, {\cal V}
- {\cal V} \, \,
\dot{{\cal Q}}^{ \dot{\beta} }_{\,\,\, \left[ b \right.} \,
\dot{{\cal Q}}^{ \dot{\al} }_{\,\,\, \left. a \right]}-
\dot{{\cal Q}}^{ \dot{\beta} }_{\,\,\, \left[ b \right.} \,{\cal V} \,
\dot{{\cal Q}}^{ \dot{\al} }_{\,\,\, \left. a \right]} - 
\dot{{\cal Q}}^{ \dot{\beta} }_{\,\,\, \left[ b \right.} \,
  \dot{{\cal Q}}^{ \dot{\al} }_{\,\,\,\left. a \right]} \,{\cal V}$
which transforms as $\overline{\bf 6}_2$.}.

Finally, by considering the following OPE from (\ref{s-threehalf})
we determine the higher spin generator
having nonzero ${\cal V}$-charge, after symmetrizing the
upper indices, 
\bea
{\cal Q}^{\left(a \right.}_{\,\,\,\al}(z) \, {\cal{W}}^{
\left. b \right)}_{\,\,\, \beta}(w)
\Bigg|_{\frac{1}{(z-w)}} =
{\cal W}^{\left( a b \right)}_{\,\,\,  \al \beta}(w),        
\label{s-2other2ope}
\eea
where the right hand side of (\ref{s-2other2ope}) 
can be written as
\bea
{\cal W}^{\left(a b \right) }_{\,\,\, \al \beta} & \equiv &
{\cal V} \, {\cal Q}^{\left( a \right.}_{\,\,\, \al} \,
{\cal Q}^{\left. b \right)}_{\,\,\, \beta}+
{\cal Q}^{\left( a \right.}_{\,\,\, \al} \,{\cal V} \,
{\cal Q}^{\left. b \right)}_{\,\,\, \beta}
+ {\cal Q}^{\left( a \right.}_{\,\,\, \al} \,
{\cal Q}^{\left. b \right)}_{\,\,\, \beta} \, {\cal V}
\nonu \\     
&-& {\cal V} \, \,
{\cal Q}^{\left( b \right.}_{\,\,\, \beta} \,
{\cal Q}^{\left. a \right)}_{\,\,\, \al}-
{\cal Q}^{\left( b \right.}_{\,\,\, \beta} \,{\cal V} \,
{\cal Q}^{\left. a \right)}_{\,\,\, \al} - 
{\cal Q}^{\left( b \right.}_{\,\,\, \beta} \,
{\cal Q}^{\left. a \right)}_{\,\,\, \al} \,{\cal V}.
\label{s-2other2}
\eea
It is obvious to see that this (\ref{s-2other2}),
which is a primary,  has
the representation ${\bf 10}_{-2}$ from the symmetric
combination of the upper two indices.
Simple counting of ${\cal V}$-charge implies that
this higher spin generator has $-2$. Furthermore, it has
linear dependence of ${\cal V}$ as in \cite{SS}
\footnote{
\label{compother1}
  We obtain
$\dot{{\cal Q}}^{\dot{\al}}_{\,\,\,\left( a \right.}(z) \,
\dot{{\cal W}}^{
\dot{\beta}}_{\,\,\, \left. b \right)}(w)
\Bigg|_{\frac{1}{(z-w)}} =
-\dot{{\cal W}}^{\dot{\al}\dot{\beta}}_{\,\,\,  \left( a b \right)}(w)$
together with
$\dot{{\cal W}}^{\dot{\al} \dot{\beta}}_{\,\,\, \left( a b \right)}  \equiv 
{\cal V} \, \dot{{\cal Q}}^{\dot{\al}}_{\,\,\, \left( a \right.} \,
\dot{{\cal Q}}^{\dot{\beta}}_{\,\,\, \left. b \right)}+
\dot{{\cal Q}}^{\dot{\al}}_{\,\,\, \left( a\right.} \,{\cal V} \,
\dot{{\cal Q}}^{\dot{\beta}}_{\,\,\, \left. b \right)}
+ \dot{{\cal Q}}^{\dot{\al}}_{\,\,\, \left( a \right.} \,
\dot{{\cal Q}}^{\dot{\beta}}_{\,\,\, \left. b \right)} \, {\cal V}     
- {\cal V}  \,
\dot{{\cal Q}}^{\dot{\beta}}_{\,\,\, \left( b \right.} \,
\dot{{\cal Q}}^{\dot{\al}}_{\,\,\, \left. a \right)}-
\dot{{\cal Q}}^{\dot{\beta}}_{\,\,\, \left( b \right.} \,{\cal V} \,
\dot{{\cal Q}}^{\dot{\al}}_{\,\,\, \left. a \right)} - 
\dot{{\cal Q}}^{\dot{\beta}}_{\,\,\, \left( b \right.} \,
\dot{{\cal Q}}^{\dot{\al}}_{\,\,\, \left. a \right)} \,{\cal V}$
corresponding to $\overline{\bf 10}_{2}$.}.

%%%%%%%%%%%%%%%%%%%%%%%%%%%%%%%%%%%%%%%%%%%
\subsection{ The $s=\frac{5}{2}$ case: ${\bf 4}_{-1},
\overline{{\bf 4 }}_1, 
\overline{\bf 4}_{-3},  {\bf 4}_3, {\bf 20}_{-1}$
and $\overline{\bf 20}_{1}$}
%%%%%%%%%%%%%%%%%%%%%%%%%%%%%%%%%%%%%%%%%%

From now on, all the higher spin generators
can be related to the corresponding multiplets in the table
$3$ of \cite{SS}.
In previous three cases,
there are some mismatches between the table
$3$ and the tables $4$ and $5$ of \cite{SS}.
As done in previous subsection, we compute the
following OPE from (\ref{s-2other}) and focus on the first order pole,
after antisymmetrizing the upper indices, 
\bea
{\cal Q}^{\left[a \right.}_{\,\,\, \beta}(z) \, {\cal{W}}^{
 \left. b \right] \dot{\al}}_{\,\,\, c \, \ga}(w)
\Bigg|_{\frac{1}{(z-w)}} =
{\cal W}^{ \left[ a b \right] \, \dot{\al} }_{\,\,\, c\, \beta \ga}(w)+
\de^{\left[ a \right.}_{c}\,
{\cal W}^{\left. b \right]  \, \dot{\al} }_{\,\,\,   \beta \ga}
-\frac{1}{4}\, \de^{\left[ b \right.}_{c}
{\cal W}^{\left. a \right]  \, \dot{\al} }_{\,\,\,  \beta \ga}\, (w),        
\label{s-fivehalfope}
\eea
where the right hand side of (\ref{s-fivehalfope})
provides the following higher spin generators
\bea
{\cal W}^{a  \, \dot{\al} }_{\,\,\,  \beta \ga} & \equiv &
{\cal V} \, {\cal Q}^a_{\,\,\,\beta} \, {\cal P}^{\dot{\al}}_{\,\,\,\ga}+
{\cal Q}^a_{\,\,\,\beta} \, {\cal V} \,{\cal P}^{\dot{\al}}_{\,\,\,\ga}+
{\cal Q}^a_{\,\,\,\beta} \, {\cal P}^{\dot{\al}}_{\,\,\,\ga}\, {\cal V} 
\nonu \\
&+&{\cal V} \, {\cal P}^{\dot{\al}}_{\,\,\,\ga} \,
{\cal Q}^a_{\,\,\,\beta} +
{\cal P}^{\dot{\al}}_{\,\,\,\ga}\, {\cal V} \, {\cal Q}^a_{\,\,\,\beta} +
{\cal P}^{\dot{\al}}_{\,\,\,\ga}\, {\cal Q}^a_{\,\,\,\beta} \,{\cal V}, 
\nonu \\
{\cal W}^{\left[ a b \right] \, \dot{\al} }_{\,\,\, c\, \beta \ga} & \equiv &
{\cal Q}^{\left[ a \right.}_{\,\,\,\beta} \,
{\cal R}^{\left. b \right]}_{\,\,\, c} \, {\cal P}^{\dot{\al}}_{\,\,\,\ga}
+{\cal R}^{\left[ b \right.}_{\,\,\, c} \,
{\cal Q}^{\left. a \right]}_{\,\,\,\beta} \,
{\cal P}^{\dot{\al}}_{\,\,\,\ga}+
{\cal Q}^{\left[ a \right.}_{\,\,\,\beta} \,
{\cal P}^{\dot{\al}}_{\,\,\,\ga}\, {\cal R}^{\left. b \right]}_{\,\,\, c} \,
\nonu \\
&+&
{\cal P}^{\dot{\al}}_{\,\,\,\ga}\,
{\cal R}^{\left[ b \right.}_{\,\,\, c} \, 
{\cal Q}^{\left. a \right]}_{\,\,\,\beta}
+{\cal R}^{\left[ b \right.}_{\,\,\, c} \,
{\cal P}^{\dot{\al}}_{\,\,\,\ga}
{\cal Q}^{\left. a \right]}_{\,\,\,\beta} +
{\cal P}^{\dot{\al}}_{\,\,\,\ga}\, {\cal Q}^{\left[ a \right.}_{\,\,\,\beta} \,
{\cal R}^{\left. b \right]}_{\,\,\, c}. 
\label{s-fivehalf}
\eea
We can see that the first generator of (\ref{s-fivehalf})
has the representation ${\bf 4}_{-1}$ with ${\cal V}$-charge
$-1$. For the second generator of (\ref{s-fivehalf}),
there are two upper antisymmetric indices with
a single lower index. We have seen the similar structure
around (\ref{s-threehalf}).
As long as the $SU(4)$ representation with ${\cal V}$-charge
is concerned, there is no difference whether
there is a  factor ${\cal V}$ in (\ref{s-threehalf})
or ${\cal P}^{\dot{\al}}_{\,\,\,\beta}$ in (\ref{s-fivehalf}).
This implies that
the above generator transforms as the representation
${\bf 20}_{-1}$ by subtracting the trace part (with a
contraction in the indices)
with ${\cal V}$-charge $-1$. They are primary
under the stress energy tensor \footnote{
\label{othercomplex} We can determine the similar
OPE, by antisymmetrizing the lower indices, 
$\dot{{\cal Q}}^{\dot{\beta}}_{\,\,\, \left[ a \right.}(z) \,
{\cal{W}}^{
b  \dot{\al}}_{\,\,\,  \left. c \right] \, \ga}(w)
\Bigg|_{\frac{1}{(z-w)}} =
-\dot{{\cal W}}^{ b \, \dot{\beta} \dot{\al} }_{
\,\,\, \left[ a c \right] \,  \ga}(w)-
\de^{b}_{\left[a \right.}\,
\dot{{\cal W}}^{  \, \dot{\beta} \dot{\al} }_{
\,\,\,    \left. c   \right] \, \ga}
+\frac{1}{4}\, \de^{ b }_{\left[ c \right.}
\dot{{\cal W}}^{\dot{\beta} \dot{\al} }_{\,\,\,  \left. a \right]  \,
\ga }\, (w)$ with two higher spin generators
$\dot{{\cal W}}^{\dot{\beta}  \dot{\al} }_{\,\,\,  a\,  \ga}  \equiv 
{\cal V} \, \dot{{\cal Q}}^{\dot{\beta}}_{\,\,\,a} \, {\cal P}^{\dot{\al}}_{\,\,\,\ga}+
\dot{{\cal Q}}^{\dot{\beta}}_{\,\,\,a} \,
{\cal V} \,{\cal P}^{\dot{\al}}_{\,\,\,\ga}+
\dot{{\cal Q}}^{\dot{\beta}}_{\,\,\,a} \, {\cal P}^{\dot{\al}}_{\,\,\,\ga}\, {\cal V} 
+{\cal V} \, {\cal P}^{\dot{\al}}_{\,\,\,\ga} \,
\dot{{\cal Q}}^{\dot{\beta}}_{\,\,\,a} +
{\cal P}^{\dot{\al}}_{\,\,\,\ga}\, {\cal V} \,
\dot{{\cal Q}}^{\dot{\beta}}_{\,\,\,a} +
{\cal P}^{\dot{\al}}_{\,\,\,\ga}\,
\dot{{\cal Q}}^{\dot{\beta}}_{\,\,\,a} \,{\cal V}$
transforming as $\overline{\bf 4}_1$ and 
$\dot{{\cal W}}^{ b \, \dot{\beta} \dot{\al} }_{\,\,\, \left[ a c \right]
\,  \ga}  \equiv 
\dot{{\cal Q}}^{\dot{\beta}}_{\,\,\,\left[ a \right.} \,
{\cal R}^{ b }_{\,\,\, \left. c \right]} \, {\cal P}^{\dot{\al}}_{\,\,\,\ga}
+{\cal R}^{ b }_{\,\,\, \left[ c \right.} \,
\dot{{\cal Q}}^{\dot{\beta}}_{\,\,\,\left. a \right]} \,
{\cal P}^{\dot{\al}}_{\,\,\,\ga}+
\dot{{\cal Q}}^{\dot{\beta}}_{\,\,\,\left[ a \right.} \,
{\cal P}^{\dot{\al}}_{\,\,\,\ga}\, {\cal R}^{b }_{\,\,\, \left. c \right]} \,
+
{\cal P}^{\dot{\al}}_{\,\,\,\ga}\,
{\cal R}^{b }_{\,\,\, \left[c \right.} \, 
\dot{{\cal Q}}^{\dot{\beta}}_{\,\,\,\left. a \right]}
+{\cal R}^{ b }_{\,\,\, \left[c \right.} \,
{\cal P}^{\dot{\al}}_{\,\,\,\ga}
\dot{{\cal Q}}^{\dot{\beta}}_{\,\,\,\left. a \right]} +
{\cal P}^{\dot{\al}}_{\,\,\,\ga}\, \dot{{\cal Q}}^{\dot{\beta}}_{\,\,\,\left[ a
\right.} \,
  {\cal R}^{ b }_{\,\,\, \left. c \right]}$ which transforms as
$\overline{\bf 20}_1$.}.

The next case can be obtained from the following OPE
result by using the higher spin generator (\ref{s-2other1}) properly
(complete antisymmetrization of the upper indices)
\bea
{\cal Q}^{\left[a \right.}_{\,\,\, \al}(z) \, {\cal{W}}^{
 \left. b c d \right] }_{\,\,\,  d\, \beta  \ga}(w)
\Bigg|_{\frac{1}{(z-w)}} =
-\de^{\left[ a \right.}_{d}\,
{\cal W}^{ \left. b c d \right] }_{\,\,\,  \beta \ga \al}(w)
+\frac{1}{4}\, \de^{\left[ d \right.}_{d}\,
{\cal W}^{ \left. b c a \right] }_{\,\,\,  \beta \ga \al}(w),
\label{s-fivehalfotherope}
\eea
where the right hand side of (\ref{s-fivehalfotherope})
contains the following higher spin generator
\bea
{\cal W}^{\left[a b c\right] }_{\,\,\,  \al \beta \ga} & \equiv &
{\cal Q}^{\left[ a \right. }_{\,\,\, \al}\, {\cal Q}^b_{\,\,\,\beta}
\, {\cal Q}^{\left. c \right]}_{\,\,\,\ga}+
{\cal Q}^{\left[ c \right.}_{\,\,\,\ga} 
\, {\cal Q}^{ a  }_{\,\,\, \al}\,
{\cal Q}^{\left. b \right]}_{\,\,\,\beta}+
{\cal Q}^{\left[ b \right.}_{\,\,\,\beta}
\, {\cal Q}^{ c }_{\,\,\,\ga}
{\cal Q}^{\left. a \right] }_{\,\,\, \al}\,
\nonu \\
&-&{\cal Q}^{\left[ a \right. }_{\,\,\, \al}\,
\, {\cal Q}^{ c }_{\,\,\,\ga} \,  {\cal Q}^{\left. b \right]}_{\,\,\,\beta}-
{\cal Q}^{\left[ b \right. }_{\,\,\,\beta}
\, {\cal Q}^{ a  }_{\,\,\, \al}\,
{\cal Q}^{\left. c \right]}_{\,\,\,\ga} -
{\cal Q}^{ \left[ c \right. }_{\,\,\,\ga}\,
{\cal Q}^{ b }_{\,\,\,\beta} \,
{\cal Q}^{\left. a \right] }_{\,\,\, \al}.
\label{s-fivehalfother}
\eea
First of all, the ${\cal V}$-charge of
(\ref{s-fivehalfother}) is given by 
$-3$.
From the tensor product of ${\bf 4} \otimes {\bf 4}
\otimes {\bf 4}$ \cite{Slansky,FKS} due to the three upper indices,
we obtain the following decomposition
$\overline{{\bf 4}} \oplus \overline{\bf 20} \oplus \overline{\bf 20}
\oplus \overline{\bf 20''}$. Then by taking the totally
antisymmetric combination of the indices,
the representation $\overline{\bf 4}_{-3}$
with ${\cal V}$-charge can be obtained and we can check
this (\ref{s-fivehalfother}) is a primary operator
\footnote{
\label{compother}
  In this case, we have
$\dot{{\cal Q}}^{\dot{\al}}_{\,\,\, \left[a \right.}(z) \, \dot{{\cal W}}^{
d\, \dot{\beta}\dot{\ga} }_{\,\,\,  \left. b c d \right]}(w)
\Bigg|_{\frac{1}{(z-w)}} =
\de^{d}_{\left[a \right.}\,
\dot{{\cal W}}^{ \dot{\beta}\dot{\ga}\dot{\al} }_{\,\,\,  \left. b c d \right]}(w)
-\frac{1}{4}\, \de^{ d }_{\left[d \right.}\,
\dot{{\cal W}}^{ \dot{\beta}\dot{\ga}\dot{\al} }_{\,\,\,  \left. b c  a\right]}(w)$
together with the higher spin generator
$\dot{{\cal W}}^{\dot{\al}\dot{\beta}\dot{\ga} }_{\,\,\,  \left[ a b c
  \right] }  \equiv 
\dot{{\cal Q}}^{\dot{\al} }_{\,\,\, \left[ a \right.}\,
\dot{{\cal Q}}^{\dot{\beta}}_{\,\,\,b}
\, \dot{{\cal Q}}^{\dot{\ga}}_{\,\,\,\left. c \right]}+
\dot{{\cal Q}}^{\dot{\ga}}_{\,\,\,\left[ c \right.} 
\, \dot{{\cal Q}}^{ \dot{\al}  }_{\,\,\, a}\,
\dot{{\cal Q}}^{\dot{\beta}}_{\,\,\,\left. b \right]}+
\dot{{\cal Q}}^{\dot{\beta}}_{\,\,\,\left[ b \right.}
\, \dot{{\cal Q}}^{ \dot{\ga} }_{\,\,\,c}
\dot{{\cal Q}}^{\dot{\al} }_{\,\,\, \left. a\right]}
-\dot{{\cal Q}}^{\dot{\al} }_{\,\,\, \left[ a \right.}\,
\, \dot{{\cal Q}}^{ \dot{\ga} }_{\,\,\,c} \,
\dot{{\cal Q}}^{\dot{\beta}}_{\,\,\, \left. b \right]}-
\dot{{\cal Q}}^{\dot{\beta} }_{\,\,\,\left[ b \right.}
\, \dot{{\cal Q}}^{ \dot{\al}  }_{\,\,\, a}\,
\dot{{\cal Q}}^{\dot{\ga}}_{\,\,\,\left. c \right]} -
\dot{{\cal Q}}^{\dot{\ga} }_{\,\,\,\left[ c \right.}\,
\dot{{\cal Q}}^{ \dot{\beta} }_{\,\,\,b} \,
\dot{{\cal Q}}^{\dot{\al} }_{\,\,\, \left. a \right]}$ transforming
as ${\bf 4}_3$.}.

%%%%%%%%%%%%%%%%%%%%%%%%%%%%%%%%%%%%%%
\subsection{ The $s=3$ case: ${\bf 1}_0, {\bf 15}_0,
  {\bf 6}_{-2}$ and ${\bf 6}_2$}
%%%%%%%%%%%%%%%%%%%%%%%%%%%%%%%%%%%%%%

Now we analyze the following OPE, from the previous
result in (\ref{s-fivehalf}),
\bea
\dot{{\cal Q}}^{\dot{\al}}_{\,\,\, a}(z) \, {\cal{W}}^{
b  \, \dot{\beta} }_{\,\,\,     \ga \de}(w)
\Bigg|_{\frac{1}{(z-w)}} =
\de^{ b }_{a}\,
{\cal W}^{ \dot{\al} \dot{\beta}  }_{\,\,\,\ga \de}(w) +
\widehat{\cal W}^{b  \, \dot{\beta} \dot{\al}  }_
        {\,\,\,   a\, \ga \de }(w),
\label{s-3ope1}
\eea
where the right hand side of (\ref{s-3ope1})
has the following higher spin generators 
\bea
{\cal W}^{ \dot{\al} \dot{\beta} }_{\,\,\,    \ga \de} & \equiv &
{\cal V} \, {\cal P}^{\dot{\al}}_{\,\,\,\ga} \, {\cal P}^{\dot{\beta}}_{\,\,\,\de}
+ {\cal P}^{\dot{\al}}_{\,\,\,\ga} \,  {\cal V} \,{\cal P}^{\dot{\beta}}_{\,\,\,\de}
+{\cal P}^{\dot{\al}}_{\,\,\,\ga} \, {\cal P}^{\dot{\beta}}_{\,\,\,\de}\,
{\cal V} 
\nonu \\
&+&{\cal V} \, {\cal P}^{\dot{\beta}}_{\,\,\,\de}\,
{\cal P}^{\dot{\al}}_{\,\,\,\ga} 
+{\cal P}^{\dot{\beta}}_{\,\,\,\de}\, {\cal V} \,
{\cal P}^{\dot{\al}}_{\,\,\,\ga} 
+{\cal P}^{\dot{\beta}}_{\,\,\,\de}\,{\cal P}^{\dot{\al}}_{\,\,\,\ga} \,
{\cal V}, 
\nonu \\
\widehat{\cal W}^{a  \, \dot{\al}\dot{\beta}  }_{\,\,\,   b\, \ga \de } & \equiv &
{\cal Q}^{ a }_{\,\,\,\ga}
\, \dot{{\cal Q}}^{\dot{\beta}}_{\,\,\,b} \, {\cal P}^{\dot{\al}}_{\,\,\,\de}
+{\cal Q}^{ a }_{\,\,\,\ga}
\, {\cal P}^{\dot{\al}}_{\,\,\,\de} \, \dot{\cal Q}^{\dot{\beta}}_{\,\,\,b}
+ {\cal P}^{\dot{\al}}_{\,\,\,\de}\,
{\cal Q}^{a }_{\,\,\,\ga}
\, \dot{{\cal Q}}^{\dot{\beta}}_{\,\,\,b} 
\nonu \\
&-&
\dot{{\cal Q}}^{\dot{\beta}}_{\,\,\,b}\,
{\cal Q}^{ a }_{\,\,\,\ga} \, {\cal P}^{\dot{\al}}_{\,\,\,\de}
- \, \dot{{\cal Q}}^{\dot{\beta}}_{\,\,\,b}
\, {\cal P}^{\dot{\al}}_{\,\,\,\de}
\, {\cal Q}^{a }_{\,\,\,\ga}
-
{\cal Q}^{a}_{\,\,\,\ga}
{\cal P}^{\dot{\al}}_{\,\,\,\de}
\, \dot{{\cal Q}}^{\dot{\beta}}_{\,\,\,b}.
\label{s-3}
\eea
For the first generator of (\ref{s-3}), there is no $SU(4)$ index
and the ${\cal V}$-charge is equal to zero.
Then we can identify this with the representation ${\bf 1}_0$.
For the second generator, 
the ${\cal V}$-charge vanishes also and it is given by
the tensor product between the representation ${\bf 4}$ and
$\overline{\bf 4}$. In the construction of \cite{SS},
we cannot find this higher spin generator. We can check
that they (\ref{s-3}) are primary operators
\footnote{ By using
the higher spin generator appearing in the footnote
\ref{othercomplex}, we obtain 
$
{\cal Q}^{a}_{\,\,\, \al}(z) \, \dot{{\cal W}}^{
\dot{\beta} \dot{\ga} }_{\,\,\,     b \,  \de}(w)
\Bigg|_{\frac{1}{(z-w)}} =
\de^{ a }_{b}\,
{\cal W}^{  \dot{\ga} \dot{\beta}   }_{\,\,\,\al \de}(w) +
\widehat{\cal W}^{a  \, \dot{\ga}
  \dot{\beta}   }_{\,\,\,   b\, \al \de }$ where the relations in
(\ref{s-3})
are used.}.

Now we describe the following OPE together with
(\ref{s-fivehalf})
\bea
\dot{{\cal Q}}^{\dot{\al}}_{\,\,\, a}(z) \, {\cal{W}}^{
\left[ b c \right] \, \dot{\beta} }_{\,\,\,  d \,   \ga \de}(w)
\Bigg|_{\frac{1}{(z-w)}} =
\de^{\left[ b \right.}_{a}\,
{\cal W}^{ \left.  c \right] \, \dot{\al} \dot{\beta}  }_{\,\,\,
d    \ga \de}(w) + \de^{\left[ c \right.}_{a}\,
\widehat{\cal W}^{ \left.  b \right] \,  \dot{\beta} \dot{\al}  }_{\,\,\,
d    \ga \de}(w) -\frac{1}{4}\,
\de^{\left[ c \right.}_{d}\,
\widehat{\cal W}^{ \left.  b \right] \,  \dot{\beta} \dot{\al}  }_{\,\,\,
a    \ga \de}(w),
\label{s-3otherope}
\eea
where the right hand side of (\ref{s-3otherope})
contains the following higher spin generator
\bea
{\cal W}^{a \, \dot{\al} \dot{\beta} }_{\,\,\,  b \,  \ga \de} & \equiv &
{\cal P}^{\dot{\al}}_{\,\,\,\ga} \, {\cal P}^{\dot{\beta}}_{\,\,\,\de}
\, {\cal R}^{a}_{\,\,\,b}+
{\cal P}^{\dot{\al}}_{\,\,\,\ga} \,  {\cal R}^{a}_{\,\,\,b}\,
{\cal P}^{\dot{\beta}}_{\,\,\,\de}
+{\cal R}^{a}_{\,\,\,b}\,
{\cal P}^{\dot{\al}}_{\,\,\,\ga} \, {\cal P}^{\dot{\beta}}_{\,\,\,\de}
\nonu \\
&+& {\cal P}^{\dot{\beta}}_{\,\,\,\de}\, {\cal P}^{\dot{\al}}_{\,\,\,\ga} \,
{\cal R}^{a}_{\,\,\,b}+
{\cal P}^{\dot{\beta}}_{\,\,\,\de}\, {\cal R}^{a}_{\,\,\,b}\,
{\cal P}^{\dot{\al}}_{\,\,\,\ga} 
+{\cal R}^{a}_{\,\,\,b}\,
{\cal P}^{\dot{\beta}}_{\,\,\,\de}\, {\cal P}^{\dot{\al}}_{\,\,\,\ga}, 
\label{s-3other}
\eea
which transforms as ${\bf 15}_0$ with a vanishing
${\cal V}$-charge and is a primary operator
\footnote{
  Again by using the higher spin generator in the footnote
\ref{othercomplex}
we determine ${\cal Q}^{a}_{\,\,\, \al}(z) \, \dot{{\cal W}}^{
c\, \dot{\beta} \dot{\ga} }_{\,\,\,  \left[ b  d \right] \,   \de}(w)
\Bigg|_{\frac{1}{(z-w)}} =
\de^{a}_{\left[ b \right.}\,
{\cal W}^{ c \, \dot{\beta} \dot{\ga}  }_{\,\,\,
\left. d \right]    \al \de}(w) - \de^{a}_{\left[ d \right.}\,
\widehat{\cal W}^{ c \,  \dot{\ga} \dot{\beta}   }_{\,\,\,
  \left. b \right]   \, \al  \de}(w)+
\frac{1}{4} \,
 \de^{c}_{\left[ d \right.}\,
\widehat{\cal W}^{ a \,  \dot{\ga} \dot{\beta}   }_{\,\,\,
  \left. b \right]   \, \al  \de}(w)$ where the relations (\ref{s-3}) and
 (\ref{s-3other})
 are used.}.

Finally, in this subsection, we consider the following
OPE with complete antisymmetric upper indices 
\bea
{\cal Q}^{\left[a \right.}_{\,\,\, \al}(z) \, {\cal{W}}^{
 \left. b c \right] \, \dot{\al} }_{\,\,\,  d \, \beta  \ga}(w)
\Bigg|_{\frac{1}{(z-w)}} =
-\de^{\left[ a \right.}_{d}\,
{\cal W}^{ \left. b c \right] \, \dot{\al}  }_{\,\,\, \beta  \al  \ga }(w)
+ \frac{1}{4}\,
\de^{\left[ c \right.}_{d}\,
{\cal W}^{ \left. b a \right] \, \dot{\al}  }_{\,\,\, \beta  \al \ga }(w),
\label{s-3other1ope}
\eea
where the right hand side of (\ref{s-3other1ope})
contains the following higher spin generator
\bea
{\cal W}^{\left[ a b\right] \,
\dot{\al}  }_{\,\,\, \beta   \ga \de} & \equiv &
{\cal Q}^{\left[ a \right.}_{\,\,\,\beta}
\, {\cal Q}^{\left. b \right]}_{\,\,\,\ga} \, {\cal P}^{\dot{\al}}_{\,\,\,\de}
+{\cal Q}^{\left[ a \right.}_{\,\,\,\beta}
 \, {\cal P}^{\dot{\al}}_{\,\,\,\de} \, {\cal Q}^{\left. b \right]}_{\,\,\,\ga}
+ {\cal P}^{\dot{\al}}_{\,\,\,\de}\,
{\cal Q}^{\left[ a \right.}_{\,\,\,\beta}
\, {\cal Q}^{\left. b \right]}_{\,\,\,\ga} 
\nonu \\
&-&
{\cal Q}^{\left[ b \right.}_{\,\,\,\ga}\,
{\cal Q}^{\left. a \right]}_{\,\,\,\beta} \, {\cal P}^{\dot{\al}}_{\,\,\,\de}
- \, {\cal Q}^{\left[ b \right.}_{\,\,\,\ga}
\, {\cal P}^{\dot{\al}}_{\,\,\,\de}
\, {\cal Q}^{\left. a \right]}_{\,\,\,\beta}
-
{\cal Q}^{\left[ a \right.}_{\,\,\,\beta}
{\cal P}^{\dot{\al}}_{\,\,\,\de}
\, {\cal Q}^{\left. b \right]}_{\,\,\,\ga}, 
\label{s-3other1}
\eea
which transforms as ${\bf 6}_{-2}$ (from the antisymmetric
combination of upper two indices) with $\cal V$-charge $-2$
and is a primary operator.
In this case, we have the conjugated version of this
higher spin generator with corresponding OPE as follows
\footnote{
\label{othercomplex1}
That is,
$\dot{{\cal Q}}^{\dot{\al}}_{\,\,\, \left[ a\right.}(z) \,
\dot{{\cal{W}}}^{
c\, \dot{\beta}\dot{\ga}  }_{\,\,\, \left. b d \right] \,  \ga}(w)
\Bigg|_{\frac{1}{(z-w)}} =
\de^{c}_{\left[a \right.}\,
\dot{{\cal W}}^{ \dot{\beta} \dot{\al} \dot{\ga}   }_
{\,\,\, \left. b     d \right]\, \ga}(w)
- \frac{1}{4}\,
\de^{c }_{\left[d \right.}\,
\dot{{\cal W}}^{ \dot{\beta} \dot{\al} \dot{\ga}   }_
{\,\,\, \left. b a \right]\,   \ga }(w)$ with
the higher spin generator 
$\dot{{\cal W}}^{ \dot{\beta} \dot{\ga} \dot{\al}  }_{\,\,\, \left[ a b
\right] \, \de}  \equiv 
\dot{{\cal Q}}^{\dot{\beta}}_{\,\,\,\left[a \right.}
\, \dot{{\cal Q}}^{\dot{\ga}}_{\,\,\, \left. b \right]}
\, {\cal P}^{\dot{\al}}_{\,\,\,\de}
+\dot{{\cal Q}}^{\dot{\beta}}_{\,\,\,\left[ a \right.}
\, {\cal P}^{\dot{\al}}_{\,\,\,\de} \,
\dot{{\cal Q}}^{\dot{\ga}}_{\,\,\, \left. b \right]}
+ {\cal P}^{\dot{\al}}_{\,\,\,\de}\,
\dot{{\cal Q}}^{\dot{\beta}}_{\,\,\,\left[ a \right.}
\, \dot{{\cal Q}}^{\dot{\ga}}_{\,\,\, \left. b \right]} 
-
\dot{{\cal Q}}^{\dot{\ga}}_{\,\,\,\left[ b \right.}\,
\dot{{\cal Q}}^{\dot{\beta}}_{\,\,\,\left. a \right]} \,
{\cal P}^{\dot{\al}}_{\,\,\,\de}
- \, \dot{{\cal Q}}^{\dot{\ga}}_{\,\,\,\left[ b \right.}
\, {\cal P}^{\dot{\al}}_{\,\,\,\de}
\, \dot{{\cal Q}}^{\dot{\beta}}_{\,\,\, \left. a \right]}
-
\dot{{\cal Q}}^{\dot{\beta}}_{\,\,\,\left[ a \right.}
{\cal P}^{\dot{\al}}_{\,\,\,\de}
\, \dot{{\cal Q}}^{\dot{\ga}}_{\,\,\, \left. b \right]}$
corresponding to the representation $\overline{\bf 6}_2$.}.

%%%%%%%%%%%%%%%%%%%%%%%%%%%%%%%%%%%%%%%%%%
\subsection{ The $s=\frac{7}{2}$ case: ${\bf 4}_{-1}$ and
  $\overline{\bf 4}_1$}
%%%%%%%%%%%%%%%%%%%%%%%%%%%%%%%%%%%%%%%%%%

By using (\ref{s-3other1}), we calculate the following OPE
and read off the first order pole
\bea
\dot{{\cal Q}}^{\dot{\al}}_{\,\,\, a}(z) \, {\cal{W}}^{
\left[ b c \right] \, \dot{\beta} }_{\,\,\,     \ga \de \ep}(w)
\Bigg|_{\frac{1}{(z-w)}} =
\de^{ \left[ b\right. }_{a}\,
{\cal W}^{ \left. c \right]
\, \dot{\al} \dot{\beta}  }_{\,\,\,\de \ga \ep}(w)
-\de^{\left[ c \right.}_{a} \, 
{\cal W}^{ \left. b \right]
\, \dot{\al} \dot{\beta}  }_{\,\,\,\de \ga \ep}(w),
\label{s-7halfope}
\eea
where   the right hand side of (\ref{s-7halfope}) contains
the higher spin generator
\bea
{\cal W}^{a  \, \dot{\al} \dot{\beta}  }_{\,\,\,   \ga \de \ep} & \equiv &
{\cal Q}^{a}_{\,\,\,\ga}\, {\cal P}^{\dot{\al}}_{\,\,\,\de}
\, {\cal P}^{\dot{\beta}}_{\,\,\,\ep}
+ {\cal P}^{\dot{\al}}_{\,\,\,\de}
{\cal Q}^{a}_{\,\,\,\ga}\, {\cal P}^{\dot{\beta}}_{\,\,\,\ep}
+ {\cal P}^{\dot{\al}}_{\,\,\,\de}
\, {\cal P}^{\dot{\beta}}_{\,\,\,\ep}\, {\cal Q}^{a}_{\,\,\,\ga}
\nonu \\
&+&
{\cal Q}^{a}_{\,\,\,\ga}\, 
{\cal P}^{\dot{\beta}}_{\,\,\,\ep}\,
{\cal P}^{\dot{\al}}_{\,\,\,\de}
+  {\cal P}^{\dot{\beta}}_{\,\,\,\ep}\,
{\cal Q}^{a}_{\,\,\,\ga}\,
{\cal P}^{\dot{\al}}_{\,\,\,\de}
+{\cal P}^{\dot{\beta}}_{\,\,\,\ep}\,
{\cal P}^{\dot{\al}}_{\,\,\,\de}
\, {\cal Q}^{a}_{\,\,\,\ga},
\label{s-7half}
\eea
which transforms as ${\bf 4}_{-1}$ from the upper index
$a$ with ${\cal V}$-charge $-1$ and is a primary operator.
Furthermore, there exists a relevant OPE with the conjugated
higher spin generator \footnote{
\label{othercomplex2}
In other words,
from the higher spin generator in the footnote
\ref{othercomplex1}, we have
${\cal Q}^{a}_{\,\,\, \al}(z) \, \dot{{\cal{W}}}^{
\dot{\beta} \dot{\de} \dot{\ga}}_{\,\,\, \left[ b c \right] \, \ep}(w)
\Bigg|_{\frac{1}{(z-w)}} =
\de^{ a }_{\left[ b \right.}\,
\dot{{\cal W}}^{ \dot{\beta}  \dot{\de} \dot{\ga}  }_{\,\,\, \left. c \right] \al
\ep}(w)
-\de^{a }_{\left[ c \right.} \, 
\dot{{\cal W}}^{ 
\dot{\beta}  \dot{\de}\dot{\ga} }_{\,\,\, \left. b \right] \al \ep}(w)$
and $\dot{{\cal W}}^{ \dot{\al} \dot{\beta} \dot{\ga}  }_{\,\,\, a \,   \de \ep}  \equiv 
\dot{{\cal Q}}^{\dot{\ga}}_{\,\,\,a}\, {\cal P}^{\dot{\al}}_{\,\,\,\de}
\, {\cal P}^{\dot{\beta}}_{\,\,\,\ep}
+ {\cal P}^{\dot{\al}}_{\,\,\,\de}
\dot{{\cal Q}}^{\dot{\ga}}_{\,\,\,a}\, {\cal P}^{\dot{\beta}}_{\,\,\,\ep}
+ {\cal P}^{\dot{\al}}_{\,\,\,\de}
\, {\cal P}^{\dot{\beta}}_{\,\,\,\ep}\,
\dot{{\cal Q}}^{\dot{\ga}}_{\,\,\,a}
+
\dot{{\cal Q}}^{\dot{\ga}}_{\,\,\,a}\, 
{\cal P}^{\dot{\beta}}_{\,\,\,\ep}\,
{\cal P}^{\dot{\al}}_{\,\,\,\de}
+  {\cal P}^{\dot{\beta}}_{\,\,\,\ep}\,
\dot{{\cal Q}}^{\dot{\ga}}_{\,\,\,a}\,
{\cal P}^{\dot{\al}}_{\,\,\,\de}
+{\cal P}^{\dot{\beta}}_{\,\,\,\ep}\,
{\cal P}^{\dot{\al}}_{\,\,\,\de}
\, \dot{{\cal Q}}^{\dot{\ga}}_{\,\,\,a}$ transforming as $\overline{\bf 4}_1$.}. 

%%%%%%%%%%%%%%%%%%%%%%%%%%%%%%%%%
\subsection{ The $s=4$ case: ${\bf 1}_0$}
%%%%%%%%%%%%%%%%%%%%%%%%%%%%%%%%%

We obtain the following OPE, by using (\ref{s-7half}),
\bea
\dot{{\cal Q}}^{\dot{\al}}_{\,\,\, a}(z) \, {\cal{W}}^{
b  \, \dot{\ga} \dot{\de} }_{\,\,\,     \beta \ep \rho}(w)
\Bigg|_{\frac{1}{(z-w)}} =
\de^{  b }_{a}\,
{\cal W}^{ 
\, \dot{\al} \dot{\ga} \dot{\de} }_{\,\,\,
\beta \ep \rho}(w),
\label{s-4ope}
\eea
where the right hand side of (\ref{s-4ope}) has the higher spin generator
\bea
{\cal W}^{\dot{\al} \dot{\beta} \dot{\ga}  }_{\,\,\,   \de \ep \rho} & \equiv &
{\cal P}^{\dot{\al}}_{\,\,\,\de}\, {\cal P}^{\dot{\beta}}_{\,\,\,\ep}
\, {\cal P}^{\dot{\ga}}_{\,\,\,\rho}+
+\, {\cal P}^{\dot{\ga}}_{\,\,\,\rho}\,
{\cal P}^{\dot{\al}}_{\,\,\,\de}\, {\cal P}^{\dot{\beta}}_{\,\,\,\ep}
+ {\cal P}^{\dot{\beta}}_{\,\,\,\ep}
\, {\cal P}^{\dot{\ga}}_{\,\,\,\rho}\, {\cal P}^{\dot{\al}}_{\,\,\,\de}
\nonu \\
&+&
{\cal P}^{\dot{\al}}_{\,\,\,\de}
\, {\cal P}^{\dot{\ga}}_{\,\,\,\rho} \,  {\cal P}^{\dot{\beta}}_{\,\,\,\ep}+
+ {\cal P}^{\dot{\beta}}_{\,\,\,\ep}\,
{\cal P}^{\dot{\al}}_{\,\,\,\de}\,
{\cal P}^{\dot{\ga}}_{\,\,\,\rho}\,
+ {\cal P}^{\dot{\ga}}_{\,\,\,\rho}\, {\cal P}^{\dot{\beta}}_{\,\,\,\ep}
\, {\cal P}^{\dot{\al}}_{\,\,\,\de}.
\label{s-4}
\eea
Again this transforms as ${\bf 1}_0$ with ${\cal V}$-charge zero
because there is no $SU(4)$ index. As described in the footnote
\ref{opeproperty}, the OPEs between the supersymmetry
generators and the ${\cal P}^{\dot{\al}}_{\,\,\,\beta}$
do not have any singular terms, we do not find any new higher
spin generators from (\ref{s-4}) \footnote{
%\label{complexother}
  Similarly, from the higher spin generator in the
  footnote \ref{othercomplex2}, there is a relation
  ${\cal Q}^{a}_{\,\,\, \al}(z) \, \dot{{\cal{W}}}^{
\dot{\beta} \dot{\ga}\dot{\de} }_{\,\,\,     b \, \ep \rho}(w)
\Bigg|_{\frac{1}{(z-w)}} =
\de^{  a }_{b}\,
{\cal W}^{ 
\, \dot{\beta} \dot{\ga} \dot{\de} }_{\,\,\,
\al \ep \rho}(w)$.}.

In this section,
the higher spin generators are obtained in
(\ref{s-one}), (\ref{s-threehalf}), (\ref{s-2}), (\ref{s-2other}),
(\ref{s-2other1}), (\ref{s-2other2}), (\ref{s-fivehalf}),
(\ref{s-fivehalfother}), (\ref{s-3}), (\ref{s-3other}),
(\ref{s-3other1}), (\ref{s-7half}), (\ref{s-4}), the footnotes
\ref{conjugat-1},  \ref{compother2},
\ref{compother1},
\ref{othercomplex},
\ref{compother},
\ref{othercomplex1}, and 
\ref{othercomplex2} explicitly.
They are written in terms of the cubic terms between the
weight-$1$ operators and  are summarized by the Table $1$
with $SU(4)$ representations and ${\cal V}$-charges.

%%%%%%%%%%%%%%%%%%%%%%%%%%%%%%%%%%%%%%%%%%%%%%%%%%%%%%%%%%%%%%%%%%%
\begin{table}[ht]
\centering % used for centering table
\begin{tabular}{|c|c| } % centered columns (4 columns)
\hline %inserts double horizontal lines
 & Higher spin generators  \\ [0.5ex] % inserts table
%heading
\hline  % inserts single horizontal line
$s=1$  & ${\cal W}({\bf 1}_0), \,
{\cal W}^{a}_{\,\,\,b}({\bf 15}_0)$ 
\\ % inserting body of the table
[1ex]
\hline
$s=\frac{3}{2}$ &  ${\cal W}^a_{\,\,\,\al}({\bf 4}_{-1}), \,
\dot{{\cal W}}^{\dot{\al}}_{\,\,\,a}(\overline{\bf 4}_{1}), \,
{\cal W}^{\left[ a b \right]}_{\,\,\,c \, \al}({\bf 20}_{-1}), \,
\dot{{\cal W}}^{a\, \dot{\al}}_{\,\,\, \left[ b c \right]}(\overline{\bf 20}_{1})$  \\
[1ex]
\hline
$s=2$ & ${\cal W}^{\dot{\al}}_{\,\,\,\beta}({\bf 1}_0), 
{\cal W}^{a \, \dot{\al}}_{\,\,\,b \,\beta}({\bf 15}_0),
 {\cal W}^{\left[ a b \right] \,
\dot{\al}}_{\,\,\, \left[ c d \right] \, \beta}({\bf 20'_{\rm{0}}}),
{\cal W}^{\left[ a b \right] c}_{c \, \al \beta}({\bf 6}_{-2}),
\dot{{\cal W}}^{c \, \dot{\al} \dot{\beta}}_{\,\,\, \left[ a b\right] c}(\overline{\bf 6}_{2}),
{\cal W}^{\left( a b \right)}_{\,\,\, \al \beta}({\bf 10}_{-2}), 
\dot{{\cal W}}^{\dot{\al} \dot{\beta}}_{\,\,\, \left( a b \right)}(\overline{\bf 10}_{2}) $ \\
[1ex]
\hline
$s=\frac{5}{2}$ &  ${\cal W}^{a \, \dot{\al}}_{\,\,\, \beta \ga}({\bf 4}_{-1}),\,
\dot{{\cal W}}^{\dot{\al} \dot{\beta}}_{\,\,\, a \, \ga}(\overline{{\bf 4 }}_1), \, 
{\cal W}^{\left[a b c\right]}_{\,\,\, \al \beta \ga}(\overline{\bf 4}_{-3}),\,
\dot{{\cal W}}^{\dot{\al} \dot{\beta} \dot{\ga}}_{\,\,\, \left[a b c\right]}({\bf 4}_3), \,
{\cal W}^{\left[a b\right] \, \dot{\al}}_{\,\,\, c \, \beta \ga}({\bf 20}_{-1}), \,
\dot{{\cal W}}^{a \, \dot{\al} \dot{\beta}}_{\,\,\, \left[b c\right] \, \ga}
(\overline{\bf 20}_{1})$ \\
[1ex]
\hline
$s=3$ & ${\cal W}^{\dot{\al} \dot{\beta}}_{\,\,\, \ga \de}({\bf 1}_0),\,
{\cal W}^{a \, \dot{\al} \dot{\beta}}_{\,\,\, b \, \ga \de}
({\bf 15}_0), \, {\cal W}^{\left[a b\right] \, \dot{\al}}_{\,\,\, \beta \ga \de}(
{\bf 6}_{-2}), \, \dot{{\cal W}}^{\dot{\al}\dot{\beta}\dot{\ga}}_{
\,\,\, \left[a b\right] \, \ga}(\overline{\bf 6}_{2})$  \\
[1ex]
\hline
$s=\frac{7}{2}$ & ${\cal W}^{a \, \dot{\al} \dot{\beta}}_{\,\,\,
\ga \de \ep}({\bf 4}_{-1}), \,
\dot{{\cal W}}^{\dot{\al}\dot{\beta}\dot{\ga}}_{\,\,\, a \,
\de \ep}({\bf 4}_{1})$  \\
[1ex]
\hline
$s=4$ & ${\cal W}^{\dot{\al}\dot{\beta}\dot{\ga}}_{\,\,\,\de\ep\rho}(
{\bf 1}_0)$  \\ 
[1ex] % [1ex] adds vertical space
\hline %inserts single line
\end{tabular}
%\label{tableone} % is used to refer this table in the text
\caption{The higher spin generators with $SU(4)$ representation
and ${\cal V}$-charge in the worldsheet theory,
corresponding to the tables $4$ and $5$ with
the level $l=1$ of
\cite{SS}.
We can observe that the two $SU(2)$ spins of the
higher spin generators are given by the number of each indices
$\al,\beta,\ga, \cdots $ and $\dot{\al}, \dot{\beta}, \dot{\ga},
\cdots $
divided by $2$. For example, the higher spin generator
with $s=4$ has
the corresponding spins
$(j_L, j_R)=(\frac{3}{2},\frac{3}{2})$.
Note that the spin $s$ is given by $s=1 +j_L+j_R$.
The ${\cal V}$-charge is given by the number of lower indices
of $SU(4)$ minus
the number of upper indices of $SU(4)$. 
} % title of Table
\end{table}
%%%%%%%%%%%%%%%%%%%%%%%%%%%%%%%%%%%%%%%%%%%%%%%%%%%%%%%%%%%%%%%%%%%%%%

%%%%%%%%%%%%%%%%%%%%%%%%%%%%%%%%%%%%%%%%%%%%%%%%%%%%%%%%%%%%%%%%%%%%%%
\section{ Some OPEs between the
generators of $PSU(2,2|4)$
  and  the lowest generators of $hs(2,2|4)$}
%1%%%%%%%%%%%%%%%%%%%%%%%%%%%%%%%%%%%%%%%%%%%%%%%%%%%%%%%%%%%%%%%%%%%%%
%%%%%%%%%%%%%%%%%%%%%%%%%%%%%%%%%%%%%%%%%%%%%%%%%%%%%%%%%%%%%%%%%%%%%

%%%%%%%%%%%%%%%%%%%%%%%%%%%%%%%%%%%%%%%%%%%%
\subsection{Primary or quasiprimary fields}
%%%%%%%%%%%%%%%%%%%%%%%%%%%%%%%%%%%%%%%%%%%%

By using the explicit OPE result in Appendix
(\ref{tjjj}), we can determine
the (quasi)primary fields
of higher spin generators.
As described before, only after checking
this (quasi)primary condition, then
the first order poles in the OPEs between the
weight-$1$ operators and the weight-$3$ operators
provide the right (quasi)primary operators of weight-$3$ we would like
to construct.

The quasiprimary operators in the Table $1$ are given by
the higher spin generators containing the quadratic ${\cal V}$
terms including the cubic ${\cal V}$ term.
The remaining higher spin generators are primary operators.

%%%%%%%%%%%%%%%%%%%%%%%%%%%%%%%%%%%%%%%%%%%%%%%%%%%%%%%%
\subsection{The OPEs between the weight-$1$
generators and the weight-$3$ generators}
%%%%%%%%%%%%%%%%%%%%%%%%%%%%%%%%%%%%%%%%%%%%%%%%%%%%%%

In section $3$, we have computed some of the 
OPEs between the conformal dimension-$1$
generators and the conformal dimension-$3$ generators
in order to determine the higher spin generators.
In Appendix $C$, we will present the remaining OPEs
between them.
We observe that the first order poles in the
right hand sides
of these OPEs (together with the symmetric or antisymmetric
combinations of the left hand sides of the OPEs) 
contain  the higher spin generators
as well as the new higher spin generators
 \footnote{
In the right hand sides of all these OPEs, the higher spin generator
${\cal W}$ in
(\ref{s-one}) does not appear at the first order poles. }.
In general, in these OPEs, there are also
fourth, third and second order poles we do not analyze
them in this paper explicitly.
In the view point of two dimensional worldsheet theory, it is important
to calculate them in order to see their algebraic structures. 

Of course, we can calculate the OPEs between
the conformal dimension-$3$ generators and analyze
the first order pole in order to determine the next higher
spin generators which consist of the quintic terms of
weight-$5$ operators.
We will not consider all these computations
in this paper although it is straightforward to do so.

%%%%%%%%%%%%%%%%%%%%%%%%%%%%%%%%%%%%%%%%%%%
\subsection{The additional generators}
%%%%%%%%%%%%%%%%%%%%%%%%%%%%%%%%%%%%%%%%%%%

We have obtained the new higher spin generators
(\ref{s-2}), (\ref{s-2other1}), the footnote \ref{compother2},
(\ref{s-3}) and Appendix (\ref{hatoperators})
\bea
\widehat{\cal W}^{a \, \dot{\al}}_{\,\,\, b \, \beta}, \qquad
\widehat{\cal W}^{\left[a b \right]}_{\,\,\, \al \beta}, \qquad
\dot{\widehat{\cal W}}^{\dot{\al}\dot{\beta}}_{\,\,\, \left[ a b \right]}, \qquad
\widehat{\cal W}^{a \, \dot{\al}\dot{\beta}}_{\,\,\, b \, \ga \de}, \qquad
\widehat{\cal W}^{a b \, \dot{\al}}_{\,\,\, c \, \beta \ga}, \qquad
\dot{\widehat{\cal W}}^{a\, \dot{\al} \dot{\beta}}_{\,\,\, b c \, \ga}.
\label{sixhat}
\eea
These also appear in the classical version of the OPEs
where there are no multiple contractions between the operators.
They appear in the computation of the higher spin generators
of $s=2, \frac{5}{2}$ and $s=3$.
Of course, we can further compute the OPEs
between the weight-$1$ operators and the above higher spin generators
(\ref{sixhat})
of weight-$3$ and expect that the first order poles of the
right hand sides of these OPEs contain the higher spin generators
in Table $1$ and the ones of (\ref{sixhat}).
At the moment it is not clear to observe
what are the roles of (\ref{sixhat}).
We need to calculate further OPEs between the weight-$1,2,3$ operators
including (\ref{sixhat}).
We do expect that when we consider the cases $l \geq 2$,
the similar additional higher spin generators
occur.

%%%%%%%%%%%%%%%%%%%%%%%%%%%%%%%%%%%%%%%%%%%%%%%%
\subsection{The next generators of $hs(2,2|4)$}
%%%%%%%%%%%%%%%%%%%%%%%%%%%%%%%%%%%%%%%%%%%%%%%%

So far, we have considered the $l=1$ case of \cite{SS}.
When $l=2$ case, we observe that the lowest spin $s=2$ higher spin
generator contains the following expression
$
{\cal V} {\cal V} {\cal V} {\cal V} {\cal P}^{\dot{\al}}_{\,\,\,\beta}+
\cdots$ corresponding to ${\bf 1}_0$ because there is no $SU(4)$ index.
According to (\ref{ZY}) and (\ref{uudotv}),
for the multiple product of ${\cal V}$ whose number is greater than
$4$, there are still various nonzero derivative terms
between the fermionic fields although
there is no nonderivative term between them
(Of course, if we consider the `classical' OPEs inside the Thielemans
package \cite{Thielemans}, then the above multiplet product of
${\cal V}$ is identically zero).
On the other hands, in the oscillator construction,
the corresponding $X$'s in \cite{SS}  appears only up to the
quartic term because the five product of fermionic fields vanishes. 
The higher spin generators
at $l=2$ consist of the quintic terms in the weight-$1$ operators
we have considered. That is, they have weight-$5$ operators.
One way to obtain these higher spin generators is to calculate the
OPEs between the weight-$3$ higher spin generators and
look at the first order pole.
It would be interesting to examine the details. 
Contrary to the construction of \cite{SS,SS1},
the multiple product of ${\cal V}$, where the number of
${\cal V}$ is greater than four, can occur due to the above
analysis.

%%%%%%%%%%%%%%%%%%%%%%%%%%%%%%%%%%%%%%%%%%%%%%%%%%%%%%%%%%%%%%%%%%%%%
%%%%%%%%%%%%%%%%%%%%%%%%%%%%%%%%%%%%%%%%%%%%%%%%%%%%%%%%%%%%%%%%%%%%%%
\section{ Conclusions and outlook}
%9%%%%%%%%%%%%%%%%%%%%%%%%%%%%%%%%%%%%%%%%%%%%%%%%%%%%%%%%%%%%%%%%%%%%%
%%%%%%%%%%%%%%%%%%%%%%%%%%%%%%%%%%%%%%%%%%%%%%%%%%%%%%%%%%%%%%%%%%%%%

The worldsheet realization of the higher spin generators of
\cite{SS} at $l=1$ is obtained.
They are summarized in the Table $1$ in addition to (\ref{sixhat}).

According to the table $3$  of \cite{SS},
there exist various ${\cal N}=8$ $AdS_5$
$PSU(2,2|4)$ multiplets with the levels $l=0, 1, 2, \cdots, \infty$.
As mentioned before, the $l=0$ case is the
five dimensional ${\cal N}=8$ gauged supergravity multiplet.
The $USp(8)$ representation in each level can be decomposed into
the $SU(4)$ with ${\cal V}$-charge.
See also \cite{FFZ} for this $l=0$ multiplet in terms of two
product of singletons.

The level $l=1$ multiplet can be interpreted as
the `massless' Konishi
multiplet in the context of ${\cal N}=4$ conformal
supermultiplet in four dimensions \cite{HST}.
According to the observation of \cite{SS1},
this multiplet can be also obtained by
the tensor product of
the above $l=0$ supergravity multiplet (characterized by
${\bf 42}_0$, ${\bf 48}_{\frac{1}{2}}$, ${\bf 27}_1$,
${\bf 8}_{\frac{3}{2}}$, ${\bf 1}_2$ with $USp(8)$
representation together with $SO(3)$ spin) with the
$SU(4)$ singlet of $SO(3)$ spin-$2$ (${\bf 1}_2$).
After then we obtain
${\bf 1}_0$, ${\bf 8}_{\frac{1}{2}}$, ${\bf 28}_1$,
${\bf 56}_{\frac{3}{2}}$, ${\bf 70}_{2}$,
${\bf 56}_{\frac{5}{2}}$, ${\bf 28}_3$, ${\bf 8}_{\frac{7}{2}}$
and ${\bf 1}_4$ where the subscript $s$ is the spin
index appearing in the Table $1$.

The physical states \cite{SS,SS1} arise in the sectors of
the master scalar field and the master gauge field
(in the five dimensional
higher spin gauge theory) corresponding to the higher spin
generators we have described in the above Table $1$.
%See also \cite{GM,GM1}. 
Note that there exists one-to-one correspondence
between the table $3$ and tables $4$ and $5$ only for
$s=\frac{5}{2}, 3, \frac{7}{2}$ and $4$
corresponding to
${\bf 56}_{\frac{5}{2}}$, ${\bf 28}_3$, ${\bf 8}_{\frac{7}{2}}$
and ${\bf 1}_4$.
That is, the representations  for $s=0, \frac{1}{2}$ (${\bf 1}_0$ and
${\bf 8}_{\frac{1}{2}}$) (and the representations
${\bf 6}$ and $\overline{\bf 6}$ for $s=1$,
the representations
${\bf 4}$ and $\overline{\bf 4}$ for $s=\frac{3}{2}$,
and the representations
${\bf 1}$ and $\overline{\bf 1}$  for $s=2$) appear
in the table $6$ of \cite{SS}. 
See also (\ref{mismatch}) for their ${\cal V}$-charges.

There are the following future directions we can study.

%%%%%%%%%%%%%%%%%%%%%%%%%%%%%%%%%%%%%%%%%%%%
$\bullet$
The complete OPEs
%%%%%%%%%%%%%%%%%%%%%%%%%%%%%%%%%%%%%%%%%%%%

In this paper,
we have focused on the construction of the
higher spin generators having weight-$3$.
We understand that there are weight-$2$ operators
in the OPEs between the weight-$1$ operators and
the weight-$3$ operators. Furthermore we did not consider
the OPEs between the weight-$1$ operators (the generators of
Lorentz symmetry and the generators of super conformal boosts)
(\ref{12generators})
and the weight-$3$ operators we have constructed in the worldsheet
theory. It would be interesting to determine the
complete OPEs between these generators
of weight-$1,2,3$ in the context of
the higher spin superalgebra $hs(2,2|4)$.
Moreover, it will be interesting how they survive
when we act them on the physical vacuum state by recalling
the footnote \ref{fiveweightone} on the weight-$1$ operators
along the line of \cite{GG2104,GG2105}.
Eventually, we would like to construct
the complete higher spin algebra which contains the
higher spin generators appearing in the tables $4$ and $5$
of \cite{SS} in closed form.

%%%%%%%%%%%%%%%%%%%%%%%%%%%%%%%%%%%%%%%%%%%%%%%%%%%%%%%%%%%
$\bullet$
In the theory of ${\cal N}=4$ super Yang-Mills coupled
to the ${\cal N}=4$ conformal supergravity
%%%%%%%%%%%%%%%%%%%%%%%%%%%%%%%%%%%%%%%%%%%%%%%%%%%%%%%%%%%

As before, in \cite{SS1},
the conserved currents
corresponding to
the higher spin gauge theory described in the table $3$ of \cite{SS}
can be described from the singleton superfield
based on \cite{Bdd,HST,FFZ} for $l=0$ and $l=1$.
Furthermore, 
in \cite{GMZ}, their tables $6$ and $7$ are related to
the four dimensional ${\cal N}=4$ conformal supergravity multiplet.
They claim that the $l=1$ case of the table $3$ of \cite{SS}
can be obtained also from the tensor product of above tables
$6$ and $7$ (See also \cite{BT} on the one loop contributions
of  ${\cal N}=4$ conformal supergravity multiplet).
It would be interesting to study precise correspondence explicitly
in the context of \cite{Maldacena,Witten,GKP}.
See also the review paper \cite{FT} for conformal supergravity and
\cite{BW} for the twistor string theory description
of conformal supergravity.

%%%%%%%%%%%%%%%%%%%%%%%%%%%%%%%%%%%%%%%%%%%%%%%%%
$\bullet$
The action of the higher spin generators on the
vacuum state
%%%%%%%%%%%%%%%%%%%%%%%%%%%%%%%%%%%%%%%%%%%%%%%%%

In the oscillator construction, it is known that
the ${\cal N}=4$ super Yang-Mills multiplet
can be identified with the multiple product of the
various oscillators acting on the physical vacuum
state \cite{Beisert}.
The similar construction in the worldsheet theory
is obtained from the multiple product of
the various zero modes of the ambitwistor
fields acting on the
Ramond ground state \cite{GG2104,GG2105}. 
As we have the complete expressions for the higher spin
generators, we can determine the precise
action on the physical vacuum state as mentioned before.

%%%%%%%%%%%%%%%%%%%%%%%%%%%%%%%%%%%%%%%%%%%%%%%%%%%%%%
$\bullet$
When the coupling of ${\cal N}=4$ super Yang-Mills
becomes nonzero
%%%%%%%%%%%%%%%%%%%%%%%%%%%%%%%%%%%%%%%%%%%%%%%%%%%%%%

As the ${\cal N}=4$ super Yang-Mills interaction is turned on,
then the higher spin generators in the tables $4$ and $5$
of \cite{SS} with $l=1, 2, \cdots, \infty$ will be no longer
conserved. As observed in \cite{SS1},
the $hs(2,2|4)$ higher spin gauge theory maybe described
by a string theory having a left-moving
and right-moving $PSU(2,2|4)$ Kac-Moody superalgebra
with a critical level $k=1$.
We have seen that this theory admits a singleton representation
\cite{GG2104,GG2105}.
Then the question is whether the affine Kac-Moody extension of the
$hs(2,2|4)$ will give us some hints in order to describe the
theory for nonzero coupling of ${\cal N}=4$ super Yang-Mills
in four dimensions beyond the free field construction of this paper.
See also the previous relevant paper \cite{DS}.

$\bullet$
Any algebraic symmetries in the DDF-like operators

In \cite{GG2104,GG2105}, the DDF-like operators \cite{DDF}
which are given by the product of the modes of
ambitwistor fields (\ref{ZY}) 
are introduced. They satisfy the nontrivial
(anti)commutator relations depending on the magnitude
of the sum of the two each modes.
The structure constants appearing in the right hand side
of these relations are given by the ones in the superalgebra
$U(2,2|4)$. They claim that the nontrivial triple products
for the specific three modes 
vanish identically.
It would be interesting to describe the above
products for any three modes and observe
whether there exist any nontrivial behaviors or not.

%%%%%%%%%%%%%%%%%%%%%%%%%%%%%%%%%%%%%%%%%%%%%%%%%%%%%%%%
$\bullet$
How to interpret the mismatch between the table $3$ and
the tables $4$ and $5$ of \cite{SS}
%%%%%%%%%%%%%%%%%%%%%%%%%%%%%%%%%%%%%%%%%%%%%%%%%%%%%%%%

There are some multiplets in table $6$ of \cite{SS}
\footnote{In addition to these, there are ${\bf 1}_0$ for $s=0$
  and ${\bf 4}_1 \oplus \overline{\bf 4}_{-1}$ for $s=\frac{1}{2}$ as
  before.}
\bea
s= 1 & : & \qquad {\bf 6}_{-2}, \qquad \overline{\bf 6}_{2},
\nonu\\
s=\frac{3}{2} & : & \qquad \overline{\bf 4}_{-3}, \qquad  {\bf 4}_{3}, 
\nonu \\
s=2 & : & \qquad {\bf 1}_{-4}, \qquad \overline{\bf 1}_{4}.
\label{mismatch}
\eea
These are the elements of the table $3$ but their
corresponding
higher spin generators do not appear in the tables $4$ and $5$.
However, it seems that for $l \geq 2$, we can check
the sum of the representations in table $4$ \cite{SS} is given by
$ 2 \cdot {\bf 1} \oplus  4 \cdot {\bf 4}  \oplus 2 \cdot
{\bf 16} \oplus 4 \cdot {\bf 24} \oplus {\bf 36}$
and this is equal to $182$ and
the sum of the representations in table $5$ is given by
$ 2 \cdot {\bf 1} \oplus 4 \cdot {\bf 4} \oplus 4 \cdot
{\bf 6} \oplus 2 \cdot {\bf 16}$ and this  is $74$.
This leads to $182+74=256$.
Then 
there is no mismatch between the table $3$ and
the tables $4$ and $5$ for $ l \geq 2$.
It is an open problem to understand how the higher spin generators
corresponding to (\ref{mismatch}) are not allowed for small spin $s$
in the oscillator construction (or in the worldsheet theory).

%%%%%%%%%%%%%%%%%%%%%%%%%%%%%%%%%%%%%%%%%%%%%%%%%%%%%%%%%%%%
$\bullet$ Can the even power of oscillators survive in the
worldsheet description?
%%%%%%%%%%%%%%%%%%%%%%%%%%%%%%%%%%%%%%%%%%%%%%%%%%%%%%%%%%%%

In the construction of \cite{SS}, the higher spin generators
with equal odd numbers of oscillators can appear only.
See also \cite{BBMS} for relevant discussion.
It is not obvious to see this restriction in
the worldsheet theory because in the OPEs between
the weight-$1, 2, 3$ operators, in general, the weight-$2,4$
operators as well as the weight-$5$ operators can appear.
See also \cite{Vasiliev} for different kinds of higher spin generators.
It would be interesting to study this direction in order to
describe the above restriction in the worldsheet theory.

\vspace{.7cm}

%%%%%%%%%%%%%%%%%%%%%%%%%%%%%%%%%%%%%%%%%%%%%%%%%%%%%%%%%%%%%%
%%%%%%%%%%%%%%%%%%%%%%%%%%%%%%%%%%%%%%%%%%%%%%%%%%%%%%%%%%%%%%%
\centerline{\bf Acknowledgments}
%%%%%%%%%%%%%%%%%%%%%%%%%%%%%%%%%%%%%%%%%%%%%%%%%%%%%%%%%%%%%%%
%%%%%%%%%%%%%%%%%%%%%%%%%%%%%%%%%%%%%%%%%%%%%%%%%%%%%%%%%%%%%%%

%We
%would like to
%thank
%Y. Hikida for the discussions.
This work was supported by
the National Research Foundation of Korea(NRF) grant
funded by the Korea government(MSIT)(No. 2020R1F1A1066893).

\newpage

\appendix

\renewcommand{\theequation}{\Alph{section}\mbox{.}\arabic{equation}}

%%%%%%%%%%%%%%%%%%%%%%%%%%%%%%%%%%%%%%%%%%%%%%%%%%%%%%%%%%%%%%%%
%%%%%%%%%%%%%%%%%%%%%%%%%%%%%%%%%%%%%%%%%%%%%%%%%%%%%%%%%%%%%%%%%%%%%
\section{ The $PSU(2,2|4)_1$ current algebra}
%%%%%AAA%%%%%%%%%%%%%%%%%%%%%%%%%%%%%%%%%%%%%%%%%%%%%%%%%%%%%%%%%%%%%%%%
%%%%%%%%%%%%%%%%%%%%%%%%%%%%%%%%%%%%%%%%%%%%%%%%%%%%%%%%%%%%

%%%%%%%%%%%%%%%%%%%%%%%%%%%%%%%%%%%%%%%%%%%%%%%%%%%%%%%%%%%%%%%%%%%%%%
\subsection{The algebra from the generators in (\ref{12generators})}
%%%%%%%%%%%%%%%%%%%%%%%%%%%%%%%%%%%%%%%%%%%%%%%%%%%%%%%%%%%%%%%%%%%%%

We present the various OPEs between
the generators in (\ref{12generators}) which did not appear in the
literature before (although some of (anti)commutator relations
between them are in \cite{GG2105}) and we take the order of
generators as in $({\cal L}^{\al}_{\,\,\,\beta},\dot{{\cal L}}^{\dot{\al}}_{\,\,\,\dot{\beta}}, {\cal R}^a_{\,\,\, b}, {\cal B}, {\cal C}, {\cal D},
{\cal Q}^a_{\,\,\,\al},\dot{{\cal Q}}^{\dot{\al}}_{\,\,\, a}, {\cal P}^{\dot{
\al}}_{\,\,\, \beta}, {\cal S}^{\al}_{\,\,\,a},
\dot{\cal S}^{a}_{\,\,\,\dot{\al}},{\cal K}^{\al}_{\,\,\,\dot{\beta}})$
as follows:
\bea
{\cal L}^{\al}_{\,\,\,\beta}(z) \, {\cal L}^{\ga}_{\,\,\,\de}(w) &=&
\frac{1}{(z-w)^2}\,  \Bigg[
\frac{1}{2} \,\de^{\al}_{\beta}\,
\de^{\ga}_{\de} - \de^{\al}_{\de} \, \de^{\ga}_{\beta} \Bigg]
+\frac{1}{(z-w)}\, \Bigg[  \de^{\al}_{\de}\,
{\cal L}^{\ga}_{\beta} - \de^{\ga}_{\beta} \, {\cal L}^{\al}_{\de}
\Bigg](w) +\cdots,
\nonu \\
{\cal L}^{\al}_{\,\,\,\beta}(z) \, {\cal Q}^{a}_{\,\,\,\ga}(w) &=&
\frac{1}{(z-w)}\, \Bigg[ \de^{\al}_{\ga} \,  {\cal Q}^{a}_{\,\,\,\beta}
-\frac{1}{2} \, \de^{\al}_{\beta}\,
{\cal Q}^{a}_{\,\,\,\ga}
\Bigg](w) + \cdots,
\nonu \\
{\cal L}^{\al}_{\,\,\,\beta}(z) \, {\cal P}^{\dot{\al}}_{\,\,\,\ga}(w) &=&
\frac{1}{(z-w)}\, \Bigg[ \de^{\al}_{\ga} \,  {\cal P}^{\dot{\al}}_{\,\,\,\beta}
-\frac{1}{2} \, \de^{\al}_{\beta}\,
{\cal P}^{\dot{\al}}_{\,\,\,\ga}
\Bigg](w) + \cdots,
\nonu \\
{\cal L}^{\al}_{\,\,\,\beta}(z) \, {\cal S}^{\ga}_{\,\,\,a}(w) &=&
\frac{1}{(z-w)}\, \Bigg[- \de^{\ga}_{\beta} \,  {\cal S}^{\al}_{\,\,\,a}
+\frac{1}{2} \, \de^{\al}_{\beta}\,
{\cal S}^{\ga}_{\,\,\,a}
\Bigg](w) + \cdots,
\nonu \\
{\cal L}^{\al}_{\,\,\,\beta}(z) \, {\cal K}^{\ga}_{\,\,\,\dot{\beta}}(w) &=&
\frac{1}{(z-w)}\, \Bigg[- \de^{\ga}_{\beta} \,  {\cal K}^{\al}_{\,\,\,\dot{\beta}}
+\frac{1}{2} \, \de^{\al}_{\beta}\,
{\cal K}^{\ga}_{\,\,\,\dot{\beta}}
\Bigg](w) + \cdots,
\nonu \\
%%%%%%%%%%%%%%%%%%%%%%%%%%%%%%%%%%%%%
\dot{{\cal L}}^{\dot{\al}}_{\,\,\,\dot{\beta}}(z) \,
\dot{{\cal L}}^{\dot{\ga}}_{\,\,\,\dot{\de}}(w) &=&
\frac{1}{(z-w)^2}\,  \Bigg[
\frac{1}{2} \,\de^{\dot{\al}}_{\dot{\beta}}\,
\de^{\dot{\ga}}_{\dot{\de}} - \de^{\dot{\al}}_{\dot{\de}} \,
\de^{\dot{\ga}}_{\dot{\beta}} \Bigg]
+\frac{1}{(z-w)}\, \Bigg[  \de^{\dot{\al}}_{\dot{\de}}\,
\dot{{\cal L}}^{\dot{\ga}}_{\dot{\beta}} -
\de^{\dot{\ga}}_{\dot{\beta}} \, \dot{{\cal L}}^{\dot{\al}}_{\dot{\de}}
\Bigg](w) +\cdots,
\nonu \\
\dot{{\cal L}}^{\dot{\al}}_{\,\,\,\dot{\beta}}(z) \,
\dot{{\cal Q}}^{\dot{\ga}}_{\,\,\,a}(w) &=&
\frac{1}{(z-w)}\, \Bigg[ -\de^{\dot{\ga}}_{\dot{\beta}} \,
\dot{{\cal Q}}^{\dot{\al}}_{\,\,\,a}
+\frac{1}{2} \, \de^{\dot{\al}}_{\dot{\beta}}\,
\dot{{\cal Q}}^{\dot{\ga}}_{\,\,\,a}
\Bigg](w) + \cdots,
\nonu \\
\dot{{\cal L}}^{\dot{\al}}_{\,\,\,\dot{\beta}}(z) \,
{\cal P}^{\dot{\ga}}_{\,\,\,\de}(w) &=&
\frac{1}{(z-w)}\, \Bigg[ -\de^{\dot{\ga}}_{\dot{\beta}}
\,  {\cal P}^{\dot{\al}}_{\,\,\,\de}
+\frac{1}{2} \, \de^{\dot{\al}}_{\dot{\beta}}\,
{\cal P}^{\dot{\ga}}_{\,\,\,\de}
\Bigg](w) + \cdots,
\nonu \\
\dot{{\cal L}}^{\dot{\al}}_{\,\,\,\dot{\beta}}(z) \, \dot{{\cal S}}^{a}_{\,\,\,\dot{\ga}}(w) &=&
\frac{1}{(z-w)}\, \Bigg[ \de^{\dot{\al}}_{\dot{\ga}} \,
{\cal S}^{a}_{\,\,\,\dot{\beta}}
-\frac{1}{2} \, \de^{\dot{\al}}_{\dot{\beta}}\,
\dot{{\cal S}}^{a}_{\,\,\,\dot{\ga}}
\Bigg](w) + \cdots,
\nonu \\
\dot{{\cal L}}^{\dot{\al}}_{\,\,\,\dot{\beta}}(z) \,
{\cal K}^{\ga}_{\,\,\,\dot{\de}}(w) &=&
\frac{1}{(z-w)}\, \Bigg[ \de^{\dot{\al}}_{\dot{\de}} \,
{\cal K}^{\ga}_{\,\,\,\dot{\beta}}
-\frac{1}{2} \, \de^{\dot{\al}}_{\dot{\beta}}\,
{\cal K}^{\ga}_{\,\,\,\dot{\de}}
\Bigg](w) + \cdots,
\nonu \\
%%%%%%%%%%%%%%%%%%%%%%%%%%%%%%%%%%%%%%%
{\cal R}^{a}_{\,\,\,b}(z) \, {\cal R}^{c}_{\,\,\,d}(w) &=&
\frac{1}{(z-w)^2}\,  \Bigg[
-\frac{1}{4} \,\de^{a}_{b}\,
\de^{c}_{d} - \de^{a}_{d} \, \de^{c}_{b} \Bigg]
+\frac{1}{(z-w)}\, \Bigg[  \de^{a}_{d}\,
{\cal R}^{c}_{\,\,\,b} - \de^{c}_{b} \, {\cal R}^{a}_{\,\,\,d}
\Bigg](w) +\cdots,
\nonu \\
{\cal R}^{a}_{\,\,\,b}(z) \, {\cal Q}^{c}_{\,\,\,\al}(w) &=&
\frac{1}{(z-w)}\, \Bigg[  -\de^{c}_{b}\,
{\cal Q}^{a}_{\,\,\,\al} + \frac{1}{4} \, \de^{a}_{b} \, {\cal Q}^{c}_{\,\,\,\al}
\Bigg](w) +\cdots,
\nonu \\
{\cal R}^{a}_{\,\,\,b}(z) \, \dot{{\cal Q}}^{\dot{\al}}_{\,\,\,c}(w) &=&
\frac{1}{(z-w)}\, \Bigg[  \de^{a}_{c}\,
\dot{{\cal Q}}^{\dot{\al}}_{\,\,\,b} - \frac{1}{4} \, \de^{a}_{b}
\, \dot{{\cal Q}}^{\dot{\al}}_{\,\,\,c}
\Bigg](w) +\cdots,
\nonu \\
{\cal R}^{a}_{\,\,\,b}(z) \, {\cal S}^{\al}_{\,\,\,c}(w) &=&
\frac{1}{(z-w)}\, \Bigg[  \de^{a}_{c}\,
{\cal S}^{\al}_{\,\,\,b} - \frac{1}{4} \, \de^{a}_{b}
\, {\cal S}^{\al}_{\,\,\,c}
\Bigg](w) +\cdots,
\nonu \\
{\cal R}^{a}_{\,\,\,b}(z) \, \dot{{\cal S}}^{c}_{\,\,\,\dot{\al}}(w) &=&
\frac{1}{(z-w)}\, \Bigg[  -\de^{c}_{b}\,
\dot{{\cal S}}^{a}_{\,\,\,\dot{\al}} + \frac{1}{4} \, \de^{a}_{b}
\, \dot{{\cal S}}^{c}_{\,\,\,\dot{\al}}
\Bigg](w) +\cdots,
\nonu \\
%%%%%%%%%%%%%%%%%%%%%%%%%%%%%%%%%%%%%%%%%%%
{\cal B}(z) \, {\cal B}(w) &=&
-\frac{1}{(z-w)^2} + \cdots,
\qquad
{\cal B}(z) \, {\cal C}(w) =
-\frac{1}{(z-w)^2} + \cdots,
\nonu \\
{\cal B}(z) \, {\cal Q}^a_{\,\,\,\al}(w)  & = & 
\frac{1}{(z-w)}\, \frac{1}{2}\, {\cal Q}^a_{\,\,\,\al}(w)
+\cdots,
\qquad
{\cal B}(z) \, \dot{{\cal Q}}_{\,\,\,a}^{\dot{\al}}(w)  = 
-\frac{1}{(z-w)}\, \frac{1}{2}\,
\dot{{\cal Q}}_{\,\,\,a}^{\dot{\al}}(w)
+\cdots,
\nonu \\
{\cal B}(z) \, {\cal S}^{\al}_{\,\,\,a}(w)  & = & 
-\frac{1}{(z-w)}\, \frac{1}{2}\, {\cal S}^{\al}_{\,\,\,a}(w)
+\cdots,
\qquad
{\cal B}(z) \, \dot{{\cal S}}^{a}_{\,\,\,\dot{\al}}(w)  = 
\frac{1}{(z-w)}\, \frac{1}{2}\,
\dot{{\cal S}}^{a}_{\,\,\,\dot{\al}}(w)
+\cdots,
\nonu \\
{\cal D}(z) \, {\cal D}(w) & = &
-\frac{1}{(z-w)^2} + \cdots,
\qquad
{\cal D}(z) \, {\cal Q}^a_{\,\,\,\al}(w)  =
\frac{1}{(z-w)}\, \frac{1}{2}\, {\cal Q}^a_{\,\,\,\al}(w)
+\cdots,
\nonu \\
{\cal D}(z) \, \dot{{\cal Q}}_{\,\,\,a}^{\dot{\al}}(w) & = & 
\frac{1}{(z-w)}\, \frac{1}{2}\,
\dot{{\cal Q}}_{\,\,\,a}^{\dot{\al}}(w)
+\cdots,
\qquad
{\cal D}(z) \, {\cal P}^{\dot{\al}}_{\,\,\,\beta}(w)  =  
\frac{1}{(z-w)}\, {\cal P}^{\dot{\al}}_{\,\,\,\beta}(w)
+\cdots,
\nonu \\
{\cal D}(z) \, {\cal S}^{\al}_{\,\,\,a}(w)   & = &  
-\frac{1}{(z-w)}\, \frac{1}{2}\, {\cal S}^{\al}_{\,\,\,a}(w)
+\cdots,
\qquad
{\cal D}(z) \, \dot{{\cal S}}^{a}_{\,\,\,\dot{\al}}(w)  = 
\frac{1}{(z-w)}\, \frac{1}{2}\,
\dot{{\cal S}}^{a}_{\,\,\,\dot{\al}}(w)
+\cdots,
\nonu \\
{\cal D}(z) \, {\cal K}^{\al}_{\,\,\,\dot{\beta}}(w)  & = & 
-\frac{1}{(z-w)}\, {\cal K}^{\al}_{\,\,\,\dot{\beta}}(w)
+\cdots,
\qquad
%%%%%%%%%%%%%%%%%%%%%%%%%%%%%%%%%%%%%%%%%%%%%%%%
{\cal Q}^{a}_{\,\,\,\beta}(z) \, \dot{{\cal Q}}^{\dot{\al}}_{\,\,\,b}(w) =
\frac{1}{(z-w)}\,   \de^{a}_{b}\,
{\cal P}^{\dot{\al}}_{\,\,\,\beta}(w) +\cdots,
\nonu \\
{\cal Q}^{a}_{\,\,\,\al}(z) \, {\cal S}^{\beta}_{\,\,\,b}(w) &=&
\frac{1}{(z-w)^2}\, \de^{a}_{b} \, \de^{\beta}_{\al}
+ \frac{1}{(z-w)}\,   \Bigg[ \de^{\beta}_{\al}\,
{\cal R}^{a}_{\,\,\,b} +\de^{a}_b \, {\cal L}^{\beta}_{\,\,\, \al}
+ \frac{1}{2} \, \de^{a}_{b}\, \de^{\beta}_{\al} \,
({\cal C}+ {\cal D}) \Bigg](w) +\cdots,
\nonu \\
{\cal Q}^{a}_{\,\,\,\al}(z) \, {\cal K}^{\beta}_{\,\,\,\dot{\ga}}(w) &=&
- \frac{1}{(z-w)}\,   
\de^{\beta}_{\al} \, \dot{{\cal S}}^{a}_{\,\,\,\dot{\ga}} (w) +\cdots,
\nonu \\
%%%%%%%%%%%%%%%%%%%%%%%%%%%%%%%%%%%%%%%%%
\dot{{\cal Q}}^{\dot{\al}}_{\,\,\,a}(z) \,
\dot{{\cal S}}^{b}_{\,\,\,\dot{\beta}}(w) &=&
\frac{1}{(z-w)^2}\, \de^{b}_{a} \, \de^{\dot{\al}}_{\dot{\beta}}
+ \frac{1}{(z-w)}\,   \Bigg[ \de^{\dot{\al}}_{\dot{\beta}}\,
{\cal R}^{b}_{\,\,\,a} +\de^{b}_a \,
\dot{{\cal L}}^{\dot{\al}}_{\,\,\,\dot{\beta}}
+ \frac{1}{2} \, \de^{b}_{a}\, \de^{\dot{\al}}_{\dot{\beta}} \,
({\cal C}- {\cal D}) \Bigg](w) +\cdots,
\nonu \\
\dot{{\cal Q}}^{\dot{\al}}_{\,\,\,a}(z) \, {\cal K}^{\beta}_{\,\,\,\dot{\ga}}(w) &=&
\frac{1}{(z-w)}\,   
\de^{\dot{\al}}_{\dot{\ga}} \, {\cal S}^{\beta}_{\,\,\,a} (w) +\cdots,
\quad
%%%%%%%%%%%%%%%%%%%%%%%%%%%%%%%%%%%%%%%%%%
{\cal P}^{\dot{\al}}_{\,\,\,\beta}(z) \, {\cal S}^{\ga}_{\,\,\,a}(w) =
-\frac{1}{(z-w)}\,   
\de^{\ga}_{\beta} \, \dot{{\cal Q}}^{\dot{\al}}_{\,\,\,a} (w)
+\cdots,
\nonu \\
{\cal P}^{\dot{\al}}_{\,\,\,\beta}(z) \,
\dot{{\cal S}}^{a}_{\,\,\,\dot{\ga}}(w) &=&
\frac{1}{(z-w)}\,   
\de^{\dot{\al}}_{\dot{\ga}} \,
{\cal Q}^{a}_{\,\,\,\beta} (w)+\cdots,
\nonu \\
{\cal P}^{\dot{\al}}_{\,\,\,\beta}(z) \,
{\cal K}^{\ga}_{\,\,\,\dot{\de}}(w) &=&
-\frac{1}{(z-w)^2}\, \de^{\ga}_{\beta} \, \de^{\dot{\al}}_{\dot{\de}}
+\frac{1}{(z-w)}\,   
\Bigg[ - \de^{\ga}_{\beta} \,
\dot{{\cal L}}^{\dot{\al}}_{\,\,\,\dot{\de}} 
+ \de^{\dot{\al}}_{\dot{\de}} \,
{\cal L}^{\ga}_{\,\,\,\beta}
+ \de^{\ga}_{\beta} \, \de^{\dot{\al}}_{\dot{\de}} \, {\cal D}
\Bigg](w)
+\cdots,
\nonu \\
%%%%%%%%%%%%%%%%%%%%%%
{\cal S}^{\al}_{\,\,\, a}(z) \,
\dot{{\cal S}}^{b}_{\,\,\,\dot{\beta}}(w) &=&
\frac{1}{(z-w)}\,   
\de^{b}_{a} \,
{\cal K}^{\al}_{\,\,\,\dot{\beta}} (w)
+\cdots.
\label{psualgebra}
\eea
The OPEs between the generators of 
$({\cal R}^a_{\,\,\, b}, {\cal V},
{\cal Q}^a_{\,\,\,\al},\dot{{\cal Q}}^{\dot{\al}}_{\,\,\, a}, {\cal P}^{\dot{
    \al}}_{\,\,\, \beta})$ with ${\cal V}\equiv 2({\cal C}-{\cal B})$
are closed by themselves. See also
Appendix (\ref{opewithuudotv}).

%%%%%%%%%%%%%%%%%%%%%%%%%%%%%%%%%%%%%%%
\subsection{ Some OPEs with different $U(1)$ generators
  in (\ref{uudotv})}
%%%%%%%%%%%%%%%%%%%%%%%%%%%%%%%%%%%%%%%

By using (\ref{uudotv}), we can rewrite some OPEs in Appendix
(\ref{psualgebra})
as follows:
\bea
{\cal U}(z) \, {\cal U}(w) &=&
-\frac{1}{(z-w)^2}\, 2 + \cdots,
\qquad
{\cal U}(z) \, {\cal B}(w) =
-\frac{1}{(z-w)^2} + \cdots,
\nonu \\
{\cal U}(z) \, {\cal C}(w) &=&
-\frac{1}{(z-w)^2} + \cdots,
\qquad
{\cal U}(z) \, {\cal D}(w) =
-\frac{1}{(z-w)^2} + \cdots,
\nonu \\
{\cal U}(z) \, {\cal Q}^a_{\,\,\, \al}(w) &= &
\frac{1}{(z-w)} \,  {\cal Q}^a_{\,\,\, \al}(w) + \cdots, 
\qquad
{\cal U}(z) \, {\cal P}^{\dot{\al}}_{\,\,\, \beta}(w) = 
\frac{1}{(z-w)} \,  {\cal P}^{\dot{\al}}_{\,\,\, \beta}(w) + \cdots, 
\nonu \\
{\cal U}(z) \, {\cal S}^{\al}_{\,\,\, a}(w) &= &
-\frac{1}{(z-w)} \,  {\cal S}^{\al}_{\,\,\, a}(w) + \cdots, 
\qquad
{\cal U}(z) \, {\cal K}^{\al}_{\,\,\, \dot{\beta}}(w) = 
-\frac{1}{(z-w)} \,  {\cal K}^{\al}_{\,\,\, \dot{\beta}}(w) + \cdots, 
\nonu \\
\dot{{\cal U}}(z) \, \dot{{\cal U}}(w) &=&
-\frac{1}{(z-w)^2}\, 2 + \cdots,
\qquad
\dot{{\cal U}}(z) \, {\cal B}(w) =
-\frac{1}{(z-w)^2} + \cdots,
\nonu \\
\dot{{\cal U}}(z) \, {\cal C}(w) &=&
-\frac{1}{(z-w)^2} + \cdots,
\qquad
\dot{{\cal U}}(z) \, {\cal D}(w) =
\frac{1}{(z-w)^2} + \cdots,
\nonu \\
\dot{{\cal U}}(z) \, \dot{{\cal Q}}_{\,\,\,a}^{ \dot{\al}}(w) &= &
-\frac{1}{(z-w)} \,  \dot{{\cal Q}}_{\,\,\,a}^{\dot{ \al}}(w) + \cdots, 
\qquad
\dot{{\cal U}}(z) \, {\cal P}^{\dot{\al}}_{\,\,\, \beta}(w) = 
-\frac{1}{(z-w)} \,  {\cal P}^{\dot{\al}}_{\,\,\, \beta}(w) + \cdots, 
\nonu \\
\dot{{\cal U}}(z) \, \dot{{\cal S}}_{\,\,\,\al}^{ a}(w) &= &
\frac{1}{(z-w)} \,  \dot{{\cal S}}_{\,\,\,\al}^{ a}(w) + \cdots, 
\qquad
\dot{{\cal U}}(z) \, {\cal K}^{\al}_{\,\,\, \dot{\beta}}(w) = 
\frac{1}{(z-w)} \,  {\cal K}^{\al}_{\,\,\, \dot{\beta}}(w) + \cdots, 
\nonu \\
{\cal V}(z) \, {\cal V}(w) &=&
\frac{1}{(z-w)^2}\, 4 + \cdots,
\qquad
{\cal V}(z) \, {\cal C}(w) =
\frac{1}{(z-w)^2}\, 2 + \cdots,
\nonu \\
{\cal V}(z) \, {\cal Q}_{\,\,\,\al}^{ a}(w) &= &
-\frac{1}{(z-w)} \,  {\cal Q}_{\,\,\,\al}^{ a}(w) + \cdots, 
\qquad
{\cal V}(z) \, \dot{{\cal Q}}^{\dot{\al}}_{ \,\,\,a}(w) = 
\frac{1}{(z-w)} \,  \dot{{\cal Q}}^{\dot{\al}}_{\,\,\, a}(w) + \cdots, 
\nonu \\
{\cal V}(z) \, {\cal S}_{\,\,\,a}^{ \al}(w) &= &
\frac{1}{(z-w)} \,  {\cal S}_{\,\,\,a}^{ \al}(w) + \cdots, 
\qquad
{\cal V}(z) \, \dot{{\cal S}}^{a}_{ \,\,\, \dot{\al}}(w) = 
-\frac{1}{(z-w)} \,  \dot{{\cal S}}^{a}_{\,\,\, \dot{\al}}(w) + \cdots.
\label{opewithuudotv}
\eea
Note that the nonzero $\cal V$-charge can be obtained
from the last four OPEs of (\ref{opewithuudotv}).

%%%%%%%%%%%%%%%%%%%%%%%%%%%%%%%%%%%%%%%%%%%%%%%%%%%%%%%%%%%%%%%%
%%%%%%%%%%%%%%%%%%%%%%%%%%%%%%%%%%%%%%%%%%%%%%%%%%%%%%%%%%%%%%%%%%%%%
\section{ The OPEs between the stress energy tensor and
$J^{I}_{\,\,\,J}\, J^{K}_{\,\,\,L}\,
  J^{M}_{\,\,\,N}$}
%%%%%AAA%%%%%%%%%%%%%%%%%%%%%%%%%%%%%%%%%%%%%%%%%%%%%%%%%%%%%%%%%%%%%%%%
%%%%%%%%%%%%%%%%%%%%%%%%%%%%%%%%%%%%%%%%%%%%%%%%%%%%%%%%%%%%

We write down the OPE between the
stress energy tensor and the cubic term as follows: 
\bea
T(z) \, J^{I}_{\,\,\,J}\, J^{K}_{\,\,\,L}\,
J^{M}_{\,\,\,N}(w) &= &
\frac{1}{(z-w)^5}\, \Bigg[ (-1)^{d_J \, d_M +1}\, \de^I_{L}
\de^K_N \de^M_J +(-1)^{(d_L +d_K)(d_I+d_J)+d_L \, d_M} \,
\de^K_J \, \de^I_N \, \de^M_L
\Bigg]
\nonu \\
& + & \frac{1}{(z-w)^4} \, \Bigg[
(-1)^{d_L\, d_M +1} \, \de^K_N \, \de^M_L \, J^I_{\,\,\,J} +
(-1)^{d_J \, d_K +1} \, \de^I_L \, \de^K_J \, J^M_{\,\,\,N} \nonu \\
& + &
(-1)^{(d_M+d_N)(d_K+d_J)+1} \, \de^I_L \, \de^M_J \, J^K_{\,\,\,N} +
(-1)^{(d_L+d_K)(d_I+d_J)+1} \, \de^K_J \, \de^I_N \, J^M_{\,\,\,L}
\nonu \\
& + &
(-1)^{(d_L+d_K)(d_I+d_J)+(d_M+d_N)(d_I+d_J)} \, \de^K_J \,
\de^M_L \, J^I_{\,\,\,N} \nonu \\
& + &
(-1)^{(d_I+d_J)(d_L+d_K)+d_J \, d_M +1} \, \de^I_N \, \de^M_J \,
J^{K}_{\,\,\,L}  +
\de^I_L \, \de^K_N \, J^M_{\,\,\,J}
  \Bigg](w)
\nonu \\
& + &
\frac{1}{(z-w)^3}\, \Bigg[ \de^I_L \, J^K_{\,\,\,J}\, J^M_{\,\,\,N} 
+  (-1)^{(d_L+d_K)(d_I+d_J)+1} \, \de^K_J \, J^I_{\,\,\,L}\, J^M_{\,\,\,N} \,
\nonu \\
& + & (-1)^{(d_I+d_J)(d_L+d_K)}  \de^I_N \, J^K_{\,\,\,L}\, J^M_{\,\,\,J} \nonu
\\ & + &
(-1)^{(d_I+d_J)(d_L+d_K+d_M+d_N)+1}\, \de^M_J \, J^K_{\,\,\,L}\, J^I_{\,\,\,N}
+ 
 \de^K_N \, J^I_{\,\,\,J}\, J^M_{\,\,\,L} \,
 \nonu \\
 & + &   (-1)^{(d_N+d_M)(d_K+d_L)+1}\,
 \de^M_L \, J^I_{\,\,\,J}\, J^K_{\,\,\,N}
  \Bigg](w)
+  \frac{1}{(z-w)^2} \, 3\, J^{I}_{\,\,\,J}\, J^{K}_{\,\,\,L}\,
J^{M}_{\,\,\,N}(w)
\nonu \\
& + &
\frac{1}{(z-w)} \, \pa \, ( J^{I}_{\,\,\,J}\, J^{K}_{\,\,\,L}\,
J^{M}_{\,\,\,N})(w)+\cdots.
\label{tjjj}
\eea
We should obtain the weight-$3$ operators
which transform as a quasiprimary.

%%%%%%%%%%%%%%%%%%%%%%%%%%%%%%%%%%%%%%%%%%%%%%%%%%%%%%%%%%%%%%%%
%%%%%%%%%%%%%%%%%%%%%%%%%%%%%%%%%%%%%%%%%%%%%%%%%%%%%%%%%%%%%%%%%%%%%
\section{ The remaining first order poles in the OPEs
described in the section $3$}
%%%%%AAA%%%%%%%%%%%%%%%%%%%%%%%%%%%%%%%%%%%%%%%%%%%%%%%%%%%%%%%%%%%%%%%%
%%%%%%%%%%%%%%%%%%%%%%%%%%%%%%%%%%%%%%%%%%%%%%%%%%%%%%%%%%%%

In this Appendix,
we present the remaining first order poles
in the OPEs between the weight-$1$ operators and 
the weight-$3$ operators.

Let us classify according to the spin of weight-$3$ operators.

%%%%%%%%%%%%%%%%%%%%%
$\bullet$  The $s=1$ case
%%%%%%%%%%%%%%%%%%%%%

In addition to the corresponding OPEs of subsection
$3.2$,
there are the following OPEs with first order poles
\bea
{\cal Q}^{a }_{\,\,\,\al}(z) \,
{\cal W}(w) \Bigg|_{\frac{1}{(z-w)}}
& = & {\cal W}^a_{\al}(w), 
\qquad
\dot{{\cal Q}}^{\dot{\al} }_{\,\,\,a}(z) \,
{\cal W}(w) \Bigg|_{\frac{1}{(z-w)}}
 =  -\dot{{\cal W}}^{\dot{\al}}_{a}(w),
\nonu \\
{\cal R}^{\left[ a \right.}_{\,\,\,b}(z) \,
{\cal W}^{\left. c \right]}_{\,\,\,d}(w) \Bigg|_{\frac{1}{(z-w)}}
& = & \de^{\left[ a \right.}_{d} \, {\cal W}^{\left. c \right]}_{\,\,\,b}(w)-
\de^{\left[ c \right.}_{b} \, {\cal W}^{\left. a \right]}_{\,\,\,d}(w).
\nonu  
\eea

%%%%%%%%%%%%%%%%%%%
$\bullet$ The $s=\frac{3}{2}$ case
%%%%%%%%%%%%%%%%%%%

There are the following first order poles in the
OPEs as well as the ones  in
the subsection $3.3$
\bea
{\cal V}(z) \,
{\cal W}^a_{\,\,\,\al}(w) \Bigg|_{\frac{1}{(z-w)}}
& = &-{\cal W}^a_{\,\,\,\al}(w),
\nonu \\
{\cal R}^{a }_{\,\,\,b}(z) \,
{\cal W}^c_{\,\,\,\al}(w) \Bigg|_{\frac{1}{(z-w)}}
&= &  -\de^{c}_{b}\, {\cal W}^a_{\,\,\,\al}(w)
+ \frac{1}{4} \, \de^{a}_{b} \, {\cal W}^c_{\,\,\,\al}(w),
\nonu \\
{\cal V}(z) \,
{\cal W}^{\left[ a b \right]}_{\,\,\, c \, \al}(w) \Bigg|_{\frac{1}{(z-w)}}
& = &-{\cal W}^{\left[ a b \right]}_{\,\,\, c \, \al}(w),
\nonu \\
{\cal R}^{\left[ a \right. }_{\,\,\,b}(z) \,
{\cal W}^{\left. c d \right]}_{\,\,\,e \, \al}(w) \Bigg|_{\frac{1}{(z-w)}}
& = & -\de^{\left[ c \right. }_{b}\,
{\cal W}^{\left. a d \right]}_{\,\,\,e \, \al}(w)+
\frac{1}{4}\, \de^{\left[ a \right.}_{b} \,
{\cal W}^{\left. c d \right]}_{\,\,\,e \, \al}(w)+
\de^{\left[ a \right. }_{e}\,
{\cal W}^{\left. c d \right]}_{\,\,\,b \, \al}(w)
-
\de^{\left[ d \right. }_{b}\,
{\cal W}^{\left. c a \right]}_{\,\,\,e \, \al}(w),
\nonu \\
{\cal V}(z) \,
\dot{{\cal W}}^{ b \dot{\al}}_{\,\,\,   \left[ a c \right]}(w)
\Bigg|_{\frac{1}{(z-w)}}
& = & \dot{{\cal W}}^{ b \dot{\al}}_{\,\,\,  \left[ a c \right]}(w),
\nonu \\
{\cal R}^{ a  }_{\,\,\, \left[ b \right.}(z) \,
\dot{{\cal W}}^{ d \,
\dot{\al}}_{\,\,\, \left. c e \right] }(w) \Bigg|_{\frac{1}{(z-w)}}
& = & \de^{a}_{\left[ c \right.}\,
\dot{{\cal W}}^{ d \, \dot{\al}}_{\,\,\, \left. b e \right] }(w)-
\frac{1}{4}\, \de^{a}_{\left[ b \right.} \,
\dot{{\cal W}}^{ d \, \dot{\al}}_{\,\,\, \left. c e \right] }(w)+
\de^a_{\left[ e \right.} \, \dot{{\cal W}}^{ d \, \dot{\al}}_{\,\,\, \left. c b  \right] }(w)
-\de^{d}_{\left[ b \right.}\, \dot{{\cal W}}^{ a \, \dot{\al}}_{\,\,\, \left. c e \right] }(w),
\nonu \\
%\eea
%
%\subsection{}
%
%\bea
{\cal V}(z) \,
\dot{{\cal W}}^{\dot{\al}}_{\,\,\,a}(w) \Bigg|_{\frac{1}{(z-w)}}
& = & \dot{{\cal W}}^{\dot{\al}}_{\,\,\,a}(w),
\nonu \\
{\cal R}^{a }_{\,\,\,b}(z) \,
\dot{{\cal W}}^{\dot{\al}}_{\,\,\,c}(w) \Bigg|_{\frac{1}{(z-w)}}
& = &\de^{a}_{c}\, \dot{{\cal W}}^{\dot{\al}}_{\,\,\,b}(w)
- \frac{1}{4} \, \de^{a}_{b} \, \dot{{\cal W}}^{\dot{\al}}_{\,\,\,c}(w).
\nonu
\eea

%%%%%%%%%%%%%%%%%%%%%
$\bullet$ The $s=2$ case
%%%%%%%%%%%%%%%%%%%%

There are the first order poles of the
following OPEs in addition to the
ones in the subsection $3.4$
\bea
{\cal R}^{a }_{\,\,\,b}(z) \,
{\cal W}^{c\, \dot{\al}}_{\,\,\, d \, \al}(w) \Bigg|_{\frac{1}{(z-w)}}
& = &\de^a_d\,{\cal W}^{c\, \dot{\al}}_{\,\,\, b \, \al}(w)-
\de^c_b \, {\cal W}^{a\, \dot{\al}}_{\,\,\, d \, \al}(w),
\nonu \\
{\cal V}(z) \,
{\cal W}^{ \left(a b\right)}_{\,\,\,\al \beta}(w) \Bigg|_{\frac{1}{(z-w)}}
& = & -2 \, {\cal W}^{\left(a b\right)}_{\,\,\,\al \beta}(w),
\nonu \\
{\cal R}^{(a }_{\,\,\,b}(z) \,
{\cal W}^{\left. c d \right)}_{\,\, \, \al \beta}(w) \Bigg|_{\frac{1}{(z-w)}}
& = & -\de^{\left( c \right.}_{b} \,
{\cal W}^{\left. a d \right)}_{\,\, \, \al \beta}(w) +\frac{1}{4}\,
\de^{\left( a \right.}_{b}\, {\cal W}^{\left. c d \right)}_{\,\, \, \al \beta}(w)
+ \de^{\left( d \right. }_{b} \, {\cal W}^{\left. c a \right)}_{\,\, \, \al \beta}(w)
\nonu \\
& - & \frac{1}{4}\, \de^{\left( a \right.}_{b} \,
{\cal W}^{\left. c d \right)}_{\,\, \, \al \beta}(w), 
\nonu \\
\dot{{\cal Q}}^{\dot{\al}}_{\,\,\,a}(z) \,
{\cal W}^{ \left(b c\right)  }_{\,\, \,  \beta \ga}(w) \Bigg|_{\frac{1}{(z-w)}}
& = & -\widehat{\cal W}^{\left(b c\right) \,
\dot{\al}}_{\,\,\, a \,  \beta \ga}(w)
+
\de^{\left( b \right.}_{a} \, {\cal W}^{ \left. c \right) \,
\dot{\al}}_{\,\,\, \beta \ga}(w)
-\de^{\left( c \right.}_{a} \, {\cal W}^{\left. b \right) \,
\dot{\al}}_{\,\,\,  \beta \ga}(w),
\nonu \\
%\eea
%
%\subsection{}
%
%\bea
{\cal Q}^{a}_{\,\,\,\al}(z) \,
{\cal W}^{  \dot{\beta} }_{\,\, \,   \ga}(w) \Bigg|_{\frac{1}{(z-w)}}
& = & {\cal W}^{ a \, \dot{\beta} }_{\,\, \,  \al  \ga}(w),
\qquad
\dot{{\cal Q}}^{\dot{\al}}_{\,\,\,a}(z) \,
{\cal W}^{  \dot{\beta} }_{\,\, \,   \ga}(w) \Bigg|_{\frac{1}{(z-w)}}
=  -\dot{{\cal W}}^{ \dot{\al}  \dot{\beta} }_{\,\, \,  a\,  \ga}(w),
\nonu  \\
{\cal R}^{\left[ a \right. }_{\,\,\, \left[ b\right.}(z) \,
{\cal W}^{\left. c e \right] \, \dot{\al}}_{\,\, \, \left. d f \right] \, \beta }(w)
\Bigg|_{\frac{1}{(z-w)}}
& = & -\de^{\left[ c \right.}_{\left[ b \right.}\,
{\cal W}^{\left. a e \right] \, \dot{\al}}_{\,\, \, \left. d f
\right] \, \beta }(w)
%+
%\frac{1}{4}\,
%\de^{\left[ a \right.}_{\left[ b \right.}\,
%    {\cal W}^{\left. c e \right] \, \dot{\al}}_{\,\, \, \left. d f
%    \right] \, \beta }(w)
+
\de^{\left[ a \right.}_{\left[ d \right.}\,
{\cal W}^{\left. c e \right] \, \dot{\al}}_{\,\, \, \left. b f
\right] \, \beta }(w)
%-
%\frac{1}{4}\,
%\de^{\left[ a \right.}_{\left[ b \right.}\,
%    {\cal W}^{\left. c e \right] \, \dot{\al}}_{\,\, \, \left. d f
%    \right] \, \beta }(w)
+\de^{\left[ a \right.}_{\left[ f \right.}\,
{\cal W}^{\left. c e \right] \, \dot{\al}}_{\,\, \, \left. d b 
\right] \, \beta }(w)\nonu \\
& - &
\de^{\left[ e \right.}_{\left[ b \right.}\,
{\cal W}^{\left. c a \right] \, \dot{\al}}_{\,\, \, \left. d f
\right] \, \beta }(w),
\nonu \\
{\cal Q}^{\left[ a \right. }_{\,\,\, \al}(z) \,
{\cal W}^{\left. b d\right] \, \dot{\beta}}_{\,\, \, \left[
c e \right] \, \ga }(w)
\Bigg|_{\frac{1}{(z-w)}} & = & -\de^{\left[ a \right.}_{\left[ c \right.}
\, {\cal W}^{ \left. b d \right] \, \dot{\beta}}_{\,\,\, \left.
e \right] \ga \al}(w) +
\de^{\left[ a \right.}_{\left[ e \right.}\,
\widehat{\cal W}^{\left. b d \right]
\, \dot{\beta}}_{\,\,\, \left.
c \right] \,  \ga \al}(w)     
-\frac{1}{4}\, \de^{\left. d \right.}_{\left[ e \right.}\,
\widehat{\cal W}^{\left. b a  \right]
\, \dot{\beta}}_{\,\,\, \left.
c \right] \, \ga \al}(w),   
\nonu \\
\dot{{\cal Q}}^{\dot{\al} }_{\,\,\, \left[ a \right.}(z) \,
{\cal W}^{\left[ b d\right] \, \dot{\beta}}_{\,\, \, \left.
c e \right] \, \ga }(w)
\Bigg|_{\frac{1}{(z-w)}} & = & \de^{\left[ b \right.}_{\left[ a \right.}
\, \dot{{\cal W}}^{ \left.  d \right] \, 
\dot{\beta} \dot{\al}}_{\,\,\, \left.
c e \right] \ga }(w) -
\de^{\left[ d \right.}_{\left[ a \right.}\,
\dot{\widehat{\cal W}}^{ \left. b \right]
\, \dot{\al}\dot{\beta}}_{\,\,\, \left.
 e c \right] \,  \ga}(w)     
+\frac{1}{4}\,\de^{\left[ d \right.}_{\left[ e \right.}\,
\dot{\widehat{\cal W}}^{ \left. b \right]
\, \dot{\al}\dot{\beta}}_{\,\,\, \left.
a c \right] \,  \ga}(w),    
\nonu \\
    {\cal V}(z) \, \dot{{\cal W}}^{\dot{\al}\dot{\beta}}_{\,\,\,
      \left( a b \right)}(w)
\Bigg|_{\frac{1}{(z-w)}} & = &
2 \,  \dot{{\cal W}}^{\dot{\al}\dot{\beta}}_{\,\,\, \left( a b
\right)}(w),
\nonu \\
{\cal R}^{a  }_{\,\,\, \left( b\right.}(z) \,
\dot{{\cal W}}^{\dot{\al}\dot{\beta}}_
{\,\, \, \left. c d \right)  }(w)
\Bigg|_{\frac{1}{(z-w)}}
& = & \de^a_{\left(c \right.} \,
 \dot{{\cal W}}^{\dot{\al}\dot{\beta}}_
{\,\, \, \left. b d \right)  }(w)
-\frac{1}{2}\, \de^{a}_{\left( b \right.}\,
\dot{{\cal W}}^{\dot{\al}\dot{\beta}}_
{\,\, \, \left. c d \right)  }(w)+
\de^a_{\left(d \right.} \,
\dot{{\cal W}}^{\dot{\al}\dot{\beta}}_
{\,\, \, \left.  c b \right)  }(w),     
\nonu \\
{\cal Q}^{a}_{\,\,\,\al}(z) \,
\dot{{\cal W}}^{ \dot{\beta}\dot{\ga}  }_{\,\, \,  \left(b c\right)}(w)
\Bigg|_{\frac{1}{(z-w)}}
& = & \dot{\widehat{\cal W}}^{a \,
\dot{\ga}\dot{\beta}}_{\,\,\, \left(b c \right)  \al}(w)
+\de^{ a }_{\left(b \right.} \,
\dot{{\cal W}}^{ \dot{\ga} \dot{\beta}}_{\,\,\,
\left. c \right) \,  \al}(w)
-\de^{a }_{\left( c \right.} \,
\dot{{\cal W}}^{\dot{\ga}\dot{\beta}}_{\,\,\,  \left. b \right)\, \al}(w),
\nonu     \\
{\cal V}(z) \, {\cal W}^{\left[ a b\right] c}_{\,\,\,c \, \al\beta}(w)
\Bigg|_{\frac{1}{(z-w)}} & = &
-2 \,  {\cal W}^{\left[ a b \right]c}_{\,\,\,c \, \al\beta}(w),
\nonu \\
{\cal R}^{\left[ a \right.}_
{\,\,\,b}(z)\,  {\cal W}^{\left. c d\right] e}_
{\,\,\,e \, \al\beta}(w)\,
\Bigg|_{\frac{1}{(z-w)}} & = &
-\de^{\left[ c \right.}_{b}\,
{\cal W}^{\left. a d \right] e}_
{\,\,\,e \, \al\beta}(w)  + \frac{1}{2}\,
\de^{\left[ a \right.}_{b}\,
{\cal W}^{\left. c d \right] e}_
   {\,\,\,e \, \al\beta}(w)
%-
%\de^{ e }_{b}\,
%{\cal W}^{\left[  a c d \right]}_
%{\,\,\,e \, \al\beta}(w),
%\nonu \\
%&+& \de^{\left[ a \right.}_{e}\,
%{\cal W}^{\left. c d \right] e}_
%{\,\,\,b \, \al\beta}(w)
-
\de^{\left[ d \right.}_{b}\,
{\cal W}^{\left.  c a  \right] e}_
{\,\,\,e \, \al\beta}(w),      
\nonu \\
\dot{{\cal Q}}^{\dot{\al}}_{\,\,\,a}(z) \,
\,  {\cal W}^{ \left[ c d \right] e }_
{\,\,\,e \, \al\beta}(w)\,
\Bigg|_{\frac{1}{(z-w)}} & = &
\de^{\left[ c \right.}_{a}\, {\cal W}^{\left. d e \right]   \, \dot{\al}}_{\,\,\,
e \, \beta \al}(w) -
\de^{\left[ d \right.}_{a}\, {\cal W}^{\left. c  e \right]   \, \dot{\al}}_{\,\,\,
e \, \al \beta}(w)-
\de^{ \left[ e \right. }_{a}\,
\widehat{{\cal W}}^{\left. c d \right] \, \dot{\al}}_{\,\,\,
    e \, \al \beta }(w)+
\frac{1}{4}\,
\de^{ \left[ e \right.  }_{e}\,
\widehat{{\cal W}}^{\left. c d \right] \,
\dot{\al}}_{\,\,\,
a \, \al \beta }(w),
\nonu \\
{\cal V}(z) \, \dot{{\cal W}}^{  c  \, \dot{\al} \dot{\beta} }_
{\,\,\, \left[ a b \right] c }(w)\,
\Bigg|_{\frac{1}{(z-w)}} & = & 2 \,
\dot{{\cal W}}^{  c  \, \dot{\al} \dot{\beta} }_
{\,\,\, \left[  a b \right] c }(w),
\nonu \\
{\cal R}^a_{\left[ b \right.}\,
 \dot{{\cal W}}^{  c  \, \dot{\al} \dot{\beta} }_
{\,\,\, \left. d e\right] c  }(w)\,
\Bigg|_{\frac{1}{(z-w)}} & = &
\de^a_{\left[ d \right.}\,
\dot{{\cal W}}^{  c  \, \dot{\al} \dot{\beta} }_
    {\,\,\, \left. b e \right] c  }(w)-\frac{1}{2}\,
\de^a_{\left[ b \right. }\,
\dot{{\cal W}}^{  c  \, \dot{\al} \dot{\beta} }_
    {\,\,\, \left. d e \right] c  }(w)
+\de^a_{\left[ e \right.}\,
\dot{{\cal W}}^{  c  \, \dot{\al} \dot{\beta} }_
    {\,\,\, \left.  d b \right] c  }(w), 
\nonu 
%&+&\de^a_{c }\,
%\dot{{\cal W}}^{  c  \, \dot{\al} \dot{\beta} }_
%    {\,\,\, \left[ b d e \right] }(w)-
%\de^c_{ \left[ b \right. }\,
%\dot{{\cal W}}^{  a  \, \dot{\al} \dot{\beta} }_
%    {\,\,\, \left. c d e \right] }(w),
%\nonu \\
%\eea
\eea
where the new higher spin generators are given by
\bea
\widehat{\cal W}^{(b c) \, \dot{\al}}_{\,\,\, a \, \beta \ga}  & \equiv &
\dot{{\cal Q}}^{\dot{\al}}_{\,\,\,a}\, {\cal Q}^{\left( b \right.}_{\,\,\,
\beta}\, {\cal Q}^{\left. c \right)}_{\,\,\,\ga}-
{\cal Q}^{\left( b \right.}_{\,\,\,
\beta}\, \dot{{\cal Q}}^{\dot{\al}}_{\,\,\,a}\,
{\cal Q}^{\left. c \right)}_{\,\,\,\ga} +
{\cal Q}^{\left( b \right.}_{\,\,\,
\beta}\, {\cal Q}^{\left. c \right)}_{\,\,\,\ga}\, \dot{{\cal Q}}^{\dot{\al}}_{\,\,\,a} \nonu \\
&-& \dot{{\cal Q}}^{\dot{\al}}_{\,\,\,a}\,
{\cal Q}^{\left( c \right.}_{\,\,\,\ga}
\, {\cal Q}^{\left. b \right)}_{\,\,\,
\beta} +
{\cal Q}^{\left( c \right.}_{\,\,\,\ga} \,
\dot{{\cal Q}}^{\dot{\al}}_{\,\,\,a}\,
{\cal Q}^{\left. b \right)}_{\,\,\,
\beta}-
{\cal Q}^{\left( c \right.}_{\,\,\,\ga}\,
{\cal Q}^{\left. b \right)}_{\,\,\,
\beta}\,
\dot{{\cal Q}}^{\dot{\al}}_{\,\,\,a},
\nonu \\
\widehat{\cal W}^{\left[b c \right] \, \dot{\al}}_{\,\,\, a \, \beta \ga}  & \equiv &
\dot{{\cal Q}}^{\dot{\al}}_{\,\,\,a}\, {\cal Q}^{\left[ b \right.}_{\,\,\,
\beta}\, {\cal Q}^{\left. c \right]}_{\,\,\,\ga}-
{\cal Q}^{\left[ b \right.}_{\,\,\,
\beta}\, \dot{{\cal Q}}^{\dot{\al}}_{\,\,\,a}\,
{\cal Q}^{\left. c \right]}_{\,\,\,\ga} +
{\cal Q}^{\left[ b \right.}_{\,\,\,
\beta}\, {\cal Q}^{\left. c \right]}_{\,\,\,\ga}\,
\dot{{\cal Q}}^{\dot{\al}}_{\,\,\,a} \nonu \\
&-& \dot{{\cal Q}}^{\dot{\al}}_{\,\,\,a}\,
{\cal Q}^{\left[ c \right.}_{\,\,\,\ga}
\, {\cal Q}^{\left. b \right]}_{\,\,\,
\beta} +
{\cal Q}^{\left[ c \right.}_{\,\,\,\ga} \,
\dot{{\cal Q}}^{\dot{\al}}_{\,\,\,a}\,
{\cal Q}^{\left. b \right]}_{\,\,\,
\beta}-
{\cal Q}^{\left[ c \right.}_{\,\,\,\ga}\,
{\cal Q}^{\left. b \right]}_{\,\,\,
\beta}\,
\dot{{\cal Q}}^{\dot{\al}}_{\,\,\,a},
\nonu \\
\dot{\widehat{\cal W}}^{b \, \dot{\al} \dot{\beta}}_{\,\,\,
\left[c d\right] \,  \ga}
& \equiv &
{\cal Q}^{b}_{\,\,\,\ga}\,
\dot{{\cal Q}}^{\dot{\beta}}_{\,\,\,\left[d \right.}\,
\dot{{\cal Q}}^{\dot{\al}}_{\,\,\, \left.c \right]}-
\dot{{\cal Q}}^{\dot{\beta}}_{\,\,\,\left[d \right.}\,
{\cal Q}^{b}_{\,\,\,\ga}\,
\dot{{\cal Q}}^{\dot{\al}}_{\,\,\, \left.c \right]}+
\dot{{\cal Q}}^{\dot{\beta}}_{\,\,\,\left[d \right.}\,
\dot{{\cal Q}}^{\dot{\al}}_{\,\,\, \left.c \right]}  \,
{\cal Q}^{b}_{\,\,\,\ga}\,
\nonu \\
&-& {\cal Q}^{b}_{\,\,\,\ga}\,
\dot{{\cal Q}}^{\dot{\al}}_{\,\,\, \left[ c \right.} \,
\dot{{\cal Q}}^{\dot{\beta}}_{\,\,\,\left. d \right]}\,+
\dot{{\cal Q}}^{\dot{\al}}_{\,\,\, \left[ c \right.}
\,    {\cal Q}^{b}_{\,\,\,\ga}\,  
\dot{{\cal Q}}^{\dot{\beta}}_{\,\,\,\left. d \right]}\,
-  \dot{{\cal Q}}^{\dot{\al}}_{\,\,\, \left[ c \right.}  \,
\dot{{\cal Q}}^{\dot{\beta}}_{\,\,\,\left. d \right]}\,
  {\cal Q}^{b}_{\,\,\,\ga}, \nonu \\
  \dot{\widehat{\cal W}}^{b \, \dot{\al} \dot{\beta}}_{\,\,\,
\left(c d\right) \,  \ga}
& \equiv &
{\cal Q}^{b}_{\,\,\,\ga}\,
\dot{{\cal Q}}^{\dot{\beta}}_{\,\,\,\left(d \right.}\,
\dot{{\cal Q}}^{\dot{\al}}_{\,\,\, \left.c \right)}-
\dot{{\cal Q}}^{\dot{\beta}}_{\,\,\,\left(d \right.}\,
{\cal Q}^{b}_{\,\,\,\ga}\,
\dot{{\cal Q}}^{\dot{\al}}_{\,\,\, \left.c \right)}+
\dot{{\cal Q}}^{\dot{\beta}}_{\,\,\,\left(d \right.}\,
\dot{{\cal Q}}^{\dot{\al}}_{\,\,\, \left.c \right)}  \,
{\cal Q}^{b}_{\,\,\,\ga}\,
\nonu \\
&-& {\cal Q}^{b}_{\,\,\,\ga}\,
\dot{{\cal Q}}^{\dot{\al}}_{\,\,\, \left( c \right.} \,
\dot{{\cal Q}}^{\dot{\beta}}_{\,\,\,\left.d \right)}\,+
\dot{{\cal Q}}^{\dot{\al}}_{\,\,\, \left( c \right.}
\,    {\cal Q}^{b}_{\,\,\,\ga}\,  
\dot{{\cal Q}}^{\dot{\beta}}_{\,\,\,\left. d \right)}\,
-  \dot{{\cal Q}}^{\dot{\al}}_{\,\,\, \left( c \right.}  \,
\dot{{\cal Q}}^{\dot{\beta}}_{\,\,\,\left. d \right)}\,
  {\cal Q}^{b}_{\,\,\,\ga}.
\label{hatoperators}
\eea

%%%%%%%%%%%%%%%%%%%%%%%%
%$\bullet$ The $s=\frac{5}{2}$ case
%%%%%%%%%%%%%%%%%%%%%%%%

%%%%%%%%%%%%%%%%%%%%%%%%%
$\bullet$ The  $s=\frac{5}{2}$ case 
%%%%%%%%%%%%%%%%%%%%%%%%%

There exist the first order poles of the
following OPEs (and the ones of the subsection $3.5$)
\bea
{\cal V}(z) \, {\cal W}^{  \left[ a b \right] \, \dot{\al} }_
{\,\,\, c \,    \beta  \ga }(w)\,
\Bigg|_{\frac{1}{(z-w)}} & = &
- {\cal W}^{  \left[ a b \right] \, \dot{\al} }_
{\,\,\, c \,    \beta  \ga }(w),
\nonu \\
{\cal R}^{\left[ a \right.}_{\,\,\,b}
\,{\cal W}^{  \left. c d \right] \, \dot{\al} }_
{\,\,\, e \,    \beta  \ga }(w)\,
\Bigg|_{\frac{1}{(z-w)}} & = &
-\de^{\left[ c \right. }_{b}\,
{\cal W}^{  \left. a d \right] \, \dot{\al} }_
{\,\,\, e \,    \beta  \ga }(w) +\frac{1}{4}\,
\de^{\left[ a \right.}_{b} \,
{\cal W}^{  \left. c d \right] \, \dot{\al} }_
{\,\,\, e \,    \beta  \ga }(w)+
\de^{\left[ a \right.}_{e} \,
{\cal W}^{  \left. c d \right] \, \dot{\al} }_
{\,\,\, b \,    \beta  \ga }(w)   
\nonu \\
&-& \de^{\left[ d \right.}_{b} \,
{\cal W}^{  \left.  c a \right] \, \dot{\al} }_
{\,\,\, e \,    \beta  \ga }(w),   
\nonu \\
{\cal V}(z) \, {\cal W}^{  a\, \dot{\al} }_
{\,\,\,   \beta  \ga }(w)\,
\Bigg|_{\frac{1}{(z-w)}} & = &
-{\cal W}^{  a\, \dot{\al} }_
{\,\,\,   \beta  \ga }(w),
\nonu \\
{\cal R}^{a }_{\,\,\,  b }(z)\,
{\cal W}^{  c\, \dot{\al} }_
{\,\,\,   \beta  \ga }(w)\,
\Bigg|_{\frac{1}{(z-w)}} & = & 
-\de^c_{b}\, {\cal W}^{  a\, \dot{\al} }_
{\,\,\,   \beta  \ga }(w)+\frac{1}{4}\,
\de^a_b \,
{\cal W}^{  c\, \dot{\al} }_
{\,\,\,   \beta  \ga }(w),
\nonu \\
{\cal Q}^{\left[ a \right.}_{\,\,\,\al}(z) \,
{\cal W}^{  \left. b \right] \, \dot{\al} }_
{\,\,\,   \beta  \ga }(w)\,
\Bigg|_{\frac{1}{(z-w)}} & = &
{\cal W}^{  \left[ a  b \right] \, \dot{\al} }_
{\,\,\,   \al \beta  \ga }(w),
\qquad
{\cal V}(z) \, \dot{{\cal W}}^{  a\, \dot{\al} \dot{\beta} }_
{\,\,\,   \left[ b c \right] \,  \ga }(w)\,
\Bigg|_{\frac{1}{(z-w)}}  = 
\dot{{\cal W}}^{  a\, \dot{\al} \dot{\beta} }_
{\,\,\,   \left[ b c \right] \,  \ga }(w),
\nonu \\
{\cal R}^{a }_{\,\,\, \left[ b \right.}(z)\,
\dot{{\cal W}}^{  c\, \dot{\al} \dot{\beta} }_
{\,\,\,   \left. d e \right] \,  \ga }(w)\,
\Bigg|_{\frac{1}{(z-w)}} & = &
\de^{a}_{\left[ d \right.}\,
\dot{{\cal W}}^{  c\, \dot{\al} \dot{\beta} }_
{\,\,\,   \left. b e \right] \,  \ga }(w)  
-\frac{1}{4} \, \de^a_{\left[ b \right.}
\dot{{\cal W}}^{  c\, \dot{\al} \dot{\beta} }_
{\,\,\,   \left. d e \right] \,  \ga }(w)+
\de^{a}_{\left[ e \right.}\,
\dot{{\cal W}}^{  c\, \dot{\al} \dot{\beta} }_
{\,\,\,   \left.  d b \right] \,  \ga }(w)
\nonu \\
& - & \de^{c}_{\left[ e \right.}\,
\dot{{\cal W}}^{  a\, \dot{\al} \dot{\beta} }_
{\,\,\,   \left.  d b \right] \,  \ga }(w),
  \nonu \\
{\cal V}(z)\,  \dot{{\cal W}}^{  \dot{\al} \dot{\beta} }_
{\,\,\,   a \, \ga }(w)\,
\Bigg|_{\frac{1}{(z-w)}} & = &
\dot{{\cal W}}^{  \dot{\al} \dot{\beta} }_
{\,\,\,   a \, \ga }(w),
\nonu \\
{\cal R}^{a }_{\,\,\, b}(z)\,
\dot{{\cal W}}^{  \dot{\al} \dot{\beta} }_
{\,\,\,   c\,  \ga }(w)\,
\Bigg|_{\frac{1}{(z-w)}} & = &
\de^a_c\, \dot{\cal W}^{  \dot{\al} \dot{\beta} }_
{\,\,\,   b\,  \ga }(w)-\frac{1}{4}\, \de^a_b \,
\dot{\cal W}^{  \dot{\al} \dot{\beta} }_
{\,\,\,   c\,  \ga }(w),
\nonu \\
\dot{{\cal Q}}^{\dot{\al}}_{\,\,\, \left[ a\right.}(z)\,
\dot{{\cal W}}^{  \dot{\beta} \dot{\ga} }_
{\,\,\,   \left. b \right]\,  \ga }(w)\,
\Bigg|_{\frac{1}{(z-w)}} & = & -
\dot{{\cal W}}^{  \dot{\al} \dot{\beta} \dot{\ga} }_
{\,\,\,   \left[ a b \right]\,  \ga }(w),
\qquad
{\cal V}(z)\,  {\cal W}^{ \left[ a b c \right] }_
{\,\,\,  \al \beta \ga }(w)\,
\Bigg|_{\frac{1}{(z-w)}}  =  -3 \,
{\cal W}^{ \left[a b c\right] }_
{\,\,\,  \al \beta \ga }(w),
\nonu \\
{\cal R}^{\left[ a \right.}_{\,\,\, b}(z)\,
{\cal W}^{ \left. c d e \right] }_{\,\,\,  \al \beta \ga }(w)\,
\Bigg|_{\frac{1}{(z-w)}} & = &
-\de^{\left[ c \right. }_{b}\,
  {\cal W}^{ \left. a d e \right] }_{\,\,\,  \al \beta \ga }(w)
-\de^{\left[ d \right. }_{b}\,
  {\cal W}^{ \left. c a  e \right] }_{\,\,\,  \al \beta \ga }(w)
-\de^{\left[ e \right. }_{b}\,
{\cal W}^{ \left. c d a \right] }_{\,\,\,  \al \beta \ga }(w) 
\nonu \\
&  + & \frac{3}{4}\, \de^{\left[ a \right. }_{b}\,
{\cal W}^{ \left. c d e \right] }_{\,\,\,  \al \beta \ga }(w),   
\nonu \\
\dot{{\cal Q}}^{\dot{\al}}_{\,\,\, a}(z)\,
{\cal W}^{ \left[ b c d  \right] }_{\,\,\,  \al \beta \ga }(w)\,
\Bigg|_{\frac{1}{(z-w)}} & = &
\de^{\left[ b \right.}_{a}\,
  {\cal W}^{ \left.  c d \right] \, \dot{\al} }_{\,\,\,
  \beta \ga \al }(w),  
\qquad
{\cal V}(z)\,  \dot{{\cal W}}^{ \dot{\al} \dot{\beta}\dot{\ga} }_
{\,\,\,  \left[ a b c \right] }(w)\,
\Bigg|_{\frac{1}{(z-w)}}  = 
3 \,  \dot{{\cal W}}^{ \dot{\al} \dot{\beta}\dot{\ga} }_
{\,\,\,  \left[ a b c \right] }(w), 
\nonu \\
{\cal R}^a_{\,\,\,\left[b \right.}(z) \,
\dot{{\cal W}}^{ \dot{\al} \dot{\beta}\dot{\ga} }_
{\,\,\,  \left. c d e\right] }(w)\,
\Bigg|_{\frac{1}{(z-w)}}
& = & \de^{a }_{ \left[ c \right.}\,
\dot{{\cal W}}^{ \dot{\al} \dot{\beta}\dot{\ga} }_
{\,\,\,  \left. b d e\right] }(w)
+\de^{a }_{ \left[ d \right.}\,
\dot{{\cal W}}^{ \dot{\al} \dot{\beta}\dot{\ga} }_
{\,\,\,  \left. c b e\right] }(w)+
\de^{a }_{ \left[ e \right.}\,
\dot{{\cal W}}^{ \dot{\al} \dot{\beta}\dot{\ga} }_
{\,\,\,  \left. c d b \right] }(w)
\nonu \\
& - &
\frac{3}{4} \, \de^{a}_{ \left[ b \right.}
\, \dot{{\cal W}}^{ \dot{\al} \dot{\beta}\dot{\ga} }_
{\,\,\,  \left. c d e\right] }(w),
\nonu \\
{\cal Q}^a_{\,\,\,\al}(z) \,
\dot{{\cal W}}^{ \dot{\al} \dot{\beta}\dot{\ga} }_
{\,\,\,  \left[b c d \right] }(w)\,
\Bigg|_{\frac{1}{(z-w)}}
& = & \de^{a}_{ \left[ b \right.}\,
\dot{{\cal W}}^{  \dot{\beta}\dot{\ga}\dot{\al} }_
{\,\,\,  \left. c d \right]\, \al }(w).
\nonu
\eea

%%%%%%%%%%%%%%%%%%
$\bullet$ The $s=3$ case
%%%%%%%%%%%%%%%%%%

We have the following OPEs
with first order poles
in addition to the ones of the subsection
$3.6$
\bea
{\cal V}(z) \,
\,  {\cal W}^{ \left[ a b  \right]\, \dot{\al} }_
{\,\,\,   \beta  \ga \de }(w)\,
\Bigg|_{\frac{1}{(z-w)}} & = & -2 \,
{\cal W}^{ \left[ a b\right] \, \dot{\al} }_
{\,\,\,   \beta  \ga \de }(w),
\nonu \\
{\cal R}^{\left[ a \right.}_{\,\,\,\left[b \right.}(z) \,
{\cal W}^{ \left. c d\right] \, \dot{\al} }_
{\,\,\,   \beta  \ga \al}(w)\,
\Bigg|_{\frac{1}{(z-w)}} & = &
-\de^{\left[ c \right.}_{b}\,
{\cal W}^{ \left. a d \right] \, \dot{\al} }_
{\,\,\,   \beta  \ga \al}(w) + \frac{1}{2}\,
\de^{\left[ a \right.}_{b}\,
{\cal W}^{ \left. c d \right] \, \dot{\al} }_
{\,\,\,   \beta  \ga \al }(w)-
\de^{\left[ d \right.}_{b}\,
{\cal W}^{ \left. c a  \right] \, \dot{\al} }_
{\,\,\,   \beta  \ga \al }(w),   
\nonu \\
{\cal V}(z) \,
\,  \dot{{\cal W}}^{ \dot{\al} \dot{\beta}\dot{\ga} }_
{\,\,\,  \left[ a b \right] \,  \de }(w)\,
\Bigg|_{\frac{1}{(z-w)}} & = &
2 \,  \dot{{\cal W}}^{ \dot{\al} \dot{\beta}\dot{\ga} }_
{\,\,\,  \left[ a b \right] \,  \de }(w),
\nonu \\
{\cal R}^a_{\,\,\,\left[b \right.}(z) \,
\,  \dot{{\cal W}}^{ \dot{\al} \dot{\beta}\dot{\ga} }_
{\,\,\,  \left. c d \right] \,  \de }(w)\,
\Bigg|_{\frac{1}{(z-w)}} & = &
\de^a_{\left[ c\right. }\,
\dot{{\cal W}}^{ \dot{\al} \dot{\beta}\dot{\ga} }_
{\,\,\,  \left. b d \right] \,  \de }(w)-\frac{1}{2}\,
\de^a_{\left[ b\right.}\,
\dot{{\cal W}}^{ \dot{\al} \dot{\beta}\dot{\ga} }_
{\,\,\,  \left. c d \right] \,  \de }(w)+
\de^a_{\left[ d\right. }\,
\dot{{\cal W}}^{ \dot{\al} \dot{\beta}\dot{\ga} }_
{\,\,\,  \left.  c b \right] \,  \de }(w),
\nonu \\
{\cal R}^{ a }_
{\,\,\,b}(z)\,  {\cal W}^{ c\,  \dot{\al} \dot{\beta}}_
{\,\,\, d \ga \de}(w)\,
\Bigg|_{\frac{1}{(z-w)}} & = & \de^a_d \,
{\cal W}^{ c\,  \dot{\al} \dot{\beta}}_
{\,\,\, b \ga \de}(w)- \de^{c}_{b} \,
{\cal W}^{ a\,  \dot{\al} \dot{\beta}}_
{\,\,\, d \ga \de}(w),
\nonu \\
{\cal Q}^a_{\,\,\,\al}(z) \,
{\cal W}^{ c\,  \dot{\al} \dot{\beta}}_
{\,\,\, d \ga \de}(w)
\,
\Bigg|_{\frac{1}{(z-w)}} & = &
\de^a_d \, {\cal W}^{ c\,  \dot{\al} \dot{\beta}}_
{\,\,\, \al \ga \de}(w)-\frac{1}{4}\, \de^{c}_{d}\,
{\cal W}^{ a\,  \dot{\al} \dot{\beta}}_
{\,\,\, \al \ga \de}(w),   
\nonu \\
\dot{{\cal Q}}^{\dot{\al}}_{\,\,\,a}(z) \,
{\cal W}^{ c   \dot{\beta} \dot{\ga}}_
{\,\,\, d \beta \ga}(w)
\,
\Bigg|_{\frac{1}{(z-w)}} & = & -\de^{c}_{a}\,
\dot{{\cal W}}^{ \dot{\al}   \dot{\beta} \dot{\ga}}_
{\,\,\, d \beta \ga}(w)+\frac{1}{4}\,
\de^{c}_d \,
\dot{{\cal W}}^{ \dot{\al}   \dot{\beta} \dot{\ga}}_
{\,\,\, a \beta \ga}(w),
\nonu \\
{\cal Q}^a_{\,\,\,\al}(z) \,
{\cal W}^{\dot{\beta} \dot{\ga}}_{\,\,\,\beta \ga}(w)
\,
\Bigg|_{\frac{1}{(z-w)}} & = &
{\cal W}^{a \dot{\beta} \dot{\ga}}_{\,\,\,\al \beta \ga}(w),
\qquad
\dot{{\cal Q}}^{\dot{\al}}_{\,\,\,a}(z) \,
{\cal W}^{\dot{\beta} \dot{\ga}}_{\,\,\,\beta \ga}(w)
\,
\Bigg|_{\frac{1}{(z-w)}}  = 
-\dot{{\cal W}}^{\dot{\al} \dot{\beta} \dot{\ga}}_{\,\,\,a \, \beta \ga}(w).
\nonu
\eea

%%%%%%%%%%%%%%%%%%%%%%%
$\bullet$ The $s=\frac{7}{2}$ case
%%%%%%%%%%%%%%%%%%%%%%%

As well as the OPEs in the subsection $3.7$
there are following OPEs with first order poles
\bea
{\cal V}(z) \,
\,  {\cal W}^{ a \, \dot{\al} \dot{\beta} }_
{\,\,\,  \ga \de \ep}(w)\,
\Bigg|_{\frac{1}{(z-w)}} & = &
-{\cal W}^{ a \, \dot{\al} \dot{\beta} }_
{\,\,\,  \ga \de \ep}(w),
\nonu \\
{\cal R}^{ a }_
{\,\,\,b}(z)\,  {\cal W}^{ c\,  \dot{\beta} \dot{\ga}}_
{\,\,\, \al \beta \ga}(w)\,
\Bigg|_{\frac{1}{(z-w)}} & = & -\de^{c}_{b}\,
{\cal W}^{ a\,  \dot{\beta} \dot{\ga}}_
{\,\,\, \al \beta \ga}(w) + \frac{1}{4}\,
\de^a_b \,  {\cal W}^{ c\,  \dot{\beta} \dot{\ga}}_
{\,\,\, \al \beta \ga}(w).
\nonu
\eea

%%%%%%%%%%%%%%%%%%
$\bullet$ The $s=4$ case
%%%%%%%%%%%%%%%%%%

Finally, we have the following first order poles
\bea
{\cal V}(z) \,
\,  \dot{{\cal W}}^{ \dot{\al} \dot{\beta} \dot{\ga}}_
{\,\,\, b \de \ep}(w)\,
\Bigg|_{\frac{1}{(z-w)}} & = &
 \dot{{\cal W}}^{ \dot{\al} \dot{\beta} \dot{\ga}}_
{\,\,\, b \de \ep}(w),
\nonu \\
{\cal R}^{ a }_
{\,\,\,b}(z)\,  \dot{{\cal W}}^{ \dot{\al} \dot{\beta} \dot{\ga}}_
{\,\,\, c \de \ep}(w)(w)\,
\Bigg|_{\frac{1}{(z-w)}} & = &
\de^a_c\, \dot{{\cal W}}^{ \dot{\al} \dot{\beta} \dot{\ga}}_
{\,\,\, b \de \ep}(w) -\frac{1}{4}\, \de^a_b \,
\dot{{\cal W}}^{ \dot{\al} \dot{\beta} \dot{\ga}}_
{\,\,\, c \de \ep}(w).
%\nonu \\
%{\cal Q}^{a}_{\,\,\,\al}(z) \,
%\,  \dot{{\cal W}}^{ \dot{\al} \dot{\beta} \dot{\ga}}_
%{\,\,\, b \de \ep}(w)\,
%\Bigg|_{\frac{1}{(z-w)}} & = & \de^a_b \,
%{\cal W}^{ \dot{\al}\dot{\beta} \dot{\ga}}_
%{\,\,\, \al \de \ep}(w),
\nonu  
\eea
%%%%%%%%%%%%%%%%%%%%%%%%%%%%%%%%%%%%%%%%%%%%%%%%%%%%%%%%%%%%%%%%
Therefore, we have calculated the first order poles
in the OPEs between five weight-$1$ operators and the weight-$3$ operators
appearing in the Table $1$.

%%%%%%%%%%%%%%%%
\section{The complete OPE between $J^I_{\,\,\, J}(z)$ and
$ J^{K}_{\,\,\,L}\, J^{M}_{\,\,\,N}\,
J^{P}_{\,\,\,Q}(w)$}
%%%%%%%%%%%%%%%%

From the defining OPE in (\ref{jjope}), we can calculate the remaining
fourth, third and second order poles of the OPE
$J^I_{\,\,\, J}(z) \, J^{K}_{\,\,\,L}\, J^{M}_{\,\,\,N}\,
J^{P}_{\,\,\,Q}(w)$. The first order pole is given by
(\ref{pole1}).

The fourth order pole can be written as
\bea
&& J^I_{\,\,\, J}(z) \, J^{K}_{\,\,\,L}\, J^{M}_{\,\,\,N}\,
J^{P}_{\,\,\,Q}(w)\Bigg|_{\frac{1}{(z-w)^4}} =
(-1)^{d_J d_P+1}\, \de^{I}_{\,\,\,L}\, \de^{K}_{\,\,\,N}\, \de^M_{\,\,\, Q}\,
\de^{P}_{\,\,\,J}
\nonu \\
&& + (-1)^{(d_N+d_M)(d_K+d_J)+d_N d_P}\, \de^{I}_{\,\,\,L} \,  \de^{M}_{\,\,\,J}\,
\de^K_{\,\,\, Q}\,
\de^{P}_{\,\,\,N}  + (-1)^{(d_L+d_K)(d_I+d_J)+1}\,
\nonu \\
&& \times \, \Bigg[
(-1)^{d_L d_P+1}\, \de^{K}_{\,\,\,J}\,  \de^I_{\,\,\, N}\, \de^{M}_{\,\,\,Q}
  \, \de^P_{\,\,\,L}+
  (-1)^{(d_N+d_M)(d_I+d_L)+d_N d_P}\, \de^{K}_{\,\,\,J} \, \de^{M}_{\,\,\, L}
\, \de^{I}_{\,\,\,Q}\,\de^{P}_{\,\,\,N}
\Bigg].
\label{ppole4}
\eea
The third order pole is summarized by 
\bea
&& J^I_{\,\,\, J}(z) \, J^{K}_{\,\,\,L}\, J^{M}_{\,\,\,N}\,
J^{P}_{\,\,\,Q}(w)\Bigg|_{\frac{1}{(z-w)^3}} =
\Bigg[ (-1)^{d_J d_M+1}\,
  \de^{I}_{\,\,\,L}\, \de^{K}_{\,\,\,N}\, \de^M_{\,\,\, J}\, J^{P}_{\,\,\,Q}
  +  \de^{I}_{\,\,\,L}\, \de^{K}_{\,\,\,N}\, \de^M_{\,\,\, Q}\, J^{P}_{\,\,\,J}
  \nonu \\
  && +
  (-1)^{(d_P+d_Q)(d_M+d_J)+1} \,
  \de^{I}_{\,\,\,L}\, \de^{K}_{\,\,\,N}\, \de^P_{\,\,\, J}\,
  J^{M}_{\,\,\,Q}+
(-1)^{(d_N+d_M)(d_K+d_J)+1} \,
  \de^{I}_{\,\,\,L}\, \de^{M}_{\,\,\,J}\, \de^K_{\,\,\, Q}\,
  J^{P}_{\,\,\,N}  \nonu \\
   && +
  (-1)^{(d_N+d_M)(d_K+d_J)+(d_P+d_Q)(d_K+d_N)} \,
  \de^{I}_{\,\,\,L}\, \de^{M}_{\,\,\,J}\, \de^P_{\,\,\, N}\,
  J^{K}_{\,\,\,Q}\nonu \\
  && +
(-1)^{(d_K+d_J)(d_N+d_M)+d_J d_P+1} \,
  \de^{I}_{\,\,\,L}\, \de^{K}_{\,\,\,Q}\, \de^P_{\,\,\, J}\,
  J^{M}_{\,\,\,N} \nonu \\
  && + (-1)^{(d_L+d_K)(d_I+d_J)+1} \, \Big( (-1)^{d_L d_M +1}\,
 \de^{K}_{\,\,\,J}\, \de^{I}_{\,\,\,N}\, \de^M_{\,\,\, L}\,
 J^{P}_{\,\,\,Q} + \de^{K}_{\,\,\,J}\, \de^{I}_{\,\,\,N}\, \de^M_{\,\,\, Q}\,
 J^{P}_{\,\,\,L}
 \nonu \\
 && + (-1)^{(d_P+d_Q)(d_M+d_L)+1} \,
 \de^{K}_{\,\,\,J}\, \de^{I}_{\,\,\,N}\, \de^P_{\,\,\, L}\,
 J^{M}_{\,\,\,Q} +
 (-1)^{(d_N+d_M)(d_I+d_L)+1} \,
 \de^{K}_{\,\,\,J}\, \de^{M}_{\,\,\,L}\, \de^I_{\,\,\, Q}\,
 J^{P}_{\,\,\,N} \nonu \\
 &&+
 (-1)^{(d_N+d_M)(d_I+d_L)+(d_P+d_Q)(d_I+d_N)}\,
  \de^{K}_{\,\,\,J}\, \de^{M}_{\,\,\,L}\, \de^P_{\,\,\, N}\,
  J^{I}_{\,\,\,Q} \nonu \\
  && + (-1)^{(d_I+d_L)(d_N+d_M)+d_L d_P+1}\,
 \de^{K}_{\,\,\,J}\, \de^{I}_{\,\,\,Q}\, \de^P_{\,\,\, L}\,
  J^{M}_{\,\,\,N}  
  \Big) \nonu \\
  &&+
  (-1)^{(d_I+d_J)(d_K+d_L)} \, \Big( (-1)^{d_J d_P+1}\,
   \de^{I}_{\,\,\,N}\, \de^{M}_{\,\,\,Q}\, \de^P_{\,\,\, J}\,
   J^{K}_{\,\,\,L} \nonu \\
   && +
   (-1)^{(d_N+d_M)(d_I+d_J)+d_N d_P} \,
   \de^{M}_{\,\,\,J}\, \de^{I}_{\,\,\,Q}\, \de^P_{\,\,\, N}\,
   J^{K}_{\,\,\,L}\Big)  \Bigg](w).
\label{ppole3}
\eea
Finally, the second order pole is described by
\bea
&& J^I_{\,\,\, J}(z) \, J^{K}_{\,\,\,L}\, J^{M}_{\,\,\,N}\,
J^{P}_{\,\,\,Q}(w)\Bigg|_{\frac{1}{(z-w)^2}} =
\Bigg[ (-1)^{d_J d_K+1}\,
\de^{I}_{\,\,\,L}\, \de^{K}_{\,\,\,J}\, J^M_{\,\,\, N}\, J^{P}_{\,\,\,Q}
+\de^{I}_{\,\,\, L}\, \de^{K}_{\,\,\,N}\, J^M_{\,\,\, J}\, J^{P}_{\,\,\,Q}
\nonu \\
&& +
(-1)^{(d_N+d_M)(d_K+d_J)+1}\,
\de^{I}_{\,\,\,L}\, \de^{M}_{\,\,\,J}\, J^K_{\,\,\, N}\, J^{P}_{\,\,\,Q}
+ (-1)^{(d_K+d_J)(d_N+d_M)}\,
\de^{I}_{\,\,\,L}\, \de^{K}_{\,\,\,Q}\, J^M_{\,\,\, N}\, J^{P}_{\,\,\,J}
\nonu \\
&& +
 (-1)^{(d_K+d_J)(d_N+d_M+d_P+d_Q)+1}\,
\de^{I}_{L}\, \de^{P}_{\,\,\,J}\, J^M_{\,\,\, N}\, J^{K}_{\,\,\,Q}
\nonu \\
&&+ (-1)^{(d_L+d_K)(d_I+d_J)+1}\,\de^{K}_{\,\,\,J}  \,
\Big( \de^{I}_{\,\,\,N}\, J^M_{\,\,\, L}\, J^{P}_{\,\,\,Q}+
(-1)^{(d_N+d_M)(d_I+d_L)+1} \,
\de^{M}_{\,\,\,L}\, J^I_{\,\,\, N}\, J^{P}_{\,\,\,Q}\nonu \\
&& +
(-1)^{(d_I+d_L)(d_N+d_M) } \, \de^{I}_{\,\,\,Q}\, J^M_{\,\,\, N}\, J^{P}_{\,\,\,L}
+ (-1)^{(d_I+d_L)(d_N+d_M+d_P+d_Q)+1} \,
\de^{P}_{\,\,\,L}\, J^M_{\,\,\, N}\, J^{I}_{\,\,\,Q}
\Big)\nonu \\
&&+
(-1)^{(d_I+d_J)(d_K+d_L)}\,J^{K}_{\,\,\,L}  \,
\Big((-1)^{d_J d_M+1}\,
\de^{I}_{\,\,\,N}\, \de^M_{\,\,\, J}\, J^{P}_{\,\,\,Q}+
\de^{I}_{\,\,\,N}\, \de^M_{\,\,\, Q}\, J^{P}_{\,\,\,J}\nonu \\
&& +
(-1)^{(d_P+d_Q)(d_M+d_J)+1}\,
\de^{I}_{\,\,\,N}\, \de^P_{\,\,\, J}\, J^{M}_{\,\,\,Q}+
(-1)^{(d_N+d_M)(d_I+d_J)+1}\, \de^{M}_{\,\,\,J}\, \de^I_{\,\,\, Q}\,
J^{P}_{\,\,\,N}  \nonu \\
&&+
(-1)^{(d_N+d_M)(d_I+d_J)+(d_P+d_Q)(d_I+d_N)} \,
\de^{M}_{\,\,\,J}\, \de^P_{\,\,\, N}\, J^{I}_{\,\,\,Q}
\nonu \\
&&+
(-1)^{(d_I+d_J)(d_N+d_M)+d_J d_P +1}
\, \de^{I}_{\,\,\,Q}\, \de^P_{\,\,\, J}\, J^{M}_{\,\,\,N}
\Big)
\Bigg](w).
\label{ppole2}
\eea
Therefore, the complete OPE is given by
Appendix (\ref{ppole4}),
Appendix (\ref{ppole3}),
Appendix (\ref{ppole2}) and (\ref{pole1}).

We can also express the various (anti)commutator relations
by using the above OPE. See the reference \cite{CFT} for explicit
formula.
Let us consider the first OPE in (\ref{s-2ope}) having an extra
generator.
It is obvious to obtain that the third order pole is
given by  $\de^{b}_a \,
{\cal{P}}^{
  \dot{\al} }_{\,\,\, \beta}$ from Appendix (\ref{ppole3})
and the second order pole is given by
$-\frac{1}{2} \, \de^{b}_a \,
\pa \, {\cal{P}}^{
  \dot{\al} }_{\,\,\, \beta} + {\cal V}^{b \, \dot{\al}}_{\,\,\, a \, \beta}$
with $
{\cal V}^{b \, \dot{\al}}_{\,\,\, a \, \beta}\equiv
-3 \, \de^{b}_{a} \, {\cal V}\, {\cal{P}}^{
  \dot{\al} }_{\,\,\, \beta} -3 \, {\cal Q}^{b}_{\,\,\,\beta}\,
\dot{{\cal Q}}^{\dot{\al}}_{\,\,\,a} + \frac{3}{2}\, \de^b_a \, \pa \,
{\cal{P}}^{
  \dot{\al} }_{\,\,\, \beta}$ from Appendix (\ref{ppole2}).
Here we intentionally
split the second order pole into the descendant of the
weight-$1$ operator $\de^{b}_a \,
{\cal{P}}^{
  \dot{\al} }_{\,\,\, \beta}$ and the (quasi)primary operator.
The first order pole is again given in (\ref{s-2ope}).

Then we obtain the following anticommutator relation
by using the formula in \cite{CFT} or performing the
two contour integrals in conformal field theory explicitly
\bea
\{ (\dot{{\cal Q}}^{\dot{\al}}_{\,\,\,a})_m, ({\cal{W}}^{
    b }_{\,\,\, \beta})_n\} & = &
\frac{1}{2} m (2 m+n)\, \de^{b}_a \,
({\cal{P}}^{
  \dot{\al} }_{\,\,\, \beta})_{m+n} + m \,
({\cal V}^{b \, \dot{\al}}_{\,\,\, a \, \beta})_{m+n} \nonu \\
& + &
\de^{b}_a\, ({\cal{W}}^{
  \dot{\al} }_{\,\,\, \beta})_{m+n} +
(\widehat{{\cal W}}^{b \, \dot{\al}}_{\,\,\, a \, \beta})_{m+n}.
\label{finalcomm}
\eea
Note that the coefficients, $\frac{1}{2} m (2 m+n)$, $m$, $1$ and $1$,
appearing in the right hand side of Appendix (\ref{finalcomm})
hold for any (anti)commutator relations we are considering in the
OPEs between the weight-$1$ operator and the weight-$3$ operator.
The nonzero central terms can appear in the corresponding (anti)commutator
relations.
We should subtract the right descendant terms with
coefficient $-\frac{1}{2}$ in the second order pole explained before
in order to use the above general behavior.
The weight-$3$ operator is not a quasiprimary operator, in general,
from Appendix (\ref{tjjj}). In order to use the formula in \cite{CFT},
we should check the quasiprimary condition on the weight-$3$ operator.

Compared to the result of
\cite{SS,Vasiliev2001},
the first three terms of Appendix (\ref{finalcomm})
should appear and the last term reflects the new generator
coming from the worldsheet symmetry algebra.
We expect that all the other (anti)commutator relations like as
Appendix
(\ref{finalcomm}) with possible central terms or new generators
can be obtained and
they (without new generators)
with some normalizations
should appear in $hs(2,2|4)$
in the work of
\cite{SS,Vasiliev2001}.
Although we observe that there are no vanishing terms of the
right hand sides in Appendix
(\ref{finalcomm}) under the restriction of wedge
modes, it is an open problem to check whether the possibility of
vanishings for the right hand sides in the (anti)commutator
relations under the wedge constraints
(when we consider other OPEs for higher weights)
arises or not.

 %%%%%%%%%%%%%%%%%%%%%%%%%%%%%%%%%%%%%%%%%%%%%%%%%%%%%%%%%%%%%%%%%%%%%%%%%%%
%%%%%%%%%%%%%%%%%%%%%%%%%%%%%%%%%%%%%%%%%%%%%%%%%%%%%%%%%%%%%%%%%%%%%%%%%%

\end{document}